\documentclass[a4paper,11pt]{article}

\usepackage{amsthm}
\usepackage{amsmath}
\usepackage{amssymb}
\usepackage{amsfonts}
\usepackage{amstext}
\usepackage{xspace}
\usepackage{graphicx}
\usepackage{mathrsfs} 
\usepackage{vmargin}

\setmarginsrb{1in}{1in}{1in}{1in}{0pt}{0pt}{0pt}{6mm}

\newtheorem{theorem}{Theorem}
\newtheorem{lemma}{Lemma}
\newtheorem{claim}{Claim}
\newtheorem{corollary}{Corollary}
\newtheorem{definition}{Definition}
\newtheorem{observation}{Observation}
\newtheorem{proposition}{Proposition}

\newtheorem{redrule}{Reduction Rule}

\newcommand{\tw}{{\mathbf{tw}}}

\newcommand{\lcac}[1]{{\mbox{\bf LCA-closure}(#1)}}
\newcommand{\forget}{\mbox{\bf forget}}

\newcommand{\pmin}{{\sc min-CMSO}}

\newcommand{\pmax}{{\sc max-CMSO}}

\newcommand{\h}[1]{\end{document}}

\newtheorem{define}{Definition}
\usepackage{boxedminipage}

\newcommand{\fd}{{\sc $p$-$\mathcal{F}$-Deletion}}
\newcommand{\ofd}{{\sc $\mathcal{F}$-Deletion}}
\newcommand{\ffd}{{\sc $\mathcal{F}$-Deletion}}

\newcommand{\defparproblem}[4]{
  \vspace{1mm}
\noindent\fbox{
  \begin{minipage}{0.96\textwidth}
  \begin{tabular*}{\textwidth}{@{\extracolsep{\fill}}lr} #1  & {\bf{Parameter:}} #3 \\ \end{tabular*}
  {\bf{Input:}} #2  \\
  {\bf{Question:}} #4
  \end{minipage}
  }
  \vspace{1mm}
}

\title{{\sc Planar-${\cal F}$ Deletion}: Approximation, Kernelization and Optimal FPT Algorithms}
\author{
{\large\sc Fedor V. Fomin \thanks{University of Bergen, Norway. \texttt{fomin@ii.uib.no}}}
\and {\large\sc Daniel Lokshtanov\thanks{University of California, USA. \texttt{dlokshtanov@cs.ucsd.edu}} }
\and {\large\sc Neeldhara Misra\thanks{Institute of Mathematical Sciences, India. \texttt{neeldhara@imsc.res.in}}}
\and {\large\sc Saket Saurabh\thanks{Institute of Mathematical Sciences, India. \texttt{saket@imsc.res.in}}}
}

\date{}
\begin{document}
\maketitle

\thispagestyle{empty}
\begin{abstract}
Let ${\cal F}$ be  a finite set of graphs. In  the \ofd{} problem,  we are given an $n$-vertex graph $G$ and an integer $k$ as input, and asked whether at most $k$ vertices can be deleted from $G$ such that the resulting graph does not contain a graph from ${\cal F}$ as a minor. \ofd{} is a generic problem and by selecting different sets of forbidden minors ${\cal F}$, one can obtain  various fundamental   problems such as {\sc Vertex Cover}, {\sc Feedback Vertex Set} or {\sc Treewidth $\eta$-Deletion}. 

In this paper we obtain a number of generic algorithmic results about \ofd{}, when ${\cal F} $ contains at least one planar graph. The highlights of our work are
 \begin{itemize}
\item A constant factor approximation algorithm for the optimization version of \ofd{};  
\item A linear time and single exponential parameterized algorithm, that is, an algorithm running in time   $O(2^{O(k)} n)$, for the parameterized version of \ofd{} where all graphs in ${\cal F}$ are connected;
\item A polynomial kernel for parameterized \ofd.
\end{itemize}
These algorithms  unify, generalize, and improve  a multitude of results in the literature. Our main results have several direct applications, but also the methods we develop on the way have applicability beyond the scope of this paper. Our results -- constant factor approximation, polynomial kernelization and FPT algorithms -- are stringed together by a common theme of  polynomial time preprocessing.

\end{abstract}

\newpage
\setcounter{page}{1}

\section{Introduction}
%!TEX root=PFD10PAGE.tex

 Let $\mathscr{G}$ be the set of all finite undirected graphs and let $\mathscr{L}$ be the family of all finite subsets of  $\mathscr{G}$.
  Thus   every element ${\cal F} \in \mathscr{L}$ is a finite set of graphs and
%  Let  ${\cal F} \in \mathscr{F}$ be a  set  containing at least one planar graph. 
%  
%Let  $\cal F$ be a finite set of graphs containing at least one planar graph. 
 throughout the paper we assume that  ${\cal F} $ is explicitly given.   In this paper we study the following \fd{} problem. 
%In  the  \fd{} problem, we are given an $n$-vertex graph
%$G$ and an integer $k$ as input, and asked whether at most $k$ vertices
%can be deleted from $G$ such that the resulting graph does not contain a graph from   ${\cal
%F}$ as a minor.  More precisely, the problem is defined as follows:

\smallskip

\defparproblem{\fd{}}{A graph $G$  and a non-negative integer $k$.}{$k$}{Does there exist $S \subseteq V(G)$, $|S| \leq k$,  
 such that $G \setminus S$ contains no graph from ${\cal F}$ as a minor?}

%\begin{center}
%
%\begin{boxedminipage}{.96\textwidth}
%
%%\textsc{$p$--\jc{}--Deletion}
% \fd{}
%
%\begin{tabular}{ r l }
%\textit{~~~~Instance:} & A graph $G$  and a non-negative integer $k$. \\
%\textit{Parameter:} & $k$\\
%\textit{Question:} & Does there exist $S \subseteq V(G)$, $|S| \leq k$, \\
%~~~~~~~~~~~~~~~~~~ & such that $G \setminus S$ contains no graph from ${\cal F}$ as a minor?
% \\
%\end{tabular}
%
%\end{boxedminipage}
%
%\end{center}

\smallskip

\noindent 
The \fd{} problem defines  a wide subclass of node (or vertex) removal problems studied from the 1970s. By the classical theorem of  Lewis and Yannakakis  \cite{LewisY80}, deciding if  removing at most $k$ vertices results with a subgraph with property $\pi$ is  NP-complete for every non-trivial property $\pi$.
By a celebrated result of Robertson and Seymour, every \fd{} problem is non-uniformly  fixed-parameter tractable (FPT). 
That is, for every $k$ there is an algorithm solving the problem  in time  $O(f(k) \cdot n^3)$ \cite{RobertsonS13}.  The 
importance of the result comes from the fact that it simultaneously gives FPT algorithms for a variety of important 
problems such as {\sc Vertex Cover}, {\sc Feedback Vertex Set}, {\sc Vertex Planarization},  etc.  
It is conceivable that meta theorems for vertex deletion problems might be formulated by addressing problems that are expressible in logics such as first order and monadic second order. However, since these capture problems that are known 
to be intractable, for example {\sc Independent Set} or {\sc Dominating Set}, we do not expect to have a theorem 
that guarantees tractability for vertex deletion problems through this route. Therefore, 
the systematic study of the \fd{} problems is the more promising way forward to obtain meta-theorems 
for vertex removal problems on general undirected graphs.
%The 
%  systematic study of the \fd{} problems is a way forward to obtain  meta-theorems  
%  for vertex removal problems on general undirected graphs.

In this paper we show that when  ${\cal F}\in \mathscr{L}$ contains at least one planar graph, it is possible to obtain 
 a number of generic results advancing known tractability borders of  \fd{}. 
  The case when $\cal F$ contains a planar graph, while being considerably more restricted than the general case, already encompasses a number of the well-studied  
 instances of \fd{}. For  example, when ${\cal F}=\{K_2\}$, a complete graph on two
vertices, this is the {\sc Vertex Cover} problem. When ${\cal F}=\{C_3\}$, a cycle on three
vertices, this is the {\sc Feedback Vertex Set} problem. Another fundamental problem, which is a special case of  \fd{}, is 
 {\sc Treewidth $\eta$-Deletion} or  {\sc
$\eta$-Transversal} which is  to delete
at most $k$ vertices to obtain a graph of treewidth at most $\eta$. Since any graph of treewidth $\eta$ excludes a 
$(\eta+1)\times (\eta+1)$ grid as a minor, we have that the set $\cal F$ of forbidden minors of treewidth 
$\eta$ graphs contains a planar graph. 
%minor of the$(\eta+1)\times (\eta+1)$ times grid and this minor is planar
%Indeed, the class of graphs of treewidth at most $\eta$ can be characterized by a
%finite set of forbidden minors, and because no graph of treewidth $\eta$ has an $(\eta+1)
%\times (\eta+1)$ grid as a minor, we can always add this grid, a planar graph, to $\cal F$. 
 {\sc Treewidth $\eta$-Deletion}  plays important role in 
 generic efficient polynomial time approximation schemes based on Bidimensionality
Theory~\cite{FominLRS11,FominLS12}. Among other examples of   \fd{} that can be found in the literature on approximation and parameterized algorithms, are 
%Examples of  other studied  special variants of  \fd{}   are
  the cases of ${\cal F}$ being 
$\{K_{2,3}, K_4\}$, $ \{K_4\}$, $\{\theta_c\}$,  and $ \{K_{3}, T_2\}$, which correspond to removing vertices to
obtain an  outerplanar graph, a series-parallel graph,  a diamond graph,   and a graph  of pathwidth one,  respectively. 
%Here $K_{i,j}$ is a
%complete bipartite graph with bipartitions of sizes $i$ and $j$, $K_i$ is a complete graph
%on $i$ vertices, and $T_2$ is the graph obtained from subdividing every edge of $K_{1,3}$
%once. % In the literature, these problems are known as {\sc Outerplanar Deletion Set} and {\sc
%%Pathwidth One Deletion Set} respectively. 
% Other deletion-based   problems also
%turn out to be special cases of \fd{}, for instance: {\sc Diamond Hitting Set}, where ${\cal F}$ is a graph with two vertices and three parallel edges, {\sc
%Pathwidth $\eta$-Deletion} (deleting at most $k$ vertices to obtain a graph of pathwidth
%at most $\eta$). 
%
% 
% 
% 

The main algorithmic contributions of our work is the following set of results for \fd{}  for the case  
when  $\cal F$ contains a planar graph:
\begin{itemize}
\setlength{\itemsep}{-2pt}
\item A constant factor approximation algorithm   for an optimization version of \fd{};  
\item A linear time and single exponential parameterized algorithm for \fd{} when all graphs in ${\cal F}$ are connected, that is, an algorithm running in time $O(2^{O(k)} n)$, where $n$ is the input size;
\item A polynomial kernel for \fd.
\end{itemize}
We use $\mathscr{F}$ to denote the subclass of $\mathscr L$ such that every ${\cal F}\in \mathscr{F}$ contains a planar graph. 
%For a finite set of graphs  $\cal{F}$, let ${\cal{G}}_{{\cal{F}},k}$ be a class of graphs such that for every $G\in {\cal{G}}_{{\cal{F}},k}$ there is a subset of vertices $S$ of size at most $k$ such that $G\setminus S$ has no minor from $\cal{F}$.
% The family ${\cal{G}}_{{\cal{F}},k}$ is minor closed and we show that the size of any minimal obstruction---the minor minimal graph that  is not contained in ${\cal{G}}_{{\cal{F}},k}$---has size polynomial in $k$.  Our results have several direct applications, and on the way to proving out results we develop new methods that have applicability beyond the scope of \fd{}.

%
%. The case of \fd{}, when ${\cal F} \in \mathscr{F}$  
%encompasses several important and 
%fundamental problems. 

%\medskip We provide {\bf \textsl{constant factor approximation, polynomial kernelization}} and {\bf\textsl{optimal FPT}} algorithms for \fd{} stringed together with a common theme of  polynomial time preprocessing.

\paragraph{Methodology.} 
All  our  results -- constant factor approximation, polynomial kernelization and   FPT algorithms for  \fd{} -- have a common theme of  polynomial time preprocessing. Preprocessing as a strategy for coping with hard problems is universally applied 
in practice and the notion of {\em kernelization} in parameterized complexity  provides a mathematical framework for analyzing the quality of preprocessing strategies. 
  In parameterized complexity each problem instance comes with a parameter $k$ and  a central notion in parameterized complexity is {\em fixed parameter tractability (FPT)}. This means, for a given instance $(x,k)$, solvability in time $f(k)\cdot p(|x|)$, where $f$ is an arbitrary function of $k$ and $p$ is a polynomial in the input size. 
  The parameterized problem is said to admit a {\it polynomial kernel} if there is a polynomial time algorithm (the degree of polynomial is independent of $k$), called a {\em kernelization} algorithm, that reduces the input instance down to an instance with size bounded by a polynomial $p(k)$ in $k$, while preserving the answer.

\begin{figure}[t]
\begin{center}
\includegraphics[width=14cm]{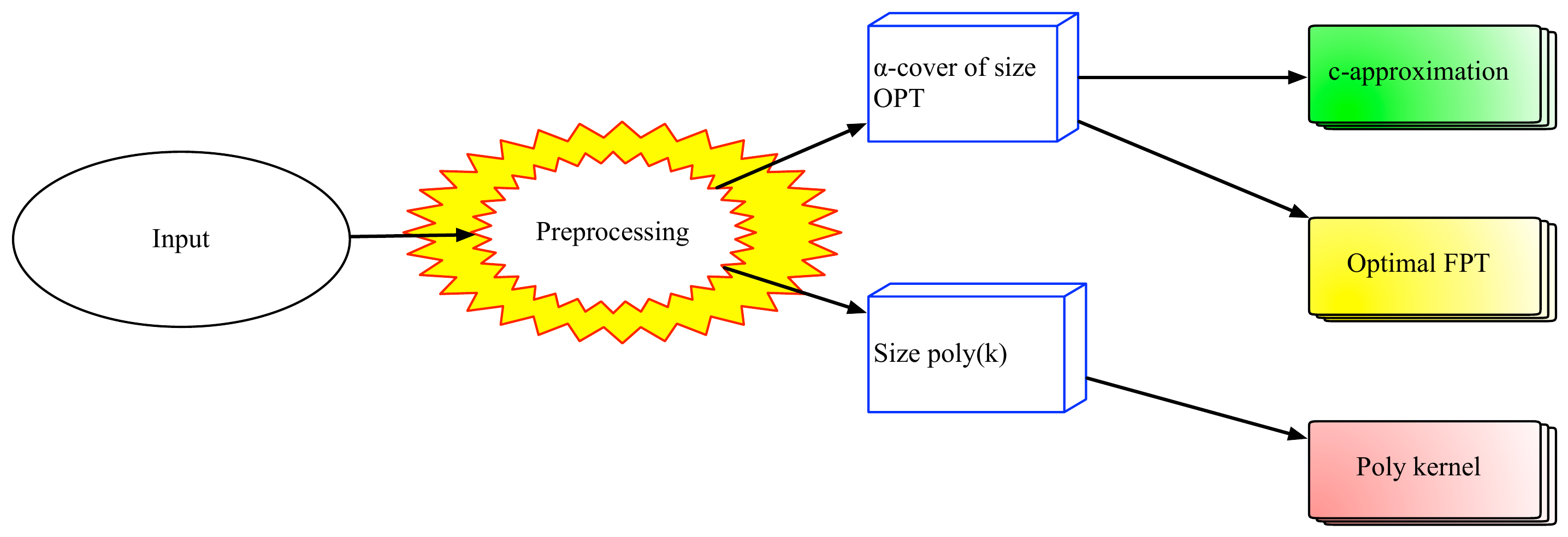}
\caption{General  view of our approach }\label{fig:diagram}
\end{center}
\end{figure}

 Thus the goal of kernelization is to apply reduction rules such that  the size of the reduced instance can be upper bounded by a function of the parameter. However, if we want to use preprocessing for approximation or  FPT algorithms, it is not necessary 
 that  the size of the reduced instance has to be upper bounded. What we need is a preprocessing procedure that allows us  to 
 navigate the solution search space efficiently.  Our first contribution is a notion of preprocessing  that is 
  geared towards  approximation and FPT algorithms. This notion relaxes the demands of kernelization and thus it is possible 
  that a larger set of problems may admit this simplification procedure, when compared to kernelization. 
 %For approximation and FPT algorithms  we provide preprocessing  with less ambitious  goals--just \emph{simplification} of the %input.  
For approximation and FPT algorithms, we use the notion of {\em $\alpha$-cover} as a measure of good preprocessing. 
For  $0<\alpha\leq 1$, we say that a vertex subset $S\subseteq V(G)$ is an \emph{$\alpha$-cover}, if 
the sum of vertex degrees     $\sum_{v \in S} d(v) $ is at least  $2 \alpha |E(G)|$.
  For example, every vertex cover of a graph is also a $1$-cover. 
  The  defining property of this preprocessing is that the equivalent simplified instance of the problem
 admits some optimal solution which is also  an $\alpha$-cover.  If we succeed with this goal, then for  an edge  selected 
 uniformly at random,   with a constant probability   at least  one of its endpoints belong to some optimal solution. Using this as a basic step, we 
 %Based on that we are able to 
 can construct approximation and FPT algorithms. 
 But how to  achieve this kind of preprocessing?
 
%One of the main techniques used in this work is the extension of the
%protrusion theory employed in \cite{H.Bodlaender:2009ng,F.V.Fomin:2010oq} for
%obtaining meta-kernelization theorems for problems on sparse graphs like
%planar and $H$-minor-free graphs, to general graphs. Bodlaender et
%al.~\cite{H.Bodlaender:2009ng} were first to use protrusion techniques (or
%rather graph reduction techniques) to obtain kernels, but the idea of using
%graph replacement for algorithms has been there for long time. The idea of
%graph replacement for algorithms dates back to Fellows and
%Langston~\cite{FellowsL89}. Arnborg et al.~\cite{ArnborgCPS93} essentially
%showed that protrusions exist for many problems on graphs of bounded
%treewidth, and gave safe ways of reducing graphs. Using this, Arnborg et
%al.~\cite{ArnborgCPS93} obtained a linear time algorithm for MSO expressible
%problems on graphs of bounded treewidth. Bodlaender and
%Fluiter~\cite{BodlaenderF96a,BodlaendervA01a,Fluiter97} generalized these
%ideas in several ways --- in particular, they applied it to some optimization
%problems. It is also important to mention the work of Bodlaender and
%Hagerup~\cite{BodlaenderH98}, who used the concept of graph reduction to
%obtain parallel algorithms for MSO expressible problems on bounded treewidth
%graphs.

To achieve our goals we use the idea of graph replacement dating back to Fellows and
Langston~\cite{FellowsL89}.  Precisely, what we use is the modern notion of ``protrusion reduction"  
that has been recently employed  in \cite{H.Bodlaender:2009ng,FominLST10} for obtaining meta-kernelization 
theorems for problems on sparse graphs like planar graphs, graphs of bounded genus \cite{BodlaenderFLPST09}, 
graphs excluding a fixed graph as a minor or induced subgraph \cite{FominLST10,FominLMPS11},  or graphs excluding a fixed graph as a topological minor~\cite{abs-1201-2780}.  In this method, we find a large protrusion -- 
a graph of small treewidth and small boundary -- and then the preprocessing rule replaces this 
protrusion by a protrusion of constant size. One repeatedly applies this until no longer possible. Finally, by using 
combinatorial arguments one upper bounds the size of the reduced induced (a graph without large protrusion). 
%In this 
%method we decompose the input graph into protrusions, that is, into graphs of small treewidth and small boundary. 
% %Our  preprocessing  can be seen as an adaptation for our purposes algebraic graph reduction techniques of Arnborg et al. \cite%{ArnborgCPS93}, see also \cite{BodlaenderFLPST09,FellowsL89_focs}. 
%%To apply preprocessing, we use decomposition  theorem into protrusions, i.e.  graphs of small treewidth and small separation. 
%Then preprocessing rules replace sufficiently large protrusions by protrusions of constant sizes. Finally, by using 
%combinatorial arguments one upper bounds the size of the reduced induced (a graph not containing protrusions) . 
The FPT algorithms use the replacement technique developed in \cite{BodlaenderFLPST09,FominLMPS11}, while for approximation algorithm we need another type of protrusion reduction. The reason why the normal protrusion replacement 
does not work for approximation algorithms is the same as why the NP-hardness reduction is not always an approximation preserving reduction. While the normal protrusion replacement works fine for preserving exact solutions, we needed a notion of protrusion reduction that also preserves approximate solutions. To this end, we develop a new notion  of {\em lossless protrusion reduction}, and show that several problems do admit lossless protrusion reductions. We exemplify the usefulness of the new concept by  obtaining constant factor approximation algorithms for \ofd{}. These FPT and approximation algorithms are obtained by showing that  solutions to the instances 
 of the problem that do not contain protrusion form an $\alpha$-cover for some fixed constant $\alpha$.   

%and show how to use lossless protrusions for approximation algorithms. 
%The problem   is that protrusion replacement technique works fine to keep exact solutions but for approximation algorithms we have to guarantee that protrusion replacement  keeps approximation. To settle this issue, we have to  develop new notion  of lossless protrusions, and show how to use lossless protrusions for approximation algorithms. 

Our final result is about kernelization for \fd{}. 
While  protrusion replacements work well for constant factor approximation and optimal FPT algorithms, we do not know how to use this technique for kernelization algorithms for \fd. The technique was developed and used successfully  for kernelization algorithms 
  on  sparse graphs~\cite{BodlaenderFLPST09,FominLST10}
  but there are several limitations of this techniques which do not allow to use it on general graphs. Even for a sufficiently simple case of \fd, namely when $\cal{F}$ is a graph with two vertices and constant number of parallel edges, to apply protrusion replacements we have to do a lot of  additional work to reduce large vertex degrees in a graph  \cite{FominLMPS11}. We do not know how to push these techniques for more complicated families  families  $\cal{F}$ and therefore, employ a different strategy. 
   The new conceptual contribution here is the notion of a \emph{near-protrusion}. Loosely speaking, a near-protrusion is a subgraph which can become a protrusion in the future, after removing some vertices of some optimal solution. The usefulness of   near-protrusions is that they allow to find an irrelevant edge, i.e.,  an edge  which removal does not change the problem. However, finding an irrelevant edge is highly non-trivial, and it requires the usage of well-quasi-ordering  for 
   graphs of bounded treewdith  and bounded boundary as a subroutine.

As far as we are equipped with new tools and concepts:   $\alpha$-cover, lossless protrusion reduction and pseudo-protrusions, we are able to 
proceed with algorithms for  \fd{}. These algorithms  unify and generalize a multitude of results in the literature. In what follows we survey 
earlier results in each direction and discuss our results.

%the following results: 
   % This brings us to randomized approximation and FPT algorithms for \fd{} and \ofd. 
   
%  \subsection{Previous Results and Our Work}
%
%  
%
%\iffalse

\vspace{-.3cm}
\paragraph{Approximation.} In the optimization version of \fd{}, we want to 
compute the minimum set $S$, which removal leaves input graph $G$   ${\cal F}$-minor-free. We denote this optimization problem by \ofd.
Characterising graph  properties  for which the corresponding vertex deletion problem can be approximated 
within a constant factor is a long standing open problem in approximation algorithms  \cite{Yannakakis94}.
In spite of long history of research, we are still far from a complete understanding. 
Constant factor approximation algorithms  for  {\sc Vertex Cover}  are known since 1970s \cite{NemT74,Bar-YehudaE81}.
 Lund and Yannakakis observed that 
 the	vertex deletion	problem	for	any	hereditary	property	with	a finite number of minimal forbidden subgraphs can be approximated with a constant ratio \cite{LundY93}. They also conjectured that 
   for every nontrivial, hereditary property with an infinite number of minimal forbidden subgraphs, the  vertex deletion problem cannot be approximated with constant ratio.  However, it appeared later that   {\sc  Feedback Vertex Set} admits  a constant factor approximation~\cite{BarYGJ98,BafnaBF99}
  and thus  the dividing line of   approximability lies somewhere else.
  On a related matter, Yannakakis  \cite{Yannakakis79} showed that  approximating  the number of vertices to delete in order to obtain  \emph{connected} graph with some property $\pi$ within factor $n^{1-\varepsilon}$ is NP-hard, see
\cite{Yannakakis79} for the definition of the property $\pi$.  This result holds for very wide class of  properties, in particular for  properties  being acyclic and  outerplanar.  There was no much progress on approximability/non-approximability of vertex deletion  problems until recent work of Fiorini et al.  \cite{Fiorini:2009ipco} who gave  a constant factor approximation algorithm for 
 \fd{} for the case when ${\cal F}$ is a diamond graph, i.e., a graph with two vertices and three parallel edges.
 
Our  first contribution   is the theorem stating that  every graph property $\pi$ expressible by a finite set of forbidden minors containing at least one planar graph, the	vertex deletion	problem for property $\pi$ admits a constant factor approximation algorithm. In other words, we prove the following theorem
\begin{theorem}\label{thm:approx_thm}
For every set  ${\cal F} \in \mathscr{F}$, 
\ofd{}
admits a  randomized constant  ratio approximation algorithm. 
 %can be approximated with a constant factor ratio.  
 \end{theorem}
%\begin{theorem}\label{thm:approx_thm}
%For every set  ${\cal F} \in \mathscr{F}$ containing a planar graph, 
%\ofd{}
%admits a  randomized constant  ratio approximation algorithm. 
% %can be approximated with a constant factor ratio.  
% \end{theorem}
Let us remark that for all known constant factor approximation algorithms of vertex deletion to a  hereditary property $\pi$, property $\pi$ is either
  characterized by an   finite number of minimal forbidden subgraphs or by finite number of forbidden minors, one of which is planar.   Theorem~\ref{thm:approx_thm} together with the result of Lund and Yannakakis, not only encompass all known  vertex deletion problems with constant factor approximation ratio but significantly extends   known tractability 
  borders for   such types of problems.

%%%%%%%Theorem~\ref{thm:approx_thm} has a number of interesting applications. ADD HERE BIDIMENSIONALITY,  what else?

%
%
%This is very interesting, in view of the fact that there is for example no
% known approximation algorithm with bounded worst-case ratio for the
% feedback-node set (or any other problem of the class), whereas the node cover problem can be easily approximated within ratio 2, but also because it would shed more  
% light into the nature of NP-complete problems from the combinatorial point of view and into their
%behaviour with respect to approximation algorithms \cite{LewisY80}.
%
%
% 
%Lund-Yannakakis
%\cite{LundY93}: The	vertex deletion	problem	for	any	hereditary	property	with	a finite number of minimal forbidden subgraphs can be approximated with a constant ratio. They also conjectured that 
%  for every nontrivial, hereditary property with an infinite number of minimal forbidden subgraphs the  vertex deletion problem cannot be approximated with constant ratio. 

\vspace{-.3cm}
\paragraph{Kernelization.}
%%Preprocessing as a strategy for coping with hard problems is universally applied 
%%in practice and the notion of {\em kernelization} provides a mathematical framework for analyzing the quality of preprocessing strategies. We consider parameterized problems, where every instance $I$ comes with a {\em parameter} $k$. Such a problem is said to admit a {\em polynomial kernel} if every instance $(I,k)$ can be reduced in polynomial time to an equivalent instance with both size and parameter value bounded by a polynomial in $k$.
The study of kernelization is a major research frontier of 
Parameterized Complexity and many important recent advances in the area
are on kernelization. These include 
general results
showing that  certain classes of parameterized problems have polynomial kernels~\cite{Alon:2010vp,H.Bodlaender:2009ng,FominLST10,KratschW12}. 
The recent development of a framework for ruling out polynomial kernels under
certain complexity-theoretic
assumptions~\cite{BDFH08,Dell:2010sh,FS08}    has added a new dimension to
the field and strengthened its connections to classical complexity.  For overviews of kernelization we  refer to surveys~\cite{Bodlaender09,GN07SIGACT}  and to the corresponding chapters in books on Parameterized Complexity 
\cite{FlumGroheBook,Niedermeierbook06}.

While the initial interest in kernelization was driven mainly by practical applications, the notion of kernelization turned out to be very important in theory as well. It is well known, see e.g. \cite{DowneyF99}, that a parameterized problem is fixed parameter tractable, or belongs to the class FPT,  if and only if it has a (not necessarily polynomial) kernel. Kernelization enables us to classify problems within the class FPT further, based on the sizes of the problem kernels. 
One of the fundamental challenges in the area is the possibility of characterising general classes of parameterized problems possessing kernels of polynomial sizes.  

Polynomial kernels for several special cases of \fd{} were studied in the literature.  
 Different kernelization techniques were invented for {\sc Vertex Cover}, eventually resulting in a
 $2k$-sized vertex kernel~\cite{AFLS07,ChenKJ01,DFRS04,Hochbaum:1994kl}.
 For the kernelization of {\sc  Feedback Vertex Set}, there has been a sequence of
 dramatic improvements starting from an $O(k^{11})$ vertex kernel by
 Buragge et al.~\cite{BEFLMR2006}, improved to  $O(k^3)$ by
 Bodlaender~\cite{Bod07}, and then finally to $O(k^2)$ by
 Thomass\'e~\cite{T09}.  A polynomial kernel for \fd{}  for class  ${\cal F}$ consisting of a graph with two vertices and several parallel edges is given in \cite{FominLMPS11}.
 Philip et al.~\cite{PhilipRS09} and Cygan et al.~\cite{CyganPPW10} obtained polynomial kernels for {\sc  Pathwidth 1-Deletion}. Our next theorem generalizes all these kernelization results.   

 \begin{theorem}\label{thm:kernel_thm}
 For every set  ${\cal F} \in \mathscr{F}$,  
 %containing a planar graph, 
\fd{} admits a polynomial kernel.
\end{theorem}

In fact, we prove  more general result---the kernelization algorithm of Theorem~\ref{thm:kernel_thm} always outputs a minor of the input graph. 
This has interesting combinatorial consequences. By Robertson and Seymour theory every non-trivial minor-closed class of graphs can be characterized by a finite set of  forbidden minors or \emph{obstructions}. 
While Graph Minors Theory insures that many interesting graph properties have finite obstructions sets, these seem to be disappointingly huge in many cases. 
There are a number of results that bound the size of the obstructions for specific minor closed families of graphs. Fellows and Langston \cite{FellowsL89_focs,FellowsL94}
suggested  a systematic method of computing the obstructions sets for many natural properties,  see also the recent work of Adler et al.  
\cite{AdlerGK08}.
 Bodendiek and Wagner  gave bounds on sizes of obstructions of genus at most $k$
\cite{BodendiekW89}, later improved by   Djidjev and   Reif
\cite{DjidjevR91}.  Gupta and Impagliazzo studied bounds     
on the size of a planar intertwine of two given planar graphs
 \cite{GuptaI91}. Lagergren \cite{Lagergren98} showed that the number of edges in every obstruction to a graph  of treewidth $k$ is at most double exponential in   $O(k^5)$.  
Dvo{\v{r}}{\'a}k et al.  \cite{Dvorak11} provide similar bound on obstructions to graphs of tree-depth at most $k$. Dinneen and  Xiong have shown that the number of vertices in connected  obstruction  for graphs with vertex cover at most  $k$  is at most  $2k+1$ \cite{Dinneen02VC}. Obstructions for graphs with feedback vertex set of size at most $k$ is discussed in the work of Dinneen et al. 
 \cite{DinneenCF01}.

For a finite set of graphs  $\cal{F}$, let ${\cal{G}}_{{\cal{F}},k}$ be a class of graphs such that for every $G\in {\cal{G}}_{{\cal{F}},k}$ there is a subset of vertices $S$ of size at most $k$ such that $G\setminus S$ has no minor from $\cal{F}$.  As a corollary of kernelization algorithm, we obtain the following combinatorial result.  
% Every minor of graph $G\in {\cal{G}}_{{\cal{F}},k}$ is also in $G\in {\cal{G}}_{{\cal{F}},k}$, and 
%by Robertson and Seymour theory class $  {\cal{G}}_{{\cal{F}},k}$has  $O(f(k))$ {obstructions}, where $f$ is some function  of $k$ only and  constants hidden in the big-Oh depend only on  $\cal{F}$.  
% By Theorem~\ref{thm:kernel_thm}, when  $\cal{F}$ contains a planar graph,   every minimal obstruction for   $ {\cal{G}}_{{\cal{F}},k}$ is of size  polynomial in $k$. 
 \begin{theorem}\label{thm:kernel_thm}
 For every set  ${\cal F} \in \mathscr{F}$,  
 %containing a planar graph, 
every minimal obstruction for   $ {\cal{G}}_{{\cal{F}},k}$ is of size  polynomial in $k$. 
\end{theorem}

%We also find the similarity in the main steps between our kernelization algorithm and the method of Fellows and Langston  \cite{FellowsL89_focs} for computing obstructions  to be very interesting. Both approaches are based on  the use of bounded treewidth search spaces and exploiting  of finite-index congruences on boundaried graphs based on test sets. This similarity indicates a possibility of strong links between polynomial kernelization and sizes of obstructions.\footnote{FF: I am not sure in this paragraph, perhaps it is too vague} 

\vspace{-.5cm}
\paragraph{Fast FPT Algorithms.} 
The study of parameterized problems proceeds in several steps. The first step is to establish if the  problem on hands is fixed parameter tractable or not. If the problem is in FPT, then the next steps are to identify if the problem admits a polynomial kernel and to find the fastest possible FPT algorithm solving the problem. 
The running time  of every FPT algorithm is   $O(f(k) n^c)$, that is, the product of a super-polynomial function $f(k) $ depending only on the parameter $k$ and polynomial $n^c$, where $n$ is the input size and $c$ is some constant.  Both steps, minimizing super-polynomial function $f(k)$ and minimizing  the exponent
$c$ of the polynomial part, are important parts in the design and analysis of parameterized algorithms. 

 The \fd{} problem was introduced by Fellows and Langston~\cite{FellowsL88}, who gave a
non-constructive algorithm running in time $O(f(k)\cdot n^{2})$ for some function
$f(k)$~\cite[Theorem $6$]{FellowsL88}.  This result was improved by Bodlaender
\cite{Bodlaender97} to  $O(f(k)\cdot n)$,  for  $f(k)=2^{2^{O(k \log k)}}$. 
%The approach of Bodlaender is to use his algorithm to compute 
There is a substantial amount of  work  on  improving the exponential function $f(k)$ for special cases of  \fd{}.
For the \textsc{Vertex Cover} problem
the existence of single-exponential algorithms is well-known
since almost the beginnings of the field of Parameterized
Complexity, the current best algorithm being by
Chen et al.~\cite{ChenKX10}. Randomized parameterized single exponential algorithm for 
\textsc{Feedback Vertex Set} was given by Becker et al.
\cite{BeckerBRG00-Ra} but 
   existence of deterministic single-exponential algorithms for
\textsc{Feedback Vertex Set} was open for a while and it took some time and 
discovery of iterative compression  \cite{ReedSmithVetta04}
 to reduce the running time to  $2^{O(k)}n^{O(1)}$~\cite{CaoCL10,ChenFLLV08,CNP+11,DehneFLRS07,GuoGHNW06,RamanSS06}.  The current  champion for
\textsc{Feedback Vertex Set} are the deterministic algorithm of 
Cao et al.  \cite{CaoCL10} with running time $O(3.83^k k n^2)$ and the randomized of Cygan et al.
with running time time~$3^kn^{O(1)}$ ~\cite{CNP+11}. Recently, Joret et al.~\cite{JoretPSST11} showed 
that  \fd{} for ${\cal F}
=\{ \theta_c\}$, where  $\theta_c$ is the graph with two vertices and $c$ parallel edges,     can be solved in time $2^{O(k)}n^{O(1)}$ for 
every fixed $c$. Philip et al.~\cite{PhilipRV10} studied {\sc Pathwidth $1$-Deletion} 
and obtained an algorithm with running time $O(7^kn^{2})$ that was later improved to $O(4.65^k n^{O(1)})$ in~\cite{CyganPPW10}.
Kim et al. \cite{KimPG12} gave a single exponential algorithm for  ${\cal F}
=\{ K_4\}$.  Unless Exponential Time Hypothesis (ETH) fails~\cite{CCF+05,ImpagliazzoPZ01}, single exponential dependence on the parameter $k$ is asymptotically the best bound one can obtain for \fd{}, and thus our next theorem provides asymptotically  optimal bounds on the exponential function of the parameter   and polynomial contribution of the input. 

We call a family ${\cal F} \in \mathscr{F}$ connected if every graph in $\cal F$ is connected. 

  \begin{theorem}\label{thm:fpt_thm1}
  For every connected  set  ${\cal F} \in \mathscr{F}$  containing a planar graph,  there is a randomized algorithm solving 
\fd{}   in time $O(c^k n)$ for some constant $c>1$. 
\end{theorem}

%We also give a deterministic algorithm 
We finally give a deterministic algorithm for \fd{}. Surprisingly, our algorithm does not use iterative compression but  is based on branching on degree sequences. 
%using branching based on degree 

  \begin{theorem}\label{thm:fpt_thm2}
  For every connected set  ${\cal F} \in \mathscr{F}$   containing a planar graph,  
\fd{}  is solvable   in time $O(c^k n\log^2{n})$  for some constant $c>1$. 
\end{theorem}

\section{Preliminaries}

%!TEX root=PlanarFDeletionKernel.tex

In this section we give various definitions which we use in the paper. We use $V(G)$ to denote the vertex set of a graph 
$G$, and $E(G)$ to denote the edge set. The degree of a vertex $v$ in $G$ is the number of edges
incident on $v$, and is denoted by $d(v)$.  A graph~$G'$ is a \emph{subgraph} of~$G$ if~$V(G') \subseteq V(G)$
and~$E(G') \subseteq E(G)$.  The subgraph~$G'$ is called an
\emph{induced subgraph} of $G$ if $E(G') = \{\{u,v\} \in E(G) \mid
u,v \in V(G')\}$. Given a subset $S\subseteq V(G)$ the subgraph
induced by $S$ is denoted by~$G[S]$.  The subgraph induced by $
V(G)\setminus S$ is denoted by $G\setminus S$.  We denote by
$N_G(S)$ the open neighborhood of $S$, i.e. the set of vertices in
$V(G)\setminus S$ adjacent to $S$. Whenever the graph $G$ is clear from the context, 
we omit the subscript in $N_G(S)$ and denote it only by $N(S)$. By $N[S]$ we denote $N(S)\cup S$. 
Let  $\cal F$ be a finite set of graphs. 
A vertex subset $S\subseteq V(G)$ of a graph $G$ is said to be a 
$\cal{F}$-{\em deletion set} if $G\setminus S$ does not contain any graphs in the family 
$\cal F$ as a minor.

\subsection{Parameterized algorithms and kernels.}
\label{paraak}
A parameterized problem $\Pi$ is a subset of $\Gamma^{*}\times \mathbb{N}$ for some finite alphabet $\Gamma$. An instance of a parameterized problem consists of $(x,k)$, where $k$ is called the parameter. We assume that $k$ is {\em given in unary} and hence $k\leq |x|$. A central notion in parameterized complexity is {\em fixed parameter tractability (FPT)} which means, for a given instance $(x,k)$, solvability in time $f(k)\cdot p(|x|)$, where $f$ is an arbitrary function of $k$ and $p$ is a polynomial in the input size. The notion of {\em kernelization} is formally defined as follows. 

\begin{definition}{\rm [\bf Kernelization]} 
Let $\Pi\subseteq \Gamma^{*}\times \mathbb{N}$ be a parameterized problem and $g$ be a computable function.
We say that $\Pi$ {\em admits a kernel of size $g$} if there exists an algorithm $K$, called  {\em kernelization algorithm}, or, in short, a {\em kernelization}, that given $(x,k)\in \Gamma^{*}\times \mathbb{N},$ outputs, in time polynomial in $|x|+k$, a pair $(x',k')\in \Gamma^{*}\times \mathbb{N}$ such that
\begin{itemize}
\item[(a)] $(x,k)\in \Pi$ 
if and only if $(x',k')\in \Pi$, and 
\item[(b)] $\max\{|x'|, k' \}\leq g(k)$.
\end{itemize}
When $g(k)=k^{O(1)}$ or $g(k)=O(k)$ then we say that 
$\Pi$ {\em admits a polynomial} or {\em linear kernel} respectively. If additionally $k' \leq k$ we say that the kernel is {\em strict}.
\end{definition}

\subsection{Treewidth.} \label{trewap}
Let $G$ be a graph. A {\em tree decomposition} of $G$ is a pair $T,\mathcal{X}=\{X_{t}\}_{t\in V(T)})$ where $T$ is a tree and ${\cal X}$ is a collection of subsets of $V(G)$ such that: 
\begin{itemize}
\item $\forall {e=uv \in E(G)}, \exists {t\in V(T)} : \{u,v\} \subseteq X_{t}$ and 
\item $\forall {v\in V(G)}, \ T[\{t\mid v\in X_{t}\}]$ is a non-empty connected subtree of $T$. 
\end{itemize}
We call the vertices of $T$ {\em nodes} and the sets in $\mathcal{ X}$ {\em bags} of the tree decomposition $(T,{\cal X})$.  The {\em width} of $(T,{\cal X})$ is equal to $\max_{}\{|X_t|-1\mid {t\in V(T)}\}$ and the {\em treewidth} of $G$ is the minimum width over all tree decompositions of $G$. 

A {\em nice tree decomposition} is a pair $(T,{\cal X})$ where $(T,{\cal X})$ is a tree decomposition such that $T$ is a rooted tree and the following conditions are satisfied: 
\begin{itemize}
\item Every node of the tree $T$ has at most two children; 
\item if a node $t$ has two children $t_1$ and $t_2$, then $X_t = X_{t_1} = X_{t_2}$; and
\item  if a node $t$ has one child $t_1$, then either $|X_t| = |X_{t_1}| + 1$  and $X_{t_1} \subset X_{t}$ (in this case we call $t_1$ {\em insert node}) or $|X_t| = |X_{t_1}| -1$ and $X_t \subset X_{t_1}$ (in this case we call $t_1$ {\em insert node}). 
\end{itemize}
It is possible to transform a given tree decomposition $(T,{\cal X})$ into a nice tree decomposition $(T',{\cal X}')$ in time $O(|V|+|E|)$~\cite{Bodlaender96}.

\subsection{Minors}
Given an edge $e=xy$ of a graph $G$, the graph  $G/e$ is obtained from  $G$ by contracting the edge $e$, that is, the endpoints $x$  and $y$ are replaced by a new vertex $v_{xy}$ which  is  adjacent to the old neighbors of $x$ and $y$ (except from $x$ and $y$).  A graph $H$ obtained by a sequence of edge-contractions is said to be a \emph{contraction} of $G$.  We denote it by $H\leq_{c} G$. A graph $H$ is a {\em minor} of a graph $G$ if $H$ is the contraction of some subgraph of $G$ and we denote it by $H\leq_{m} G$. We say that a graph $G$ is {\em $H$-minor-free} when it does not contain $H$ as a minor. We also say that a graph class ${\cal G}$ is {\em $H$-minor-free} (or, excludes $H$ as a minor) when all its members are $H$-minor-free. It is well-known~\cite{RobertsonS-V} that if $H \leq_m G$ then $tw(H) \leq tw(G)$. We will also use the following fact about excluding planar graphs as minors.

\begin{proposition}\label{prop:planar_exclude_treewidth} There is a constant $c$ such that for every planar $H$ and graph $G$ with $tw(G) \geq 2^{c|V(H)|^3}$, $H$ is a minor of $G$.
\end{proposition}

\subsection{$t$-Boundaried graphs and Gluing.}
A $t$-boundaried graph is a graph $G$ and a set $B \subset V(G)$ of size at most $t$ with each vertex $v \in B$ having a label $\ell_G(v) \in \{1, \ldots, t\}$. Each vertex in $B$ has a unique label. We refer to $B$ as the {\em boundary} of $G$. For a $t$-boundaried $G$ the function $\delta(G)$ returns the boundary of $G$. Two $t$-boundaried graphs $G$ and $H$ are isomprphic if there is a bijection $f$ from $V(G)$ to $V(H)$ such that $uv \in E(G) \iff f(u)f(v) \in E(H)$, for every $v \in \delta(G)$ we have $f(v) \in \delta(H)$ and $\ell_G(v) = \ell_H(f(v))$. Specifically $f$ is an isomorphism between $G$ and $H$ in the normal graph sense, but additionally $f$ respects the labels of the border vertices. Observe that a $t$-boundaried graph may have no boundary at all. A graph $G$ is isomorphic to a $t$-boundaried graph $H$ of there is an isomorphism between $G$ and $H$. 

Two $t$-boundaried graphs $G_1$ and $G_2$ can be {\em glued} together to form a graph $G = G_1 \oplus G_2$. The gluing operation takes the disjoint union of $G_1$ and $G_2$ and identifies the vertices of $\delta(G_1)$ and $\delta(G_2)$ with the same label. If there are vertices $u_1$, $v_1 \in \delta(G_1)$ and $u_2$, $v_2 \in \delta(G_2)$ such that $\ell_{G_1}(u_1) = \ell_{G_2}(u_2)$ and $\ell_{G_1}(v_1) = \ell_{G_2}(v_2)$ then $G$ has vertices $u$ formed by unifying $u_1$ and $u_2$ and $v$ formed by unifying $v_1$ and $v_2$. The new vertices $u$ and $v$ are adjacent if $u_1v_1 \in E(G_1)$ or $u_2v_2 \in E(G_2)$.

The {\em boundaried gluing operation} $\oplus_{\delta}$ is similar to the normal gluing operation, but results in a $t$-boundaried graph rather than a graph. Specifically $G_1 \oplus_\delta G_2$ results in a $t$-boundaried graph where the graph is $G = G_1 \oplus G_2$ and a vertex is in the boundary of $G$ if it was in the boundary of $G_1$ or $G_2$. Vertices in the boundary of $G$ keep their label from $G_1$ or $G_2$. Both for gluing and boundaried gluing we will refer to $G_1 \oplus G_2$ or $G_1 \oplus_\delta G_2$ as the {\em sum} of $G_1$ and $G_2$, and $G_1$ and $G_2$ are the {\em terms} of the sum.

For a $t$-boundaried graph $G$ and boundary vertex $v \in \delta(G)$, {\em forgetting} $v$ results in a $t$-boundaried graph identical to $G$, except that $v$ is no longer a boundary vertex. All other boundary vertices keep their labels. Forgetting a non-boundary vertex leaves the graph unchanged, as does forgetting a vertex that is not in the vertex set of $G$. Forgetting a set $S \subseteq \delta(G)$ of vertices means forgetting all vertices in the set. The function $\forget(G, S)$ returns the $t$-boundaried graph resulting from forgetting $S$ in $G$.

We will frequently need to construct $t$-boundaried graphs from subgraphs of a graph $G$. For a graph $G$ and two disjoint vertex sets $P$ and $B$ we define $G_P^B$ to be the $t$-boundaried graph $G[P \cup B]$ with boundary $B$. The labelling of the border $B$ is chosen in a manner independent of $P$ - such that if $P_1$, $P_2$ and $B$ are disjoint then $G_{P_1}^B \oplus_\delta G_{P_2}^B = G_{P_1 \cup P_2}^B$.

%!TEX root=PlanarFDeletionKernel.tex

\subsection{Monadic Second Order Logic (MSO)}
\label{countmsop}
The syntax of MSO on graphs includes the logical connectives $\vee$, $\land$, $\neg$, 
$\Leftrightarrow $,  $\Rightarrow$, variables for 
vertices, edges, sets of vertices and sets of edges, the quantifiers $\forall$, $\exists$ that can be applied 
to these variables, and the following five binary relations: 
\begin{enumerate}

\item $u\in U$ where $u$ is a vertex variable 
and $U$ is a vertex set variable;\item  $d \in D$ where $d$ is an edge variable and $D$ is an edge 
set variable; \item  $\mathbf{inc}(d,u)$, where $d$ is an edge variable,  $u$ is a vertex variable, and the interpretation 
is that the edge $d$ is incident on the vertex $u$; \item $\mathbf{adj}(u,v)$, where  $u$ and $v$ are 
vertex variables $u$, and the interpretation is that $u$ and $v$ are adjacent; \item  equality of variables representing vertices, edges, set of vertices and set of edges. 
\end{enumerate}

Many common graph-theoretic notions such as vertex degree, connectivity,
planarity, being acyclic, and so on, can be expressed in MSO, as can be
seen from introductory
expositions~\cite{BorieParkerTovey1992,HandbookGraphGrammars1997Ch5}. 
%Of
%particular interest to us are minimization graph
%problems,where we are given a
% \subseteq {\cal G}_{g} \times \mathbb{N}
%graph $G$ as input and the objective is to compute a minimum 
%vertex/edge set $S$ such that the
%MSO-expressible predicate $P_\Pi(G,S)$ is satisfied. 

\paragraph{$H$ minor-models.}

Recall that a $t$-boundaried graph $H$ is a minor of a $t$-boundaried graph $G$ if (a $t$-boundaried graph isomorphic to) $H$ can be obtained from $G$ by deleting vertices or edges or contracting edges, but never contracting edges with both endpoints being boundary vertices. Let $V(H) = \{h_1, \ldots, h_c\}$, and let $B_G := \{b_1^G, \ldots b_t^G\}$ and $B_H := \{b_1^H, \ldots b_t^H\}$ denote $\delta(G)$ and $\delta(H)$ respectively. Then, the formulation that $H \leq_m G$ is given by $\phi(G,H,B_G,B_G)$:

\begin{align}
\label{cmso:boundariedminor}
\nonumber\phi(G,H,B_G,B_H)\equiv\exists X_1,\ldots,X_c\subseteq V(G)[\\
\nonumber & \bigwedge_{i \neq j} (X_i\cap X_j=\emptyset) \wedge \bigwedge_{1 \leq i \leq c}Conn(G,X_i)\wedge\\
\nonumber & \bigwedge_{(h_i,h_j) \in E(H)}\exists x \in X_i \wedge y \in X_j [(x,y) \in E(G)] \wedge \\
\nonumber & \bigwedge_{(b_i^H \in B_H)} \exists x \in X_i [x = b_i^G]  \\
%\nonumber & \forall u,v \in B_G [ u \in X_i \implies v \notin X_i] \\
]\end{align}

\subsection{Finite Integer Index and Protrusions}
\label{fiind}
%\paragraph{Finite Integer Index}
For a parameterized problem  $\Pi$ and two $t$-boundaried graphs $G_1,G_2\in{\cal G}$, we say that $G_1 \equiv _{\Pi} G_2$ if there exists a constant $c$ such that for every $t$-boundaried graph $G$ and for every integer $k$, $(G_1 \oplus G, k) \in \Pi$ if and only if $(G_2 \oplus G, k+c) \in \Pi$. For every $t$, the relation $\equiv_\Pi$ on $t$-boundaried graphs is an equivalence relation, and we call $\equiv_\Pi$ the {\em canonical equivalence relation} of $\Pi$. We say that a problem $\Pi$ has {\em Finite Integer Index} if for every $t$, $\equiv_\Pi$ has finite index on $t$-boundaried graphs. Thus, if $\Pi$ has finite integer index then for every $t$ there is a finite set ${\cal S}$ of $t$-boundaried graphs for every $t$-boundaried graph $G_1$ there exists $G_2 \in \cal{S}$ such that $G_2 \equiv _{\Pi} G_1$. Such a set ${\cal S}$ is called a {\em set of representatives for $(\Pi, t).$}  We will repeatedly make use of the following proposition.

\begin{proposition}[\cite{BodlaenderFLPST09}]\label{prop:cfdelisfii} For every connected ${\cal F} \in \mathscr{F}$, \ffd{} has finite integer index. \end{proposition}

%\todo{Daniel: In the current meta kernel paper i think we say ALL. We should point out the bug and thank bart in acknowledgements?}

\paragraph{Protrusions and Protrusion Replacement}
For a graph $G$ and $S\subseteq V(G)$, we define $\partial_G(S)$ as the set of vertices in $S$ that have a neighbor in $V(G) \setminus S$. For a set $S \subseteq V(G)$ the {\em neighbourhood} of $S$ is $N_G(S) = \partial_G(V(G) \setminus S)$. When it is clear from the context, we omit the subscripts. A {\em $r$-protrusion} in a graph $G$ is a set $X \subseteq V$ such that $|\partial(X)|\leq r$ and $\tw(G[X]) \leq r$. If $G$ is a graph containing a $r$-protrusion $X$ and $X'$ is a $r$-boundaried graph, the act of {\em replacing} $X$ by $X'$ means replacing $G$ by $G_{V(G) \setminus X}^{\partial(X)} \oplus X'$.

A {\em protrusion replacer} for a parameterized graph problem $\Pi$ is a family of algorithms, with one algorithm for every constant $r$. The $r$'th algorithm has the following specifications. There exists a constant $r'$ (which depends on $r$) such that given an instance $(G, k)$ and an $r$-protrusion $X$ in $G$ of size at least $r'$, the algorithm runs in time $O(|X|)$ and outputs an instance $(G', k')$ such that $(G',k') \in \Pi$ if and only if $(G,k) \in \Pi$, $k' \leq k$ and $G'$ is obtained from $G$ by replacing $X$ by a $r$-boundaried graph $X'$ with less than $r'$ vertices. Observe that since $X$ has at least $r'$ vertices and $X'$ has less than $r'$ vertices this implies that $|V(G')| < |V(G)|$. The following proposition is the driving force of~\cite{BodlaenderFLPST09} and the starting point for our algorithms.

\begin{proposition}[\cite{Fluiter97,BodlaenderFLPST09}]\label{prop:fiiReplace} Every parameterized problem with finite integer index has a protrusion replacer. \end{proposition}
Together, Propositions~\ref{prop:cfdelisfii} and~\ref{prop:fiiReplace} imply that for every connected ${\cal F} \in \mathscr{F}$, \ffd{} has a protrusion replacer.

Bodlaender et al.~\cite{H.Bodlaender:2009ng} were first to use protrusion techniques (or rather graph reduction techniques) to obtain kernels, but the idea of graph replacement for algorithms dates back to Fellows and Langston~\cite{FellowsL89}. Arnborg et al.~\cite{ArnborgCPS93} essentially showed that protrusion replacement is possible and useful for many problems on graphs of bounded treewidth, and gave safe ways of reducing graphs. Using this, Arnborg et al.~\cite{ArnborgCPS93} obtained a linear time algorithm for MSO expressible problems on graphs of bounded treewidth. Bodlaender and Fluiter~\cite{BodlaenderF96a,BodlaendervA01a,Fluiter97} generalized these ideas in several ways --- in particular, they lifted graph reduction techniques to optimization problems and proved Proposition~\ref{prop:fiiReplace}.  It is also important to mention the work of Bodlaender and Hagerup~\cite{BodlaenderH98}, who used the concept of graph reduction to obtain parallel algorithms for MSO expressible problems on bounded treewidth graphs.

\subsubsection{Least Common Ancestor-Closure of Sets in Trees.}
For a rooted tree $T$ and vertex set $M$ in $V(T)$ the least common ancestor-closure ({\em LCA-closure}) $\lcac{M}$ is obtained by the following process. Initially, set $M' = M$. Then, as long as there are vertices $x$ and $y$ in $M'$ whose least common ancestor $w$ is not in $M'$, add $w$ to $M'$. When the process terminates, output $M'$ as the LCA-closure of $M$. The following folklore lemma summarizes two basic properties of LCA closures.
\begin{lemma}\label{lem:lcaClosure} Let $T$ be a tree, $M \subseteq V(T)$ and $M' = \lcac{M}$. Then $|M'| \leq 2|M|$ and for every connected component $C$ of $T \setminus M'$, $|N(C)| \leq 2$. 
\end{lemma}  
\begin{proof}
To prove that $|M'| \leq 2|M|$ make a tree $T'$ with vertex set $M'$, and for every vertex $v \in M'$ adding an edge to the lowermost ancestor of $v$ in $M'$ in the tree $T$. Observe that in $T'$ all leaves are from $M$, since every vertex in $M' \setminus M$ is the least common ancestor of two vertices below it in $T$. Furthermore, for the same reason every vertex in $M' \setminus M$ has at least two decendants in $T'$. A standard counting argument for trees shows that the number of vertices with at least two decendants is at most the number of leaves. Hence $|M' \setminus M| \leq |M|$ and so $|M'| \leq 2|M|$.

We now prove that $|N(C)| \leq 2$. Suppose not, and let $r$ be the root of $P$. At most one of $C$'s neighbours is the parent of $r$ and hence at least two of $C$'s neighbours, say $u$ and $v$ are children of vertices in $C$. The vertices $u$ and $v$ are both in $M'$, and they are both descendents of $r$. But then the least common ancestor of $u$ and $v$ must lie in $C$ and hence is not in $M'$, contradicting the construction of $M'$. So we conclude that $|N(P)| \leq 2$.
\end{proof}

\section{A Randomized Algorithm for ``connected" \fd{}}\label{sec:firstAlgo}
In this section we give a randomized algorithm for \fd{} when every graph in ${\cal F}\in \mathscr{F}$ is connected. Recall that we call a family $\cal F$ connected if all the graphs in $\cal F$ is connected. We will show that for every connected ${\cal F}$ the algorithm runs in polynomial time, with the exponent of the polynomial depending on the family ${\cal F}$. If the input graph has a ${\cal F}$-deletion set  of size at most $k$, the algorithm will detect a ${\cal F}$-deletion set  of size at most $k$ with probability at least $\frac{1}{c^k}$. Here the constant $c$ depends on ${\cal F}$. The algorithm has no false positives - we show that if it reports that a ${\cal F}$-deletion set  of size at most $k$ exists then $G$ indeed has such a set. 

In the following sections we will progressively improve the algorithm; first we give an implementation of the algorithm with expected running time $O(n \cdot OPT)$. Then we show how to modify the (sped up) algorithm so that it not only decides whether $G$ has a ${\cal F}$-deletion set  of size at most $k$, but also outputs a solution. We show that if $G$ has a ${\cal F}$-deletion set  of size at most $k$, the algorithm will output a solution of size $k$ with probability at least $\frac{1}{c^k}$. We then proceed to show that this algorithm in fact outputs constant factor approximate solutions with constant probability, yielding a constant factor approximation for \fd{} for connected $\cal F$ in expected $O(n \cdot OPT)$ time. The main structure of the improved algorithm remains the same as the one described here.

The first building block of our algorithm is a simple algorithm to reduce the input instance to an equivalent instance that does not contain any large protrusions with small border.
\begin{lemma}\label{lem:naiveReplace}
For every ${\cal F} \in \mathscr{F}$ and constants $r$ and $r'$ such that \fd{} has a protrusion replacer that reduces $r$-protrusions of size $r'$, there is an algorithm that  takes as input an instance $(G,k)$ of \fd{}, runs in $n^{O(r')}$ time and outputs an equivalent instance $(G',k')$ such that $|V(G')| \leq V(G)$, $k' \leq k$ and $G'$ has no $r$-protrusion of size at least $r'$.
\end{lemma}
\begin{proof}
It is sufficient to give a $n^{O(r')}$ time algorithm to find a $r$-protrusion $X$ in $G$ of size at least $r'$, if such a protrusion exists. If we had such an algorithm to find a protrusion we could keep looking for $r$-protrusions $X$ in $G$ of size at least $r'$, and if one is found replacing them using the protrusion replacer. Since each replacement decreases the number of vertices by one we converge to an instance $(G', k')$ with the desired properties after at most $n$ iterations.

To find an $r$-protrusion of size at least $r'$ observe that if such a protrusion exists, then there must be at least one such protrusion $X$ such that $G[X \setminus \partial(X)]$ has at most $r'$ connected components. Indeed, if $G[X \setminus \partial(X)]$ has more than $r'$ connected components then let $X'$ be $\partial(X)$ plus the union of any $r'$ components of $G[X \setminus \partial(X)]$. Now $X'$ is an $r$-protrusion of size at least $r'$ and  $G[X' \setminus \partial(X')]$ has at most $r'$ components. To find a $r$-protrusion $X$ of size at least $r'$ on at most $r'$ components, guess $\partial(X)$ and then guess which components of $G \setminus \partial(X)$ are in $X$. The size of the search space is bounded by $n^r \cdot n^{r'}$ and for each candidate $X$ we can test whether it is a protrusion in linear time using Bodlaender's linear time treewidth 
algorithm~\cite{Bodlaender96}.
\end{proof}

The second building block of our algorithm is a lemma whose proof we postpone until the end of this section. The lemma states that for any ${\cal F} \in \mathscr{F}$, if $G$ contains no large protrusions with small border then any feasible solution to \fd{} is incident to a linear fraction of the edges of $G$. Recall that an $\alpha$-cover in $G$ is a set $S$ such that $\sum_{v \in S} d(v) \geq \alpha \cdot \sum_{v \in V(G)} d(v) = 2\alpha \cdot m$.
\begin{lemma}\label{lem:noProtRatio} For every ${\cal F} \in \mathscr{F}$ there exist constants $r$ and $\alpha$ such that if a graph $G$ has no $r$-protrusion of size at least $r'$, then every ${\cal F}$-deletion set  $S$ of $G$ is a $\frac{\alpha}{r'}$-cover of  $G$.
\end{lemma}
We now combine Lemmata~\ref{lem:naiveReplace} and~\ref{lem:noProtRatio} to give a randomized algorithm for \fd{} for all  ${\cal F} \in \mathscr{F}$ such that each graph in $\cal F$ is connected. 
%In Algorithm~\ref{alg:slowRandFPT}, let $r$ be the constant as guaranteed by Lemma~\ref{lem:noProtRatio} and let $r'$ be the smallest integer such that the protusion replacer for \ffd{} reduces $r$-protrusions of size $r'$.
\begin{figure}[ht]
\begin{center}
\begin{boxedminipage}{.96\textwidth}

\noindent 
{\bf Randomized-FPT-beta}(($G$,$k$))\\
%Let $S:=\emptyset$, $i:=0$ $G_{i}=G$ and 
Set $G_{current}:=G$ and $k_{current}:=k$.  \\
%Select the constant $r$ as guaranteed by Lemma~\ref{lem:noProtRatio}. \\
%Select $r'$ so that the protusion replacer for \ffd{} reduces $r$-protrusions of size $r'$. \\
While $(G_{current}$ is not ${\cal F}$-free$)$ do as follows:
\begin{enumerate}
\item If $k_{current} \leq 0$ return that $G$ does not have a $k$-sized ${\cal F}$-deletion set .
\item~\label{alg1step:reduce} Apply Lemma~\ref{lem:naiveReplace} on $(G_{current},k_{current})$ and obtain an equivalent instance $(G',k')$. 
\item~\label{alg1step:select} Pick a vertex $u \in V(G)$ at random with probability $\frac{d(u)}{2m}$. Set $G_{current}:=G'\setminus \{u\}$ and $k_{current}:=k'-1$ 
\end{enumerate}
Return that $G$ has a $k$-sized ${\cal F}$-deletion set .
\end{boxedminipage}
\caption{In Algorithm {\bf Randomized-FPT-beta}, let $r$ be the constant as guaranteed by Lemma~\ref{lem:noProtRatio} and let $r'$ be the smallest integer such that the protusion replacer for \ffd{} reduces $r$-protrusions of size $r'$.\label{fig:randFPTslow}}
\end{center}
\end{figure}

\begin{lemma}\label{lem:simpleAlg} Algorithm~\ref{fig:randFPTslow} runs in polynomial time, if $(G,k)$ is a ``no'' instance it outputs ``no'' and if  $(G,k)$ is a ``yes'' instance it outputs ``yes'' with probability at least $\frac{1}{c^k}$ where $c$ is a constant depending only on ${\cal F}$. 
\end{lemma}
\begin{proof}
Since each iteration runs in polynomial time and reduces the number of vertices in $G_{current}$ by at least one, Algorithm~\ref{fig:randFPTslow} runs in polynomial time. Furthermore, Step~\ref{alg1step:reduce} reduces the instance to an equivalent instance with $k'\leq k_{current}$ and Step~\ref{alg1step:select} only decreases $k_{current}$ when it puts a vertex into the solution. Hence when the algorithm outputs ``yes'' then a $k$-sized ${\cal F}$-deletion set  exists. It remains to show the last part of the statement.

We say that an iteration of Step~\ref{alg1step:select} is {\em successful} if there exists a ${\cal F}$-deletion set  $S$ of $G'$ with $|S| \leq k'$ such that the vertex $u$ selected in this step is in $S$. If the step is successfull then $S \setminus \{u\}$ is a ${\cal F}$-deletion set  of $G'$ of size at most $k'-1$. Thus, if the input graph $G$ has a $k$-sized ${\cal F}$-deletion set  and all the iterations of Step~\ref{alg1step:select} are successful then the algorithm maintains the invariant that $G_{current}$ has a ${\cal F}$-deletion set  of size at most $k_{current}$, and thus after at most $k$ iterations it terminates and outputs that $(G,k)$ is a ``yes'' instance. When Step~\ref{alg1step:select} is executed the graph $G'$ has no $r$-protrusions of size at least $r'$. Thus by Lemma~\ref{lem:noProtRatio} every ${\cal F}$-deletion set  set of $G'$ is an $\frac{\alpha}{r'}$-cover for a constant $\alpha$ depending only on ${\cal F}$. Hence the probability that $u$ is in a minimum size ${\cal F}$-deletion set  of $G'$ is at least $\frac{\alpha}{r'}$. We conclude that the probability that the first $k$ executions of Step~\ref{alg1step:select} are successful is at least $(\frac{\alpha}{r'})^k$ concluding the proof.
\end{proof}

Repeating the algorithm presented in Figure~\ref{fig:randFPTslow} $O(c^k)$ times yields a $O(2^{O(k)}n^{O(1)})$ time algorithm for \fd{} for all connected ${\cal F} \in \mathscr{F}$. However we are not entirely done with the proof of Lemma~\ref{lem:simpleAlg}, as it remains to prove Lemma~\ref{lem:noProtRatio}. In order to complete the proof we need to define protrusion decompositions.

\subsection{Protrusion Decompositions and Proof of Lemma~\ref{lem:noProtRatio}}
We recall the notion of a protrusion decomposition defined in~\cite{BodlaenderFLPST09} and show that if a graph $G$ has a set $X$ such that $\tw(G\setminus X)\leq d$, then it admits a protrusion decomposition for an appropriate value of the parameters. We then use this result to prove Lemma~\ref{lem:noProtRatio}.

\begin{define}{\rm [{\bf Protrusion Decomposition][\cite{BodlaenderFLPST09}]}}
A graph $G$ has an $(\alpha,\beta)${\em -protrusion decomposition} if $V(G)$ has a partition ${\cal P}=\{R_{0},R_{1},\ldots,R_{t}\}$ where 
\begin{itemize}
\item $\max\{t, |R_{0}|\}\leq \alpha$, 
\item each $N_{G}[R_i]$, $i\in\{1,\ldots,t\}$ is a $\beta$-protrusion of $G$, and
\item for all $i > 1$, $N[R_i] \subseteq R_0$.
\end{itemize}
We call the sets $R^{+}_{i}=N_{G}[R_i]$, $i\in\{1,\ldots,t\}$ {\em protrusions} of ${\cal P}$.
\end{define}

We now show that for every ${\cal F} \in \mathscr{F}$ every graph with an ${\cal F}$-deletion set  $X$ has an $(\alpha, \beta)$-protrusion decomposition where $\beta$ is constant and $\alpha = O(|N[X]|)$.

\begin{lemma}[Protrusion Decomposition Lemma]
\label{lem:prottfd}
If a $n$-vertex graph $G$ has a vertex subset $X$ such that $\tw(G\setminus X) \leq b$, then $G$ admits a $((4|N[X]|)(b+1), 2(b+1))$-protrusion decomposition. Furthermore, if we are given the set $X$ then this protrusion decomposition can be computed in linear time. Here $b$ is a constant. 
\end{lemma}
\begin{proof} 
We give a proof for the case when $X$ is explicitly given to us. The proof will automatically imply the existence of a $(4(b+1)|N(X)|,2b+2)$-protrusion decomposition of $G$ for the case when we are just guaranteed the existence of $X$. The algorithm starts by computing a nice tree decomposition $(T,{\cal B})$ of $G \setminus X$ with width at most $b$. Notice that since $b$ is a constant this can be done in linear time~\cite{Bodlaender96}. 

For every $v \in N(X)$ add a node $u$ in $T$ such that $v \in B_u$ to a set $M'$. We have that $M' \leq |N(X)|$. Let $M'$ be the set of marked nodes and set $M = \lcac{M'}$. By Lemma~\ref{lem:lcaClosure}, $M \leq 2|M'| \leq 2|N(X)|$. Let $Q_1, Q_2 \ldots Q_t$ be the connected components of $T \setminus Q$. Since $T$ is a binary tree $T \setminus M$ has at most $2|M|+1$ connected components, so $t \leq 4|N(X)|+1$. By Lemma~\ref{lem:lcaClosure} we have that for every $i \leq t$, $|N_T(Q_i)| \leq 2$. 

Define $R_0 = $X$ \cup \bigcup_{u \in M} B_u$ and for each $1 \leq i \leq t$ set $R_i = \bigcup_{u \in {Q_i}} B_u \setminus R_0$. Since every vertex of $G \setminus X$ appears in a bag of the tree-decomposition, $R_0, \ldots R_t$ forms a partition of $V(G)$. By construction we have that for every $i \geq 1$, $N(R_i) \subseteq R_0$ and $\tw(G[N[R_i]]) \leq b$. Furthermore, since $|N_T(Q_i)| \leq 2$ we have $|N(R_i)| \leq 2(b + 1)$. Thus $R_0 \ldots R_t$ form a $(\alpha, \beta)$-protrusion decomposition of $G$ where $\beta \leq 2(b+1)$ and $\alpha \leq \max(|R_0|,t) \leq (4|N[X]|)(b + 1)$. It is easy to implement a procedure that computes $R_0 \ldots R_t$ in this way in linear time. 
\end{proof}
We are now in a position to prove Lemma~\ref{lem:noProtRatio}
\begin{proof}[Proof of Lemma~\ref{lem:noProtRatio}.] We need to prove that for every ${\cal F} \in \mathscr{F}$ there exist constants $r$ and $\alpha$ such that if a graph $G$ has no $r$-protrusion of size at least $r'$, then every minimal ${\cal F}$-deletion set  $S$ of $G$ is a $\frac{\alpha}{r'}$-cover of $G$. By Proposition~\ref{prop:planar_exclude_treewidth} there exists a constant $\eta$ depending only on ${\cal F}$ such that $\tw(G \setminus S) \leq \eta$. By Lemma~\ref{lem:prottfd}, $G$ has a $((4|N[S]|)(\eta+1), 2(\eta+1))$-protrusion decomposition $R_0 \ldots R_t$. Set $r = 2(\eta+1)$ and suppose $G$ has no $r$-protrusions of size at least $r'$. Then $t \leq (4|N[S]|)(\eta+1)$, $|R_0| \leq (4|N[S]|)(\eta+1)$ and so $|V(G)| = |R_0| + \sum_i |R_i| \leq (4|N[S]|)(\eta+1)(r' + 1) \leq (8|N[S]|)(\eta+1)r'$. Since $\tw(G \setminus S) \leq \eta + 1$ it follows that $G \setminus S$ is $(\eta + 1)$-degenerate and so $\sum_{v \in V(G)\setminus S} d(v) \leq (8|N[S]|)(\eta+1)^2r'$. Set $\alpha = \frac{1}{18(\eta+1)^2}$ and observe that
$$\sum_{v \in V(G)} d(v) \leq \sum_{v \in S} d(v) + \sum_{v \in V(G) \setminus S} d(v) \leq  \sum_{v \in S} d(v) + (8|N[S]|)(\eta+1)^2r' \leq \frac{r}{\alpha} \cdot \sum_{v \in S} d(v).$$
The last inequality follows from the fact that there are no isolated vertex in $S$. 
\end{proof}

\section{Fast Protrusion Replacement}
%!TEX root=PlanarFDeletionApproxandAlgo.tex
What makes the polynomial factor of Algorithm~\ref{fig:randFPTslow} large is the algorithm of Lemma~\ref{lem:naiveReplace} to remove all large enough protrusions with small border size. In this section we give much faster algorithms that reduce ``almost all'' large protrusions with small border. We then show that reducing almost all protrusions instead of all protrusions is sufficient to obtain the conclusion of Lemma~\ref{lem:noProtRatio}. The ``fast protrusion reduction'' algorithms we design in this section are applicable to any problem that uses protrusion reducer, and hence they are useful well beyond the scope of this paper. We give two algorithms for fast protrusion replacement, a randomized algorithm and a slightly slower deterministic algorithm.

%%% ----------------------------------------------------------------------------
%%% ----------------------------------------------------------------------------
%%% The Randomized Fast Protrusion Replacer.
%%% ----------------------------------------------------------------------------
%%% ----------------------------------------------------------------------------

\paragraph{The Randomized Fast Protrusion Replacer.}
We now describe an algorithm that we call the {\em Randomized Fast Protrusion Replacer (RFPR)}. The algorithm works for parameterized graph problems $\Pi$ that have a protrusion replacer, takes as input an instance $(G, k)$ and outputs another instance $(G', k')$. Just as a normal protrusion replacer, the RFPR is actually a family of algorithms with one algorithm for each value of the integer $r$. We describe how the algorithm proceeds for a fixed value of $r$. Let $r'$ be the smallest integer such that the protrusion replacer for $\Pi$ replaces $r$-protrusions of size at least $r'$. 

The RFPR proceeds as follows. We select a random partition of $V(G)$ into $r+1$ sets $X_1, X_2, \ldots X_{r+1}$. For every $i \leq r+1$ we compute the connected components of $G[X_i]$ and add these components to a collection ${\cal C}'$. This results in a partition of $V(G)$ into ${\cal C}' = C_1', C_2', \ldots C_{t'}'$. Now, discard every component $C_i'$ such that $N(C_i') > r$ and every component $C_i$ such that $\tw(G[N[C_i]]) > r$. Discarding all of these components can be done in linear time - the only computationally hard step is to check whether the treewidth of the components is at most $r$, this can be done in linear time using Bodlaenders's algorithm~\cite{Bodlaender96}. Let ${\cal C}^* = C_1^*, \ldots , C_{t^*}^*$ be the remaining components.

For every $C_i^* \in {\cal C}^*$, $N[C_i^*]$ is a $r$-protrusion in $G$. However some, if not all of the components in ${\cal C}^*$ could have less than $r'$ vertices and so the protrusion replacer can't reduce them. However it could be possible to group some components in ${\cal C}^*$ with the same neighbourhood together such that their union is a protrusion that is large enough to be reduced. From ${\cal C}^*$ we will compute a collection ${\cal R}$ of disjoint vertex sets such that for every $R \in {\cal R}$, $N[R]$ is an $r$-protrusion in $G$ of size at least $r'$. Our aim is to compute such a set with $|{\cal R}|$ being large. For every component $C_i^* \in {\cal C}^*$ of size at least $r'$ we add $C_i^*$ to ${\cal R}$ and remove $C_i^*$ from ${\cal C}^*$. Let ${\cal C} = C_1 \ldots C_{t}$ be the remaining components. All components in ${\cal C}$ have size at most $r'$. Set $R_{big}$ to be the number of components $C_i^* \in {\cal C}^*$ on at least $r'$ vertices that are added to ${\cal R}$. 

Now we partition ${\cal C}$ into groups according to the neighbourhood of the components. Specifically we compute a partition of ${\cal C}$ into ${\cal Z}_1, \ldots {\cal Z}_q$ such that for every pair $C_i \in {\cal C}$, $C_{i'} \in {\cal C}$ such that $N(C_i) = N(C_{i'})$, $C_i$ and $C_{i'}$ are in the same ${\cal Z}_j$, while for every pair $C_i \in {\cal C}$, $C_{i'} \in {\cal C}$ such that $N(C_i) \neq N(C_{i'})$ we have $C_i \in {\cal Z}_j \rightarrow C_{i'} \notin {\cal Z}_j$. Such a partition can be computed in time $O(nr)$ because every component in ${\cal C}$ has at most $r$ neighbours; First we sort the neighbor lists of each component according to some ordering of the vertex set, for example an arbitrary labelling of the vertices from $1$ to $n$. Then we do $r$ stable bucket sorts on ${\cal C}$ sorting the components first on their first neighbour, then their second neighbor, etc.

Having computed the partitioning ${\cal Z}_1, \ldots {\cal Z}_q$ we now compute ${\cal R}$ as follows. As long as there is a ${\cal Z}_i$ such that $\sum_{{C_j} \in {{\cal Z}_i}} |C_j| \geq r'$ select a minimal collection ${\cal Z} \subseteq {\cal Z}_i$ such that $\sum_{{C_j} \in {\cal Z}} |C_j| \geq r'$. Add $\bigcup_{{C_j} \in {\cal Z}} C_j$ to ${\cal R}$ and remove the components of ${\cal Z}$ from ${\cal Z}_i$. This procedure can easily be implemented in linear time. This concludes the construction of ${\cal R}$.

Given ${\cal R}$ we proceed as follows, for a set $R \in {\cal R}$ we run the protrusion replacer for $\Pi$ on $(G,k)$ with protrusion $N[R]$. The protrusion replacer outputs an equivalent instance $(G^*,k^*)$ with $|V(G^*)| < |V(G)|$. Here $G^*$ is a graph where $R$ has been replaced by a smaller protrusion $R'$. Since all the sets in ${\cal R}$ are disjoint, the other sets in ${\cal R}$ are now $r$-protrusions in $G^*$ of size at least $r'$. Thus we can run the protrusion replacer on all the sets in ${\cal R}$. This takes time $\sum_{R \in {\cal R}} O(|R|) = O(n)$. Let $(G', k')$ be the instance obtained after running the protrusion replacer on all the sets in ${\cal R}$. The RFPR outputs the instance $(G', k')$. We collect a few simple facts about the RFPR 
in the following lemma.

\begin{lemma}\label{lem:rfprFacts} Given an instance $(G,k)$, the RFPR runs in time $O(n+m)$, computes a collection ${\cal R}$ of protrusions and and outputs an equivalent instance $(G',k')$, such that $|V(G')| \leq |V(G)| - |{\cal R}|$. Furthermore ${\cal R} \geq R_{big} + \left\lceil \sum_{i \leq q} \frac{\sum_{C \in {\cal Z}_i} |C| - (r' - 1)}{2(r' - 1)} \right\rceil$.
\end{lemma}

\begin{proof}
The instances $(G,k)$ and $(G',k')$ are equivalent because $(G',k')$ is obtained from $(G,k)$ by repetitive applications of a protrusion replacer. In the description of the algorithm we made sure that each individual stage of the algorithm runs in linear time. Finally, each application of the protrusion replacer reduces the size of the graph by at least one. We apply the protrusion replacer $|{\cal R}|$ times. Hence $|V(G')| \leq |V(G)| - |{\cal R}|$.

Finally, when the RFPR selects a minimal collection ${\cal Z} \subseteq {\cal Z}_i$ such that $\sum_{{C_j} \in {\cal Z}} |C_j| \geq r'$, since each $C_j \in {\cal Z}_i$ has size at most $r'$ it follows that  $\sum_{{C_j} \in {\cal Z}} |C_j| \leq 2(r'-1)$. Thus every time we add a set to ${\cal R}$, $\sum_{C \in {\cal Z}_i} |C|$ decreases by at most $2(r'-1)$. At the end when we can not add more sets to ${\cal R}$ we have that for every $i$, $\sum_{C \in {\cal Z}_i} |C| \leq r'$. This proves the last part of the statement of the lemma.
\end{proof}

%%% ----------------------------------------------------------------------------
%%% ----------------------------------------------------------------------------
%%% Analyzing the Randomized Fast Protrusion Replacer.
%%% ----------------------------------------------------------------------------
%%% ----------------------------------------------------------------------------

\paragraph{Analyzing the Randomized Fast Protrusion Replacer.}
We now analyze how many vertices the Fast Protrusion Replacer reduces the instance by. To that end we need to define the notion protrusion covers.
\begin{define}\label{def:protrusionCover}
An $(a, b, r)$-{\em protrusion cover} in a graph $G$ is a collection ${\cal Z} = Z_1, \ldots, Z_t$ of sets such that for every $i$, $N[Z_i]$ is a $r$-protrusion in $G$ and $a \leq |Z_i| \leq b$, and for every $i \neq j$, $Z_i \cap Z_j = \emptyset$ and there are no edges from $Z_i$ to $Z_j$. The {\em size} of ${\cal Z}$ is $|{\cal Z}|$.
\end{define}

\begin{lemma}\label{lem:randomRFPRCover}
Let $\Pi$ be a problem that has a protrusion replacer which replaces $r$-protrusions of size at least $r'$, and let $s \geq r' \cdot 2^r$. If $G$ is a graph with a $(s, 6s, r)$-protrusion cover ${\cal X}$, then if the RFPR is run on $(G,k)$, with probabilty at least $1 - e^{-\frac{|{\cal X}|}{8(r+1)^{6s}}}$ the output instance $(G', k')$ satisfies $|V(G)|-|V(G')| \geq \frac{|{\cal X}|}{4(r+1)^{6s}}$.
\end{lemma}

\begin{proof}
By Lemma~\ref{lem:rfprFacts} the RFPR computes a set ${\cal R}$ of protrusions and $|V(G)|-|V(G')| \geq |{\cal R}|$. Thus it is sufficient to show that with high probablility, ${\cal R} \geq \frac{|{\cal X}|}{4(r+1)^{6s}}$. Define $\overline{X} = V(G) \setminus \bigcup_{X \in {\cal X}} X$. Since no edge goes between different sets in ${\cal X}$ we have that for every $X \in {\cal X}$, $N(X) \subseteq \overline{X}$. The only randomized step of the RFPR is the initial partitioning of $V(G)$ into sets $X_1, \ldots X_{r+1}$. We may think of this partitioning step as selecting a random coloring of $V(G)$ with colors from $\{1, \ldots, r+1\}$.

We say that a set $X$ in ${\cal X}$ {\em succeeds} if all vertices in $X$ are colored with the same color, and no vertex of $N(X)$ is colored with that color. Since every set $X \in {\cal X}$ has at most $r$ neighbours we have that the probability that $X$ succeeds given any coloring of $\overline{X}$ is at least $\frac{1}{(r+1)^{|X|}}$. Hence the expected number of sets $X \in {\cal X}$ that succeed is at least $\frac{|{\cal X}|}{(r+1)^{6s}}$. Suppose $t$ sets succeed. We prove that the set ${\cal R}$ constructed by the Randomized Fast Protrusion Replacer has size at least $t/2$.

For each set $X$ that succeeds, the connected components of $X$ are added to ${\cal C}'$, and since they all have treewidth at most $r$ and have at most $r$ neighbors, none of them are discarded. Hence the connected components of $X$ are all added to ${\cal C}^*$. Since $|Z| \geq r' \cdot 2^r$, if we group the connected components of $Z$ by their neighbourhood, at least one group has combined size at least $r'$. If this group contains a connected component on at least $r'$ vertices then this component is added to ${\cal R}$ directly and $X$ contributes one to $R_{big}$. If this group does not contain any components of size at least $r'$ then the group is added in its entirety to some set ${\cal Z}_i$. In this case the group contributes at least $r'$ to $\sum_{C \in {{\cal Z}_i}} |C|$. By Lemma~\ref{lem:rfprFacts},  ${\cal R} \geq R_{big} + \left\lceil \sum_{i \leq q} \frac{\sum_{C \in {\cal Z}_i} |C| - (r' - 1)}{2(r' - 1)} \right\rceil$. Hence the total number of sets added to ${\cal R}$ is at least $t/2$.

Since the neighbourhoods of different sets in ${\cal X}$ may overlap there are dependencies between which sets succeed. However, given any coloring of $\overline{X}$ the success of different sets in ${\cal X}$ is independent, since whether $X$ succeeds or not depends only on the color of vertices in $X$ and the color of vertices in $N(X) \subseteq \overline{X}$. Thus for every coloring of $\overline{X}$ the number of sets that succeed is a sum of independent $0$-$1$ variables taking value $1$ with probability at least $\frac{1}{(r+1)^{|X|}}$. Standard Chernoff bounds for the binomial distribution show that if $T$ is a sum of $n$ independent $0$-$1$ variables taking value $1$ with probabily $p$, then $P[X \leq np/2] < e^{-\frac{np}{8}}$. Plugging this in for the number of sets in ${\cal X}$ that succeed yields that the probability that $|{\cal R}| \leq \frac{|{\cal X}|}{4(r+1)^{6s}}$ is at most $e^{-\frac{|{\cal X}|}{8(r+1)^{6s}}}$.
\end{proof}

%%% ----------------------------------------------------------------------------
%%% ----------------------------------------------------------------------------
%%% The Deterministic Fast Protrusion Replacer
%%% ----------------------------------------------------------------------------
%%% ----------------------------------------------------------------------------

\paragraph{The Deterministic Fast Protrusion Replacer}
We prove that the RFPR can be made deterministic at the cost of a $\log n$ factor in the running time. The only randomized step of the RFPR is the initial step where the vertices of $G$ are partitioned into $r+1$ sets $X_1, \ldots X_{r+1}$. We may think of this partitioning step as selecting a random coloring of $V(G)$ with colors from $\{1, \ldots, r+1\}$. The main difference between the randomized and the deterministic Fast Protrusion Replacer is how this coloring is chosen. The Deterministic Fast Protusion Replacer only partitions $V(G)$ in two sets $X_1$ and $X_2$ - this corresponds to coloring the vertices with colors $1$ and $2$. To describe the colorings the Deterministic Fast Protrusion Replacer (DFPR) uses we use the notion of {\em universal sets}.

\begin{definition}[\cite{NaorSS95}]
A \emph{$(n,t)$-universal set} ${\cal P}$ of a ground set $U$ on $n$ elements is a collection ${\cal P}$ of subsets of $U$ such that for every set $S \subseteq U$ and set $S' \subseteq S$ there is a set $P \in {\cal P}$ such that $P \cap S = S'$.
\end{definition}

\begin{theorem}[\cite{NaorSS95}]
\label{propuniversalsets}
There is a deterministic algorithm with running time $O(2^{t+o(t)} n\log n)$ that constructs an $(n,t)$-universal set ${\cal P}$ such that $|{\cal P}|=2^{t+o(t)}\log n$.
\end{theorem}

The DFPR has two parameters, $r$ and $s$, instead of just one parameter $r$. It constructs a $(n,6s+r)$-universal set ${\cal P}$ in time $O(2^{6s+r + o(6s+r)} n\log n) = O(2^{20s}n \log n)$ and selects the first set $P \in {\cal P}$. It sets $X_1 = P$, $X_2 = V(G) \setminus P$ and then it proceeds just as the RFPR would. For a fixed set $P \in {\cal P}$ this takes linear time and will reduce $(G,k)$ to an equivalent instance $(G',k')$. Choosing different sets $P \in {\cal P}$ results in different output instances $(G',k')$. The DFPR tries all possible choices for $P \in {\cal P}$ and then finally outputs the instance $(G',k')$ that maximizes $|V(G)| - |V(G')|$. The total time taken by the DFPR is $O((2^{20s}n \log n) + |{\cal P}| \cdot O(n+m) = O((2^{20s}(n+m)\log n)$. This proves the following lemma.
\begin{lemma}\label{lem:dfprFacts} Given an instance $(G,k)$, the DFPR runs in time $O((2^{20s}(n+m)\log n)$, computes a collection ${\cal R}$ of protrusions and outputs an equivalent instance $(G',k')$, such that $|V(G')| \leq |V(G)| - |{\cal R}|$. 
\end{lemma}
We now give a lemma analogous to Lemma~\ref{lem:randomRFPRCover} for the DFPR. 
\begin{lemma}\label{lem:deterministicDFPRCover}
Let $\Pi$ be a problem that has a protrusion replacer which replaces $r$-protrusions of size at least $r'$, and let $s \geq r' \cdot 2^r$. If $G$ is a graph with a $(s, 6s, r)$-protrusion cover ${\cal X}$, then if the RFPR is run on $(G,k)$, the output instance $(G', k')$ satisfies $|V(G)|-|V(G')| \geq \frac{|{\cal X}|}{2^{20s}\log n}$.
\end{lemma}
\begin{proof}
In the proof of Lemma~\ref{lem:randomRFPRCover} we showed that $|V(G)|-|V(G')|$ is lower bounded by the number of sets that succeeds. Since each set $X \in {\cal X}$ has size at most $6s$ and $|N[X]| \leq 6s+r \leq 7s$ it follows that for every $X \in {\cal X}$ there is some coloring set $P \in {\cal P}$ that makes $X$ succeed. Hence there is a coloring  $P \in {\cal P}$ that makes at least $\frac{|{\cal X}|}{|{\cal P}|} \geq \frac{|{\cal X}|}{2^{r+6s+o(r+6s)}\log n}$ sets succed. In the proof of Lemma~\ref{lem:randomRFPRCover} we showed that $|V(G)|-|V(G')|$ is at least half the number of succeeding sets. Since $2 \cdot 2^{r+6s+o(r+6s)}\log n \leq 2^{20s}$ we have $|V(G)|-|V(G')| \geq \frac{|{\cal X}|}{2^{20s}\log n}$.
\end{proof}

We now proceed to prove that if $G$ has a protrusion decomposition such that a linear fraction of the vertices appear in large enough $r$-protrusions then with high probability the Randomized Fast Protrusion Replacer will reduce $G$ by a linear fraction of its vertices. To that end we need to have a closer look at the relationship between protrusion decompositions and protrusion covers.

%%% ----------------------------------------------------------------------------
%%% ----------------------------------------------------------------------------
%%% Protrusion Covers from Protrusion Decompositions.
%%% ----------------------------------------------------------------------------
%%% ----------------------------------------------------------------------------

\paragraph{Protrusion Covers from Protrusion Decompositions.}
First we prove that in a graph of small treewidth we can always find protrusion covers with large size. 

\begin{lemma}
\label{lem:partitiontw} 
There exists a constant $c$ such that for any integers $n \geq s > b \geq 2$ and $n$-vertex graph $G$ of treewidth $b$, $G$ has a $(s, 6s, 2(b+1))$ cover of size at least $\frac{n}{122s}$.
\end{lemma}

\begin{proof}
Let $(T,{\cal B})$ be a nice tree-decomposition of $G$ of width $b$. For a subset $Q \subseteq V(T)$ by $P(Q)$ we denote $\cup_{q \in Q} B_q$. For a rooted tree $T$, and a vertex $v \in T$, a component $C$ of $T \setminus \{v\}$ is said to be below $v$ if all vertices of $C$ are descendants of $v$ in $T$. We start by constructing a set $S \subseteq V(T)$ and a collection $Q_1, \ldots, Q_{|S|}$ of connected components of $T \setminus S$ using the following greedy procedure.

Let $r$ be the root of $T$. In the beginning $S=\emptyset$ and $T^r = T$. We maintain a loop invariant that $T^r$ is the connected component of $T \setminus S$ that contains $r$. Now, at step $i$ of the greedy procedure we pick a lowermost vertex $v_i$ in $V(T^r)$ such that there is a connected component $Q_i$ of $T^r \setminus \{v_i\}$ {\em below} $v_i$ such that $|P(Q_i)| \geq 3s + 7(b+1)$. Now we add $v_i$ to $S$ and update $T^r$ accordingly. The procedure terminates when no vertex $v$ in $T^r$ has this property. In particular, if for any $v \in T^r$, every component $Q$ of $T^r \setminus \{v\}$ below $v$, $|P(Q)| < 3s + 7(b+1)$, the procedure terminates. Since $(T,{\cal B})$ is a nice tree decomposition, we have that for any vertex $v \in T_r$ and parent $u$ of $v$, if $C_v$ and $C_u$ are the components of $T^r \setminus \{v\}$ and $T^r \setminus \{u\}$ maximizing $|P(C_v)|$ and $|P(C_u)|$ respectively, then $|P(C_u)| \leq 2|P(C_v)|$. Hence we know that for every component $Q$ of $T \setminus S$, $|P(Q)| < 6s + 14(b+1) \leq 20s$. This bound holds both for the components included in the collection $Q_1, \ldots, Q_{|S|}$ and the ones that do not. 

Having constructed $S$ and $Q_1, \ldots, Q_{|S|}$ we let $S' = \lcac{S}$. By Lemma~\ref{lem:lcaClosure} we have $|S'| \leq 2|S|$. Let $S^* = S' \setminus S$. Since $|S^*| \leq |S|$, at most $\frac{|S|}{2}$ of the components $Q_1, \ldots, Q_{|S|}$ contain at least two vertices of $S^*$. This implies that at least $\frac{|S|}{2}$ of the components $Q_1, \ldots, Q_{|S|}$ contain at most one vertex of $S^*$. Without loss of generality, let $Q_1, \ldots, Q_{|S|/2}$ contain at most one vertex of $S^*$ each. For every $i \leq |S|/2$, if $Q_i$ contains no vertex of $S^*$ then $Q_i' = Q_i$ is a component of $Q \setminus S'$ with $|P(Q_i')| \geq 3s + 7(b+1) \geq s + 2(b+1)$. If $Q_i$ contains one vertex $v$ of $S^*$, since $v$ has degree at most $3$ and $|P(Q_i)| \geq 3s+b$, $Q_i \setminus \{v\}$ has at least one component $Q_i'$ with $|P(Q_i')| \geq s + 2(b+1)$. Thus we have constructed a set $S'$ and a collection of components  $Q_1', \ldots, Q_{|S|/2}'$ of $T \setminus S'$ of size at least $s + 2(b+1)$. By Lemma~\ref{lem:lcaClosure} every $Q_i'$ has at most two neighbors in $T$.

We make a collection ${\cal Z}$ as follows. For every $i \leq |S|/2$ let $Z_i = P(Q_i') \setminus P(S')$. Since $Q_i'$ has at most two neighbors in $T$ it follows that $N[Z_i]$ is a $2(b+1)$-protrusion and that $|Z_i| \geq s + 2(b+1) - 2(b+1) = s$. We have already shown that $|Q_i'| \leq 20s$ so $|Z_i| \leq 20s$ as well. Hence ${\cal Z}$ is in fact a $(s, 6s, 2(b+1))$-protrusion cover of $G$. It remains to lower bound $|{\cal Z}|$. We have that $|{\cal Z}| = |S|/2$. Furthermore we have that $S$, together with the connected components of $T \setminus S$ cover $T$. Since every bag has size at most $(b+1) \leq s$, $T \setminus S$ has at most $2|S|+1 \leq 3|S|$ connected components and for every component $Q$ of $T \setminus S$, $|P(Q)| \leq 20s$ we have that $|S|(b+1) + 3|S| \cdot 20s \geq n$. Since $s \geq b+1$ this implies that $|S| \geq \frac{n}{122s}$.
\end{proof}

\begin{lemma}\label{lem:decToCover} If $G$ has an $(\alpha, \beta)$-protrusion decomposition, then for every $s > \beta$, $G$ has a $(s, 6s, 3(\beta+1))$-protrusion cover of size at least $\frac{n}{122s} - \alpha$.
\end{lemma}
\begin{proof}
Let $R_0, \ldots R_t$ be an $(\alpha, \beta)$-protrusion decomposition of $G$. At most $\alpha$ vertices are in $R_0$, and at most $\alpha \cdot s$ vertices are in sets $R_i$ for $i \geq 1$ such that $|R_i| < s$. For each $i \geq 1$ such that $|R_i| \geq s$ we apply Lemma~\ref{lem:partitiontw} and obtain a $(s,6s,2(\beta+1))$-protrusion cover ${\cal Z}_i$ in $G[R_i]$. We let ${\cal Z}$ be the union of all the ${\cal Z}_i$'s constructed in this manner. For every $Z \in {\cal Z}_i$, $N_{G[{R_i}]}[Z_i]$ is a $2(\beta+1)$-protrusion in $G[R_i]$. However $Z$ might have neighbors also in $R_0$. The number of neighbors of $Z$ in $R_0$ is at most $\beta$ and hence $N[Z]$ is a $3(\beta+1)$-protrusion in $G$. We conclude that ${\cal Z}$ is a $(s, 6s, 3(\beta+1))$-protrusion cover in $G$. The size of ${\cal Z}$ is at least $\frac{n - \alpha - \alpha \cdot s}{122s} \geq \frac{n}{122s} - \alpha$.
\end{proof}

%We are now ready to prove our main results on Fast Protrusion Replacement

%%% ----------------------------------------------------------------------------
%%% ----------------------------------------------------------------------------
%%% The Fast Protrusion Replaceer Theorems
%%% ----------------------------------------------------------------------------
%%% ----------------------------------------------------------------------------

\paragraph{The Fast Protrusion Replacer Theorems}
We are now ready to prove our main results on Fast Protrusion Replacement.

\begin{theorem}[Randomized Fast Protrusion Replacer Theorem]\label{thm:RFPR} Let $\Pi$ be a problem that has a protrusion replacer that replaces $r$ protrusions of size at least $r'$, and let $s$ and $\beta$ be constants such that $r \geq 3(\beta + 1)$ and $s \geq 2^r \cdot r'$. Given an instance $(G,k)$ as input, the RFPR will run in time $O(n+m)$ and produce an equivalent instance $(G', k')$ with $|V(G')| \leq |V(G)|$ and $k' \leq k$. If additionally $G$ has a $(\alpha, \beta)$-protrusion decomposition such that $\alpha \leq \frac{n}{244s}$, then with probability at least $1 - e^{-\frac{n}{2000s(r+1)^{6s}}}$ we have $ |V(G)| - |V(G')| \geq \frac{n}{1000(r+1)^{6s}}$.
\end{theorem}

\begin{proof}
The first part of the statement follows directly from Lemma~\ref{lem:rfprFacts}. If $G$ has a $(\alpha, \beta)$-protrusion decomposition such that $\alpha \leq \frac{n}{244s}$, then by Lemma~\ref{lem:decToCover}, $G$ has a $(s, 6s, 3(\beta + 1))$-protrusion cover ${\cal X}$ of size at least $\frac{n}{122s} - \alpha \geq \frac{n}{244s}$. Plugging ${\cal X}$ into Lemma~\ref{lem:randomRFPRCover} yields that with probability at least $1 - e^{-\frac{|{\cal X}|}{8s(r+1)^{6s}}} \geq 1 - e^{-\frac{n}{2000s(r+1)^{6s}}}$ we have $ |V(G)| - |V(G')| \geq \frac{|{\cal X}|}{4(r+1)^{6s}} \geq \frac{n}{1000(r+1)^{6s}}$.
\end{proof}

\begin{theorem}[Deterministic Fast Protrusion Replacer Theorem]\label{thm:DFPR} Let $\Pi$ be a problem that has a protrusion replacer that replaces $r$ protrusions of size at least $r'$, and let $s$ and $\beta$ be constants such that $r \geq 3(\beta + 1)$ and $s \geq 2^r \cdot r'$. Given an instance $(G,k)$ as input, the DFPR will run in time $O(2^{20s} \cdot (n+m)\log n)$ and produce an equivalent instance $(G', k')$ with $|V(G')| \leq |V(G)|$ and $k' \leq k$. If additionally $G$ has a $(\alpha, \beta)$-protrusion decomposition such that $\alpha \leq \frac{n}{244s}$ then we have $ |V(G)| - |V(G')| \geq \frac{n}{244 \cdot 2^{20s}\log n}$.
\end{theorem}

\begin{proof}
The first part of the statement follows directly from Lemma~\ref{lem:dfprFacts}. If $G$ has a $(\alpha, \beta)$-protrusion decomposition such that $\alpha \leq \frac{n}{244s}$, then by Lemma~\ref{lem:decToCover}, $G$ has a $(s, 6s, 3(\beta + 1))$-protrusion cover ${\cal X}$ of size at least $\frac{n}{122s} - \alpha \geq \frac{n}{244s}$. Plugging ${\cal X}$ into Lemma~\ref{lem:deterministicDFPRCover} yields that $ |V(G)| - |V(G')| \geq \frac{|{\cal X}|}{2^{20s}\log n} \geq \frac{n}{244 \cdot 2^{20s}\log n}$.
\end{proof}

It can be shown that Theorem~\ref{thm:RFPR} could replace the simple protrusion reduction algorithm of Lemma~\ref{lem:naiveReplace} and make thus Algorithm~\ref{fig:randFPTslow} run in linear time. However we are first going to refine Algorithm~\ref{fig:randFPTslow} even further so that it becomes simultaneously single exponential parameterized algorithm and an approximation algorithm for \ffd{} for all connected ${\cal F} \in \mathscr{F}$. To that end we develop the notion of lossless protrusion replacement.

\section{Lossless Protrusion Replacement}\label{apxprotrusion}
In this section we develop the notion of lossless protrusion replacement. We consider ${\sc CMSO}$ vertex subset problems. In a {\sc min-CMSO} vertex subset problem, $\Pi$, we are given a graph $G$ as input. The objective is to find a set $S \subseteq V(G)$ minimizing $|S|$ such that such that the CMSO-expressible predicate $P_\Pi(G,S)$ is satisfied. Similarly, in a {\sc max-CMSO}  vertex subset  problem, $\Pi$, we are given a graph $G$ as input. The objective is to find a set $S \subseteq V(G)$ maximizing 
$|S|$ such that the CMSO-expressible predicate $P_\Pi(G,S)$ is satisfied. Given a {\sc min-CMSO} ({\sc max-CMSO}) vertex subset problem, $\Pi$ and an input graph $G$ to $\Pi$, by $OPT(G)$ we denote the size of the smallest (largest) set $S$ such that the CMSO-expressible predicate $P_\Pi(G,S)$ is satisfied. Next we define the notion of a lossless protrusion replacer. A lossless protrusion replacer is essentially a protrusion replacer that reduces protrusions in such a way that any feasible solution to the reduced instance can be changed into a feasible solution of the original instance without changing the gap between the feasible solution and the optimum. The notion of lossless protrusion replacement is central in our approximation algorithms.

\begin{definition}[Lossless Protrusion Replacer]
A lossless protrusion replacer for {\sc min-CMSO} ({\sc max-CMSO}) vertex subset problem $\Pi$ is a family of algorithms, with one algorithm for every constant $r$. The $r$'th algorithm has the following specifications. There exists a constant $r'$ (which depends on $r$) such that given an instance $G$ and an $r$-protrusion $X$ in $G$ of size at least $r'$, the algorithm runs in time $O(|X|)$ and outputs an instance $G'$  with the following properties. 
\begin{itemize}
\item $G'$ is obtained from $G$ by replacing $X$ by a $r$-boundaried graph $X'$ with less than $r'$ vertices and thus $|V(G')| < |V(G)|$. 
\item $OPT(G') \leq OPT(G)$.
\item There is an algorithm that runs in $O(|X|)$ time and given a feasible solution $S'$ to $G'$ outputs a set $X^* \subseteq X$ such that $S=(S'\setminus X') \cup X^*$ is a feasible solution to $G$ and $|S|\leq |S'| + OPT(G)-OPT(G')$.
\end{itemize}
\end{definition}

We would like to give sufficient conditions for a problem to have a lossless protrusion replacer. An ideal setting would be that every graph optimization problem that has finite integer index when parameterized by the size of the optimal solution has a lossless protrusion replacaer. Unfortunately such a theorem seems to be out of reach, and it is quite possible that this is not true. However, in~\cite{BodlaenderFLPST09} a sufficient condition is given for a ${\sc CMSO}$ vertex subset problem to have finite integer index. This condition is called {\em strong monotonicity} and it is proved that every ${\sc CMSO}$ vertex subset problem that is stronly monotone has finite integer index and hence has a protrusion replacer. It turns out that strong monotonicity is a sufficient condition for a ${\sc CMSO}$ vertex subset problem to not only have a protrusion replacer, but also a lossless protrusion replacer. We now prove this fact. 

Let $\Pi$ be a \pmin{} problem and ${\cal F}_t$ be the set of pairs $(G,S)$ where $G$ is a $t$-boundaried graph and $S \subseteq V(G)$. For a $t$-boundaried graph $G$ we define the function $\zeta_G : {\cal F}_t \rightarrow \mathbb{N} \cup \{\infty\}$ as follows. For a pair $(G',S') \in {\cal F}_t$, if there is no set $S \subseteq V(G)$ such that $P_\Pi(G \oplus G', S \cup S')$ holds, then $\zeta_G((G',S')) = \infty$. Otherwise $\zeta_G((G',S'))$ is the size of the smallest $S \subseteq V(G)$ such that $P_\Pi(G \oplus G', S \cup S')$ holds. If $\Pi$ is a \pmax{} problem then we define $\zeta_G((G',S'))$ to be the size of the largest $S \subseteq V(G)$ such that $P_\Pi(G \oplus G', S \cup S')$ holds. If there is no set $S \subseteq V(G)$ such that $P_\Pi(G \oplus G', S \cup S')$ holds, then $\zeta_G((G',S')) = \infty$.

\begin{define}[\cite{BodlaenderFLPST09}]\label{def:minstrongmonmin}
A \pmin{} problem $\Pi$ is said to be \emph{strongly monotone} if there exists a function $f : \mathbb{N} \rightarrow \mathbb{N}$ such that the following condition is  satisfied. For every $t$-boundaried graph $G$, there is a subset $S\subseteq V(G)$ such that for every $(G',S')\in {\cal F}_t$ such that $\zeta_G((G',S'))$ is finite, $P_\Pi(G\oplus G',S\cup S')$ holds and $|S|\leq \zeta_G((G',S'))+f(t)$. 
\end{define}

\begin{define}[\cite{BodlaenderFLPST09}]\label{def:minstrongmonmax}
A \pmax{} problem $\Pi$ is said to be \emph{strongly monotone} if there exists a function $f : \mathbb{N} \rightarrow \mathbb{N}$ such that the following condition is  satisfied. For every $t$-boundaried graph $G$, there is a subset $S\subseteq V(G)$ such that for every $(G',S')\in {\cal F}_t$ such that $\zeta_G((G',S'))$ is finite, $P_\Pi(G\oplus G',S\cup S')$ holds and $|S|\geq \zeta_G((G',S'))-f(t)$. 
\end{define}

\begin{theorem}\label{thm:losslessProtrusion}
Every {\sc min-CMSO} or {\sc max-CMSO} vertex subset problem $\Pi$, that is also strongly monotone admits a lossless protrusion replacer. 
\end{theorem}
Before proving the theorem we will need an auxiliary lemma.
\begin{lemma}
\label{lem:twoar}
If a graph $G$ contains an $r$-protrusion $X$ where $|X|>c>0$, then it also contains a $(2r+1)$-protrusion $Y$ where $c<|Y|\leq 2c$. Moreover, given $X$ we can compute $Y$ and a tree decomposition of $Y$ of width $\leq 2r$ in $O(|X|)$ time.
\end{lemma}
\begin{proof}
Let $(T,{\cal X})$ be a nice tree decomposition of $G[X]$ rooted at a node $r$. We can compute  $(T,{\cal X})$ from $G[X]$ in time $O(|X|)$ using Bodlaender's algorithm~\cite{Bodlaender96}.
If $|X|\leq 2c$, we are done. Given a vertex $x$ of the rooted tree $T$, we denote by ${\cal D}_{T}(x)$ the subset of $V(T)$ containing $x$ and all its descendants in $T$. Let $B_{T}$ be the set containing each vertex $x$ of $T$ with the property that the vertices appearing in $\bigcup_{y\in {\cal D}_{T}(x)}X_y$ (i.e. the vertices of the nodes corresponding to $x$ and its descendants) are more than $c$. As $|X|\geq 2c$, $B_{T}$ is a non-empty set. We choose $b$ to be  a member of $B_{T}$ whose descendants do not belong in $B_{T'}$. This choice of $b$ ensures that $c<|\bigcup_{y\in {\cal D}(b)}X_y|\leq 2c$. We define $Y=\partial_{G}X\cup \bigcup_{y\in {\cal D}_{T}(b)}X_y$. As $G[Y]$ is an induced subgraph of $X$ it follows that $\tw(G[Y]) \leq r$. Furthermore $\partial_{G}(Y)\subseteq  \partial_{G}X \cup X_{b}$, therefore $Y$ is a $(2r+1)$-protrusion of $G$.
\end{proof}
\begin{proof}[Proof of Theorem~\ref{thm:losslessProtrusion}]
We prove the theorem for \pmin{} problems; the proof for \pmax{} problems is similar. Let $\Pi$ be a monotone \pmin{} problem. We define a partial order $\leq_\Pi$ on pairs $(G, S)$ such that $G$ is a $t$-boundaried graph and $S \subseteq V(G)$. We say that $(G, S) \leq_\Pi (G',S')$ if for every $(G_3, S_3)$, $P_\Pi(G \oplus G_3, S \cup S_3) \rightarrow P_\Pi(G' \oplus G_3, S' \cup S_3)$. We say that that $(G, S) \equiv_\Pi (G',S')$ if $(G, S) \leq_\Pi (G',S')$ and $(G', S') \leq_\Pi (G,S)$. Clearly $\equiv_\Pi$ is an equivalence relation and since $P_\Pi$ is a CMSO-expressible predicate it follows from~\cite{BoriePT92,CourcelleM93} that for every fixed $t$, $\equiv_\Pi$ has finitely many equivalence classes. Thus there exists finite set ${\cal S}$ of pairs $(G_R,S_R)$ such that for every $(G,S)$ there is a $(G_R,S_R) \in {\cal S}$ such that $(G,S) \equiv (G_R,S_R)$. We say that a pair $(G, S)$ is {\em bad} if there is no $(G', S')$ such that $P_\Pi(G \oplus G', S \cup S')$ holds. A pair that is not bad is called {\em useful}. Let ${\cal U}$ be the set of all useful pairs in ${\cal S}$.

For a graph $G$ and pair $(G_R,S_R) \in {\cal U}$ define $\gamma_G(G_R,S_R)$ to be the size of the smallest set $S \subseteq V(G)$ such that $(G_R, S_R) \leq_\Pi (G,S)$. If no such set $S$ exists, $\gamma_G(G_R,S_R) = \infty$. We now prove that for any $G$, the maximum finite value of $\gamma_G$ and the minimum (finite) value of $\gamma_G$ differs by at most $f(t)$. Let $S \subseteq V(G)$ be the set such for every $(G',S')\in {\cal F}_t$ such that $\zeta_G((G',S'))$ is finite, $P_\Pi(G\oplus G',S\cup S')$ holds and $|S|\leq \zeta_G((G',S'))+f(t)$. Consider a useful pair $(G_R,S_R) \in {\cal U}$ such that $\gamma_G(G_R,S_R)$ is finite. Then there exists a set $S' \subseteq V(G)$ of size $\gamma_G(G_R,S_R)$ such that $(G_R,S_R) \leq_\Pi (G, S')$. Since $(G, S') \leq_\Pi (G, S)$ and $S'$ is the smallest set such that $(G_R,S_R) \leq_\Pi (G, S')$ it follows that $|S'| \leq |S|$. On the other hand since $(G_R,S_R)$ is useful there exists some $(G^*,S^*)$ such that $P_\Pi(G_R \oplus G^*,S_R \cup S^*)$ holds. Then $P_\Pi(G \oplus G^*,S' \cup S^*)$ holds as well and hence $\zeta_G((G^*,S^*)) \leq |S'|$. Since $\zeta_G((G^*,S^*))$ is finite it follows that $|S| \leq \zeta_G((G^*,S^*))+f(t) \leq |S'|+f(t)$. But this means that $|S|-f(t) \leq \gamma_G(G_R, S_R) \leq |S|$ and so the finite values of $\gamma_G$ differ by at least $f(t)$. By the pigeon hole principle there exists a finite collection ${\cal R}$ of $t$-boundaried graphs such that for any $t$-boundaried $G$ there is a $G_R \in {\cal R}$ and a constant $c_R \geq 0$ such that for every useful pair $(G',S')$, $\gamma_G(G',S') = \gamma_{G_R}(G',S') + c_R$. We call ${\cal R}$ a {\em set of representatives} for $(\Pi, t)$.

For every integer $c$ we define a relation $\prec_c$ on $t$-boundaried graphs. We say that $G_1 \prec_c G_2$ if for every useful pair $(G,S)$, $\gamma_{G_1}(G,S) + c = \gamma_{G_2}(G,S)$. Observe that if $G_1 \prec_c G_2$ then $G_2 \prec_{-c} G_1$. Also, we have just shown that for every $G$ there is a $G_R \in {\cal R}$ and constant $c_R \geq 0$ such that $G_R \prec_{c_R} G$. We now show that if $G \prec_c G'$ then for any $t$-boundaried graph $G_3$ and feasible solution $S$ to $\Pi$ on $G \oplus G_3$, there is a set $X^* \subseteq V(G')$ depending only on $S \cap V(G)$ and $G$ such that $S' = X^* \cup S \setminus V(G)$ is also a feasible solution to $\Pi$ on $G' \oplus G_3$ and $|S'| \leq |S| + c$. 

Let $G \prec_c G'$ and consider a $t$-boundaried $G_3$ and a feasible solution $S$ of $\Pi$ on $G \oplus G_3$. Let $S_G = S \cap V(G)$ and $S_3 = S \setminus S_G$. $(G,S_G)$ is a useful pair and so there is a pair $(G_R,S_R) \in {\cal U}$ such that $(G_R,S_R) \equiv_\Pi (G,S_G)$. Thus $\gamma_G(G_R,S_R) \leq |S_G|$ and hence $\gamma_{G'}(G_R,S_R) \leq |S_G|+c$. There is a set $X^* \subseteq V(G')$ such that $(G_R,S_R) \leq_\Pi (G',X^*)$ and $|X^*| \leq |S_G|+c$. The set $X^*$ depends solely on $(G_R,S_R)$ which depends solely on $S \cap V(G)$ and $G$. Furthermore, since $(G_R,S_R) \leq_\Pi (G',X^*)$ we have that $S' = X^* \cup S_3$ is also also a feasible solution to $\Pi$ on $G' \oplus G_3$ and $|S'| \leq |S_G|+c+|S_3| \leq |S|+c$.

We can now describe the lossless protrusion replacer for the problem $\Pi$. For parameter $r$ consider the set ${\cal R}$ of representatives for $(\Pi, 2(r+1))$. Let $r'$ be the size of the largest graph in ${\cal R}$ plus one. The lossless protrusion replacer for $\Pi$ will reduce $r$-protrusions of size at least $r'$. Given an $r$-protrusion $X$ of size at least $r'$ we find a $2(r+1)$ protrusion $Y \subseteq X$ such that $r' \leq |Y| \leq 2r'$. This can be done in $O(|X|)$ time by Lemma~\ref{lem:twoar}. Consider now the $2(r+1)$-boundaried graph $G_{Y\setminus \delta(Y)}^{\delta Y}$. There exists a $2(r+1)$-boundaried graph $G_R \in {\cal R}$ and constant $c_R \geq 0$ such that $G_R \prec_{c_R} G_{Y\setminus \delta(Y)}^{\delta Y}$. Furthermore since $|Y| \geq r'$ we have that $|V(G_R)| < |Y|$. The protrusion replacer outputs the graph $G'$ obtained by replacing $Y$ by $G_R$ in $G$.

For every subset $S_R \subseteq V(G_R)$ such that the pair $(G_R, S_R)$ is useful, the protrusion replacer stores a subset $S_Y \subset Y$ such that $(G_R, S_R) \leq_\Pi (G_{Y\setminus \delta(Y)}^{\delta Y},S_Y)$. Since $G_R \prec_{c_R} G_{Y\setminus \delta(Y)}^{\delta Y}$ there is such a set $S_Y$ of size at most $|S_R| + c$. Now, for any feasible solution $S$ in $G'$ let $S_R = S \cup V(G_R)$. The pair $(G_R, S_R)$ is useful and so the lossless protrusion replacer outputs the set $X^* = S_Y$ which it has stored for $S_R$. Now $S' = S_Y \cup (S \setminus V(G_R))$ is a feasible solution to $G$ because $(G_R, S_R) \leq_\Pi (G_{Y\setminus \delta(Y)}^{\delta Y},S_Y)$. Furthermore, since $|S_Y| \leq |S_R| + c$ we have that $|S'| \leq |S_Y| + |S \setminus V(G_R)| \leq |S_R| + c + |S \setminus V(G_R)| \leq |S|+c$. Thus it remains to prove that $c \leq OPT(G)-OPT(G')$, or in other words that $OPT(G') \leq OPT(G)-c$.

However $G_{Y\setminus \delta(Y)}^{\delta Y} \prec_{-c_R} G_R$, and hence for an optimal solution $S$ of $G = G_{Y\setminus \delta(Y)}^{\delta Y} \oplus G_{V(G)\setminus Y}^\delta(Y)$ there is a feasible solution $S'$ in $G_R \oplus G_{V(G)\setminus Y}^\delta(Y)$ of size at most $|S|-c_R$. Hence $OPT(G') \leq OPT(G)-c$ and the theorem follows.
\end{proof}

Inserting a lossless protrusion replacer instead of a normal protrusion replacer into the Fast Protrusion Replacer algorithms directly yields the following theorems.

\begin{theorem} \label{thm:RFPRLossless} Let $\Pi$ be a minimization (maximization) problem that has a lossless protrusion replacer that replaces $r$ protrusions of size at least $r'$, and let $s$ and $\beta$ be constants such that $r \geq 3(\beta + 1)$ and $s \geq 2^r \cdot r'$. Given an instance $G$ as input, the Randomized Fast Protrusion Replacer will run in time $O(n+m)$ and produce an instance $G'$ with $|V(G')| \leq |V(G)|$. Given any feasible solution $S'$ to $G'$ a feasible solution $S$ of $G$ of size at most (at least) $|S'| - OPT(G') + OPT(G)$ can be computed in $O(n+m)$ time. If additionally $G$ has a $(\alpha, \beta)$-protrusion decomposition such that $\alpha \leq \frac{n}{244s}$, then with probability at least $1 - e^{-\frac{n}{2000s(r+1)^{6s}}}$ we have $ |V(G)| - |V(G')| \geq \frac{n}{1000(r+1)^{6s}}$. 
\end{theorem}

\begin{theorem}\label{thm:DFPRLossless} Let $\Pi$ be a minimization (maximization) problem that has a lossless protrusion replacer that replaces $r$ protrusions of size at least $r'$, and let $s$ and $\beta$ be constants such that $r \geq 3(\beta + 1)$ and $s \geq 2^r \cdot r'$. Given an instance $G$ as input, the Deterministic Fast Protrusion Replacer will run in time $O(2^{20s} \cdot (n+m)\log n)$ and produce an instance $G'$ with $|V(G')| \leq |V(G)|$. Given any feasible solution $S'$ to $G'$ a feasible solution $S$ of $G$ of size at most (at least) $|S'| - OPT(G') + OPT(G)$ can be computed in $O(n+m)$ time. If additionally $G$ has a $(\alpha, \beta)$-protrusion decomposition such that $\alpha \leq \frac{n}{244s}$ then we have $ |V(G)| - |V(G')| \geq \frac{n}{244 \cdot 2^{20s}\log n}$.
\end{theorem}

%\section{Combinatorial Facts around \fd{}}
%\input{combinatorialfd.tex}

\section{Approximation and Fast Parameterized Algorithm for \fd{}}
%!TEX root=PFDMerge-KernelAlgo.tex

%\section{Approximation Algorithms for \fd{}}
We are now ready to give the linear time, lossless variant of Lemma~\ref{lem:naiveReplace}. Throughout this section $OPT(G)$ is the size of the smallest ${\cal F}$-deletion set  of $G$, for the set ${\cal F}$ currently under consideration. First we give an auxiliary lemma analyzing an execution of the Lossless RFPR on a graph with an ${\cal F}$-deletion set $S$. 
\begin{lemma}\label{lem:ReduceLinearFraction} For every connected ${\cal F} \in \mathscr{F}$, there exist constants $\rho$, $r$, $s$, $c < 1$ and $\gamma > 0$ such that if we run the Lossless RFPR with parameters $r$, $s$ on a graph $G$ which has a ${\cal F}$ deletion set $S$ which is not a $\rho$-cover, then with probability at least $1-e^{-\gamma n}$ the output instance $G'$ satisfies $V(G') \leq |V(G)|(1-c)$.
\end{lemma}
\begin{proof}
If $G$ has a ${\cal F}$-deletion set  $S'$ which is not an $\rho$-cover, it also has a inclusion minimal ${\cal F}$-deletion set  $S$ which is not an $\rho$-cover. Such a minimal $S$ contains no isolated vertices and hence satisfies $N[S] \leq 2\sum_{v\in S} d(v) \leq 2\rho m$. 

By Proposition~\ref{prop:planar_exclude_treewidth}, there exists a constant $b$ such that $\tw(G \setminus S) \leq b$. By Lemma~\ref{lem:prottfd}, $G$ has a $(4(b+1)|N[S]|,2(b+1))$-protrusion decomposition. Set $\beta = 2(b+1)$, $r=3(\beta+1)$ and $r'$ to be the smallest integer such that the lossless protrusion replacer will replace $r$-protrusions of size at least $r'$. Set $s = 2^r \cdot r'$. The protrusion decomposition of $G$ is a $(4(b+1)|N[S]|,\beta)$-protrusion decomposition. By Theorem~\ref{thm:RFPRLossless} there exist constants $0 < c < 1$ and $0 < \gamma$ such that if we run the Lossless RFPR on $G$ and $4(b+1)|N[S]| \leq \frac{n}{244s}$ then with probability at least $1-e^{-\gamma n}$, the output graph $G'$ satisfies $|V(G)|-|V(G')| \geq c|V(G)|$. We show that there is a constant $\rho < \frac{1}{3}$ such that if $S$ is not a $\rho$-cover, then $|N[S]| \leq \frac{n}{1000(b+1)s}$. 

Since $\tw(G \setminus S) \leq b$ we have that $G \setminus S$ is $(b+1)$-degenerate. If $S$ is not a $\rho$-cover then $m \leq n(b+1) + \sum_{v \in S} d(v) \leq n(b+1) + 2\rho m$. Rearranging yields that $N[S]\leq 2\rho m \leq n\frac{2\rho(b+1)}{1-2\rho} \leq n\rho 6(b+1)$. Choosing $\rho = 6000(b+1)^2s$ yields that $|N[S]| \leq \frac{n}{1000(b+1)s}$. Hence, if $S$ is not a $\rho$-cover then with probability at least $1-e^{-\gamma n}$ the output instance $G'$ of the Lossless RFPR satisfies $V(G') \leq |V(G)|(1-c)$.
\end{proof}

\begin{lemma}\label{lem:smartReduce} For every connected ${\cal F} \in \mathscr{F}$ there is an algorithm that given a graph $G$, takes $O(n+m)$ time and outputs a graph $G'$ such that $V(G') \leq V(G)$ and $OPT(G') \leq OPT(G)$. Given a ${\cal F}$-deletion set  $S'$ of $G'$ the algorithm can compute an ${\cal F}$-deletion set  $S$ of $G$ of size $|S'|+OPT(G)-OPT(G')$ in time $O(n+m)$. Furthermore there exist a constant $0 < \rho < 1$ such that with probability at least $\frac{1}{2}$, every ${\cal F}$-deletion set  $S'$ of $G'$ is a $\rho$-cover of $G$.
\end{lemma}

\begin{proof}
By Lemma~\ref{lem:ReduceLinearFraction} there exist constants $\rho$, $r$, $s$, $c < 1$ and $\gamma > 0$ such that if we run the Lossless RFPR with parameters $r$, $s$ on a graph $G$ which has a ${\cal F}$ deletion $S$ which is not a $\rho$-cover, then with probability at least $1-e^{-\gamma n}$ the output instance $G'$ satisfies $V(G') \leq |V(G)|(1-c)$. We set these constants as guaranteed by Lemma~\ref{lem:ReduceLinearFraction}.

The algorithm sets $G_1 := G$, $i = 1$ and enters a loop that proceeds as follows. The algorithm runs the Lossless RFPR on $G_i$ with parameters $r$ and $s$, let the output of the Lossless RFPR be $G_{i+1}$. If $|V(G_{i+1})| > |V(G_i)|(1-c)$ the algorithm halts and outputs $G_i$. Otherwise, the algorithm increments $i$ and returns to the beginning of the loop.

The total time spent by the algorithm is upper bounded by a geometric series, and so the running time of the algorithm is $O(n+m)$. Similarly, by repeatedly applying Theorem~\ref{thm:RFPRLossless} we can in linear time transform any ${\cal F}$-deletion set  $S_i$ of $G_i$ back into a ${\cal F}$-deletion set  $S$ of $G$ of size at most $|S'|+OPT(G)-OPT(G')$. It remains to prove that when the algorithm terminates, with probability at least $\frac{1}{2}$ we have that every ${\cal F}$-deletion set  $S'$ of $G'$ is an $\rho$-cover of $G$.

The algorithm makes $t = O(\log n)$ calls to the Lossless RFPR. For $i \leq t+1$ let $n_i = |V(G_i)|$. In call $i$, by Lemma~\ref{lem:ReduceLinearFraction}, if $G_i$ has an ${\cal F}$-deletion set  $S$ which is not a $\rho$-cover then the probability that $V(G_{i+1}) > V(G_i)(1-c)$ is at most $e^{-\gamma n_i}$. By the union bound the probability that this occurs at some step $i$ is $\sum_{i \leq t} e^{-\gamma n_i}$. The $n_i$'s are a decreasing geometric series and so for a sufficiently large (constant) $N$ we have that if $n_t \geq N$ then $\sum_{i \leq t} e^{-\gamma n_i} \leq 2e^{-\gamma n_t} \leq 1/2$.

Finally, if $n_t \leq N$ then any non-empty set $S$ is a $\frac{1}{N^2}$ cover, and so if $\rho > \frac{1}{N^2}$ we can adjust $\rho$ to $\frac{1}{N^2}$. This proves the lemma.
\end{proof}

We are now ready to give the algorithm which is the main engine behind both our $2^{O(k)}n$ time algorithm and the quadratic approximation algorithm for \ffd{} for connected sets ${\cal F} \in \mathscr{F}$.

\begin{figure}[ht]
\begin{center}
\begin{boxedminipage}{.96\textwidth}

\noindent 
{\bf Randomized-${\cal F}$-Deletion}($G$)\\
Set $G_1 :=G$ and i := 1\\
While ($G_i$ is not ${\cal F}$-free$)$ do as follows:
\begin{enumerate}\setlength\itemsep{-.7mm}
\item\label{alg2step:reduce} Apply Lemma~\ref{lem:smartReduce} on $G_i$ and obtain a new graph $G_i'$ 
\item\label{alg2step:select} Pick a vertex $u_i \in V(G_i')$ at random with probability $\frac{d_{G_i'}(u)}{2|E(G_i')|}$. Set $G_{i+1}:=G_i'\setminus \{u_i\}$.
\item Increment $i$ by $1$.
\end{enumerate}
Set $S_i = \emptyset$ \\
For $j = i$ downto $2$:
\begin{enumerate}\setlength\itemsep{-.7mm}
\item Set $S_{j-1}' := S_j \cup \{u_{j-1}\}$.
\item Apply Lemma~\ref{lem:smartReduce} on $G_j'$ and $S_j'$ and obtain a set $S_j$.
\end{enumerate}
Output $S := S_1$.
\end{boxedminipage}
\caption{Randomized Algorithm for \ffd{} for connected ${\cal F} \in \mathscr{F}$ \label{fig:randfast}}
\end{center}
\end{figure}

We say that a {\em round} of Algorithm~\ref{fig:randfast} is an iteration of the while-loop. Round $x$ is the iteration when the value of $i$ is $x$. The algorithm {\em suceeds} in round $i$ if $OPT(G_i') = OPT(G_{i+1})+1$ and it {\em fails} in round $i$ otherwise. The {\em number of rounds} of a run of Algorithm~\ref{fig:randfast} is the maximum value $i$ takes. We make a series of observations about Algorithm~\ref{fig:randfast}. For every $i$ we have that $|V(G_i')| \leq |V(G_i)|$ and $|V(G_{i+1})| < |V(G_i')|$.  Hence we make the following observation. 
\begin{observation} Algorithm~\ref{fig:randfast} terminates after at most $n$ rounds. \end{observation}
The next observation follows directly from Lemma~\ref{lem:smartReduce}.
\begin{observation} The time taken in each round and each iteration of the for loop is $O(n+m)$. \end{observation}
Next we prove that the algorithm always outputs feasible solutions.
\begin{observation} Algorithm~\ref{fig:randfast} outputs an ${\cal F}$-deletion set of $G$. \end{observation}
\begin{proof}
Let $t$ be number of rounds. We have that $G_t$ is {\cal F}-free and so $S_t = \emptyset$ is a ${\cal F}$-deletion set  of $G_t$. If $S_j$ is a ${\cal F}$-deletion set  of $G_j$ then $S_j' = S_j \cup \{u_{j-1}\}$ is a ${\cal F}$-deletion set  of $G_{j-1}'$. Then, by Lemma~\ref{lem:smartReduce}, $S_{j-1}$ is a ${\cal F}$-deletion set  of $G_{j-1}$. Hence, by downward induction on $j$, $S_1$ is a ${\cal F}$-deletion set  of $G_1 = G$.
\end{proof}
Next we upper bound the size of the output solution $S$.
\begin{lemma}\label{lem:RandSizeBoundSoln} Let $p$ be the number of rounds in which Algorithm~\ref{fig:randfast} fails. Then the size of the output solution $S$ is $|S| = OPT(G) + p$. \end{lemma}
\begin{proof}
For every $x$, define $f_x$ to be the number of rounds $i \geq x$ such that the algorithm fails in round $i$. Let $t$ be the be number of rounds. We prove by downward induction on $i$ that $|S_i| = OPT(G_i) + f_i$. Since $|S_t| = |f_t| = |OPT(G_t)|= 0$ this clearly holds for $t$. Consider now some $i < t$ such that the equation holds for $i+1$. 

If the algorithm succeeded in round $i$ we have that $|S_i'| = |S_{i+1}| + 1$, that $OPT(G_i') = OPT(G_{i+1}) + 1$ and that $f_i = f_{i+1}$ hence $|S_i'| = |S_{i+1}| + 1 = OPT(G_{i+1}) + f_{i+1} + 1 = OPT(G_i') + f_i$. On the other hand if the algorithm fails in round $i$ we have $|S_i'| = |S_{i+1}| + 1$, that $OPT(G_i') = OPT(G_{i+1})$ and that $f_i = f_{i+1} + 1$. Then $|S_i'| = |S_{i+1}| + 1 = OPT(G_{i+1}) + f_{i+1} + 1 = OPT(G_i') + f_i$. Hence in both cases we have that $|S_i'| = OPT(G_i') + f_i$. By Lemma~\ref{lem:smartReduce} we have that $|S_i| = |S_i'| + OPT(G_i) - OPT(G_i') = OPT(G_i)+f_i$. This concludes the proof.
\end{proof}
Now we lower bound the success probability of any round $i$ of Algorithm~\ref{fig:randfast}.
\begin{lemma} There is a constant $p > 0$ such that the probability that Algorithm~\ref{fig:randfast} succeeds in any given round $i$ is at least $p$.
\end{lemma}
\begin{proof} By Lemma~\ref{lem:smartReduce} there is a constant $\rho$ such that with probability $1/2$, every ${\cal F}$-deletion set  of $G_i'$ is a $\rho$-cover. Let $S^*$ be an optimal ${\cal F}$-deletion set  of $G_i'$. If $u_i \in S^*$ then $S^* \setminus u_i$ is an optimal ${\cal F}$-deletion set  of $G_i' \setminus u_i = G_{i+1}$. So if $u_i \in S^*$ then the algorithm succeds in round $i$. If $S^*$ is a $\rho$-cover of $G_i'$ then the probability that $u_i \in S^*$ is at least $\rho$. Hence the probability that every ${\cal F}$-deletion set  of $G_i'$ is a $\rho$-cover {\em and} $u_i \in S^*$ is at least $p = \rho/2$.
\end{proof}

In each round Algorithm~\ref{fig:randfast} succeeds probability at least $p$. In a round $i$ where the algorithm succeeds we have that $OPT(G_{i+1}) < OPT(G)$. Since the algorithm terminates when $OPT(G_i)=0$ we get the following observation.

\begin{observation} There exists a constant $p > 0$ such that the expected number of rounds of a run of Algorithm~\ref{fig:randfast} is at most $\frac{1}{p}OPT(G)$. \end{observation}

Since the number of rounds where  Algorithm~\ref{fig:randfast} fails is at most the total number of rounds it follows form Lemma~\ref{lem:RandSizeBoundSoln} that the expected size of the output solution $|S|$ is at most $OPT(G) + \frac{1}{p}OPT(G)$. This proves the following lemma.

\begin{lemma}\label{lem:randConApx} For every connected ${\cal F} \in \mathscr{F}$, Algorithm~\ref{fig:randfast} runs in time $O(n+m)$, expected time $O((n+m)OPT(G))$ and outputs an {\cal F} solution $S$ with $E[|S|] = c \cdot OPT(G)$ for some constant $c$.
\end{lemma}

While Lemma~\ref{lem:randConApx} only gives constant factor approximation algorithms for \ffd{} for {\em connected} 
${\cal F} \in \mathscr{F}$, we can use this approximation algorithm to make an approximation algorithm for all ${\cal F} \in \mathscr{F}$.

\begin{theorem}\label{lem:randConApx} For every ${\cal F} \in \mathscr{F}$, \ffd{} has a constant factor approximation running in time $O(nm)$ and expected time $O((n+m)OPT(G))$. It outputs a feasible solution $S$ with expected size $c \cdot OPT(G)$ for a constant $c$.
\end{theorem}

\begin{proof}
By Proposition~\ref{prop:planar_exclude_treewidth}, for every ${\cal F} \in \mathscr{F}$ there is a constant $\eta$ such that for any ${\cal F}$-deletion set  $S$ of $G$ we have $\tw(G \setminus S) \leq \eta$. Since {\sc Treewidth $\eta$-Deletion} is a \ffd{} problem for a connected ${\cal F'} \in \mathscr{F}$ it follows from Lemma~\ref{lem:randConApx} has a constant factor approximation with the desired running time. We run this algorithm and find a set $S'$ such that $\tw(G \setminus S') \leq \eta$. We have that $E[|S'|] = O(OPT(G))$, where OPT(G) refers to the size of the smallest ${\cal F}$-deletion set  in $G$. Since $\tw(G \setminus S') \leq \eta$ we can solve \ffd{} on $G \setminus S$ in linear time and find a set $S^*$ of size $OPT(G \setminus S') \leq OPT(G)$. We return $S = S' \cup S^*$, $S$ is a ${\cal F}$-deletion set  of $G$ with expected size $O(OPT(G))$.
\end{proof}

Interestingly we can also use Algorithm~\ref{fig:randfast} to give a fast randomized FPT algorithm for \fd{}. 

\begin{theorem}\label{thm:FastRandFPT} For every connected ${\cal F} \in \mathscr{F}$, \fd{} has a randomized $O(c^kn)$ time algorithm. Given a yes instance the algorithm finds a solution and outputs it with probability $1/2$. If the algorithm outputs a solution, it is a feasible solution of size at most $k$.
\end{theorem}

\begin{proof}
We modify Algorithm~\ref{fig:randfast} in the following way; if $G_{k+1}$ is not ${\cal F}$-free then output ``no'' and halt. If the size of the output solution is more than $k$ then output ``no'' instead. The algorithm runs for at most $k+2$ rounds so the total running time is at most $O(nk)$. If it outputs a solution $S$ then $S$ is an ${\cal F}$ deletion of size at most $k$. We prove that if $G$ has an ${\cal F}$ deletion of size at most $k$ then the algorithm will output a solution with probability at least $\frac{1}{c^{k+2}}$ for a constant $c$. Repeating this algorithm $O((1/c)^k)$ times and outputting a solution if either iteration does then proves the theorem.

In each round, the probability that Algorithm~\ref{fig:randfast} succeeds is at least $p$ for some constant $p$. Thus the probability that Algorithm~\ref{fig:randfast} succeeds in all its rounds before it terminates (after at most $k+2$ rounds) is at least $p^{k+2}$. If the algorithm succeeds in all rounds and outputs a solution then this solution is optimal and hence has size at most $k$ if $(G,k)$ is a yes instance. Finally, if the algorithm succeeds for $k+2$ rounds then $OPT(G_1) > OPT(G_2)  \ldots OPT(G_{k+2})$ and so $OPT(G_1) \geq k+1$. Hence, if $(G, k)$ is a ``yes'' instance and the Algorithm~\ref{fig:randfast} succeeds in all of its rounds then it will output a solution of size at most $k$ before terminating. This concludes the proof.
\end{proof}

\section{Parameterized Algorithms for \fd{}}
%!TEX root=PlanarFDeletionApproxandAlgo.tex
%\subsection{A Deterministic FPT Algorithm for \fd{}}
We now give a deterministic $O(c^kn\log^2 n)$ time FPT Algorithm for \fd{} for all connected ${\cal F} \in \mathscr{F}$.

\begin{lemma}\label{lem:DetReduceLinearFraction} For every connected ${\cal F} \in \mathscr{F}$, there exist constants $\rho$, $r$, $s$, $c < 1$ such that if we run the DFPR with parameters $r$, $s$ on an instance $(G,k)$ such that $G$ has a ${\cal F}$ deletion $S$ which is not a $\rho$-cover, then the output instance $(G',k')$ satisfies $|V(G)|-|V(G')| \geq \frac{c|V(G)|}{\log |V(G)|}$.
\end{lemma}

\begin{proof}
If $G$ has a ${\cal F}$-deletion set  $S'$ which is not an $\rho$-cover, it also has a inclusion minimal ${\cal F}$-deletion set  $S$ which is not an $\rho$-cover. Such a minimal $S$ contains no isolated vertices and hence satisfies $N[S] \leq 2\sum_{v\in S} d(v) \leq 2\rho m$. 

By Proposition~\ref{prop:planar_exclude_treewidth}, there exists a constant $b$ such that $\tw(G \setminus S) \leq b$. By Lemma~\ref{lem:prottfd}, $G$ has a $(4(b+1)|N[S]|,2(b+1))$-protrusion decomposition. Set $\beta = 2(b+1)$, $r=3(\beta+1)$ and $r'$ to be the smallest integer such that the protrusion replacer will replace $r$-protrusions of size at least $r'$. Set $s = 2^r \cdot r'$. The protrusion decomposition of $G$ is a $(4(b+1)|N[S]|,\beta)$-protrusion decomposition. By Theorem~\ref{thm:DFPR} there exist constants $0 < c < 1$ and $0 < \gamma$ such that if we run the DRFPR on $G$ and $4(b+1)|N[S]| \leq \frac{n}{244s}$ then the output graph $G'$ satisfies $|V(G)|-|V(G')| \geq \frac{c|V(G)|}{\log n}$. We show that there is a constant $\rho < \frac{1}{3}$ such that if $S$ is not a $\rho$-cover, then $|N[S]| \leq \frac{n}{1000(b+1)s}$. 

Since $\tw(G \setminus S) \leq b$ we have that $G \setminus S$ is $(b+1)$-degenerate. If $S$ is not a $\rho$-cover then $m \leq n(b+1) + \sum_{v \in S} d(v) \leq n(b+1) + 2\rho m$. Rearranging yields that $N[S]\leq 2\rho m \leq n\frac{2\rho(b+1)}{1-2\rho} \leq n\rho 6(b+1)$. Choosing $\rho = 6000(b+1)^2s$ yields that $|N[S]| \leq \frac{n}{1000(b+1)s}$. Hence, if $S$ is not a $\rho$-cover then the output instance $G'$ of the RFPR satisfies $|V(G)|-|V(G')| \geq \frac{c|V(G)|}{\log n}$.
\end{proof}

\begin{lemma}\label{lem:DetSmartReduce} For every connected ${\cal F} \in \mathscr{F}$ there is an algorithm that given an instance $(G,k)$, takes $O((n+m)\log^2 n)$ time and outputs an equivalent instance graph $(G',k')$ such that $V(G') \leq V(G)$ and $OPT(G') \leq OPT(G)$. Furthermore there exist a constant $0 < \rho < 1$ such that every ${\cal F}$-deletion set  $S'$ of $G'$ is a $\rho$-cover of $G$.
\end{lemma}

\begin{proof}
By Lemma~\ref{lem:DetReduceLinearFraction} there exist constants $\rho$, $r$, $s$, $c < 1$ such that if we run the DFPR with parameters $r$, $s$ on an instance $(G,k)$ such that $G$ has a ${\cal F}$ deletion $S$ which is not a $\rho$-cover, then the output instance $(G',k')$ satisfies $|V(G)|-|V(G')| \geq \frac{c|V(G)|}{\log |V(G)|}$. We set these constanst as guaranteed by Lemma~\ref{lem:DetReduceLinearFraction}.

The algorithm sets $(G_1, k_1) := (G, k)$, $i = 1$ and enters a loop that proceeds as follows. The algorithm runs the DFPR on $(G_i, k_i)$ with parameters $r$ and $s$, let the output of the DFPR be $(G_{i+1}, k_{i+1})$. If $|V(G_i)|-|V(G_{i+1})| < \frac{c|V(G)|}{\log |V(G)|}$ the algorithm halts and outputs $G_i$. Otherwise, the algorithm increments $i$ and returns to the beginning of the loop.

One iteration of the loop takes time $O((|V(G_i)| + |E(G_i)|)\log |V(G_i)|)$. Furthermore, every $\log n$ consecutive iterations of the loop reduces the number of vertices by a linear fraction. Hence the total running time us bounded by $O(n \log^2 n)$. Let $(G',k')$ be the instance we output. By Lemma~\ref{lem:DetReduceLinearFraction} we have that every ${\cal F}$-deletion set  $S'$ of $G'$ is an $\rho$-cover of $G$.
\end{proof}

We will say that an instance $(G,k)$ is {\em irreducible} if running the algorithm of Lemma~\ref{lem:DetSmartReduce} when run on $(G,k)$ just outputs $(G,k)$ unchanged. Observe that if we run the algorithm of Lemma~\ref{lem:DetSmartReduce} when run on an instance $(G,k)$, the instance $(G',k')$  output by the algorithm is irreducible. A direct consequence of Lemma~\ref{lem:DetSmartReduce} is that in an irreducible instance $(G,k)$ every ${\cal F}$-deletion set  $S$ in $G$ is a $\rho$-cover.

We now give a deterministic algorithm for \fd{}.\ for connected ${\cal F} \in \mathscr{F}$. The intuition behind this algorithm is that vertices of high degree seem more useful for a solution than the vertices of low degree. Towards this we introduce the notion of buckets. We partition the vertex set of $G$ into sets that we refer to as {\em buckets}, in the following fashion. For every $j\geq 1$ define 
$$B_j =\Big\{v\in V(G)~\Big|~\frac{n}{2^{j}} < d(v)  \leq \frac{n}{2^{j-1}}\Big\}.$$ 

We set constants $\eta > 0$ and $d > 0$ such that $\frac{4d + 3 \eta}{2} < \rho$. For the presentation of the algorithm we fix a ${\cal F}$-deletion set  set $X$ of size at most $k$. Next we define a notion of big and good for buckets.

\begin{definition}
A bucket $B_i$ is said to be {\em big} if  $|B_i| > i\eta$ and it is said to be {\em good} if 
$|B_i \cap X| \geq d|B_i|.$
\end{definition}
%A bucket $B_i$ is said to be {\em big} if $$|B_i| > i\eta$$ and it is said to be {\em good} if $$|B_i \cap S| \geq d|B_i|.$$
The next lemma says that if $(G,k)$ is a irreducible yes instance to \fd{} then it has a bucket that is both big and good simulatenouly. 
\begin{lemma} 
\label{lem:goodbigbucket}
For any connected ${\cal F} \in \mathscr{F}$, let $(G,k)$ be a irreducible yes instance to \fd{}. Then $G$ has a bucket that is both big and good.
\end{lemma}
\begin{proof} 
Since $(G,k)$ a irreducible yes instance to \fd{} every optimal ${\cal F}$-hitting set $X$ is a $\rho$-cover for $G$, that is, $\sum_{v \in V(G)} d(v) \leq \rho \sum_{v \in X} d(v)$. For a contradiction, assume that $G$ does not have a bucket that is both big and good. 
$$\sum_{v \in X} d(v) = \sum_{i=1}^{\log n} \sum_{v \in B_i \cap X} d(v)$$

$$= \sum_{ \{i | B_i \mbox{{\small is not good}} \}} \sum_{v \in B_i \cap X} d(v) + \sum_{ \{i | B_i \mbox{{\small is not big}} \}} \sum_{v \in B_i \cap X} d(v)$$

$$ \leq  d \cdot 4m + \sum_{ \{i | B_i \mbox{{\small is not big}} \}} i \eta \cdot  \left(\frac{n}{2^i} \right) $$

$$ \leq  d \cdot 4m + 3 \eta n = 2m\frac{4d + 3 \eta}{2} < 2m\rho $$
Which contradicts that $X$ is a $\rho$-cover.
\end{proof}

\begin{figure}[t]
\begin{center}

\begin{boxedminipage}{.96\textwidth}
\noindent 
{\bf Algorithm-FPT-Det}($G$,$k$)\\
\begin{description}
\setlength{\itemsep}{-2pt}
\item[Step 1:] Check whether $G$ is  $\cal F$-free, if yes then {\bf return}(true). Else if $k\leq 0$ and $G$ is not 
$\cal F$-free return that $G$ does not have a $k$-sized $\cal F$-hitting set. 
\item[Step 2:] Apply Lemma~\ref{lem:DetSmartReduce} on $(G,k)$ and obtain an equivalent irreducible instance $(G^*,k^*)$.  
\item[Step 3:] Let $B_{j}$, $j\in\{a,b,\ldots,\ell\}$, be the good buckets for $G^*$. For every  
good bucket $B_j$, and for every subset $S\subseteq B_j$ of size at least $d|B_j|$ check whether \\
{\bf Algorithm-FPT-Det}($G^*\setminus \{S\}$,$k-|S|$) returns true. If any of these calls return true 
then {\bf return}(true) else {\bf return}(false).  
\end{description}
\end{boxedminipage}
\end{center}

\caption{A $2^{O(k)}n\log^2 n$ deterministic FPT algorithm for \fd{}.\label{fig:detfptalgo}}
\end{figure}

\begin{theorem}
Let ${\cal F} \in \mathscr{F}$ be a connected obstruction set. There exists a determintistic algorithm for \fd{} running in time $O(c_h^kn \log^2 n)$ on a $n$ vertex graph. The constant $c_h$ only depends on ${\cal F}$.  
\end{theorem}
\begin{proof}
The deterministic algorithm for \fd{} is described in details in Figure~\ref{fig:detfptalgo}.  Given a graph $G$, the algorithm essentially applies Lemma~\ref{lem:DetSmartReduce} to obtain $G^*$ and then recursively tries to compute the solution to the problem by branching on all large subsets of all the good buckets. The correctness follows directly from Lemma~\ref{lem:goodbigbucket}. Next we analyze the running time of the algorithm. Suppose for the sake of analysis that all buckets are big, and let $a_i$ be the size of bucket $i$. Then we have that
 
$$T(k) \leq  \sum_ {i = 1} ^ { \log n} {a_i \choose k} T(k - da_i) $$

$$T(k) \leq \sum _ {i = 1} ^ { \log n} 2^{a_i}T(k - da_i) $$

Assuming $T(k) = x^k$, substitute recursively to get:

$$T(k) \leq \sum _ {i = 1} ^ { \log n} 2^{a_i}x^{(k - da_i)} $$

$$T(k) \leq x^k \sum _ {i = 1} ^ { \log n} \left( \frac{2}{x^d} \right)^{a_i} $$

If $\frac{2}{x^d} < 1$ then each term of the sum is maximized when the exponent is as small as possible. We will choose $x$ (based on d) such that $\frac{2}{x^d} < 1$ holds. Since $a_i \geq \eta i$ for any big bucket we have that 
$$T(k) \leq x^k \sum _ {i = 1} ^ { \log n} \left( \frac{2}{x^d} \right)^{\eta i} $$
The sum above is a geometric series and converges to a value that is at most $1$ for $x = c$, for a suitably small choice of $c$ depending only on $d$ and $\eta$, which depended only on ${\cal F}$. This bounds the running time by $c^k$. Further, if not all buckets are big the sum above should only be done over the big buckets, yielding the same result.
\end{proof}

%\section{Applications}
%\input{achieve.tex}
%%%%%% Kernels
\section{Kernelization for \fd{}}
In this section we give polynomial kernel for \fd{} problem. We start with some combinatorial results on 
folios and Well Quasi Ordering (WQO) of $t$-Boundaried Graphs and then move to the kernelization 
steps of our algorithm.

\subsection{Minors, Folios and Well Quasi Ordering of $t$-Boundaried Graphs}
\paragraph{Minors and folios of $t$-boundaried graphs.}
We can define a minor relation for $t$-boundaried graphs just as for normal graphs. We say that a $t$-boundaried graph $H$ is a minor of a $t$-boundaried graph $G$ if (a $t$-boundaried graph isomorphic to) $H$ can be obtained from $G$ by deleting vertices or edges or contracting edges, but never contracting edges with both endpoints being boundary vertices. Here, when we contract an edge between a boundary vertex $u$ and a non-boundary vertex $v$ the resulting vertex is a boundary vertex with label $\ell_G(u)$. If $H$ is a minor of $G$ we say that $H \leq_m G$.

For a $t$-boundaried graph $G$ the {\em folio} of $G$ is the set ${\bf folio}(G) = \{H~:~H \leq_m G\}$. The $\delta$-{\em folio} of $G$ is the set $\delta$-${\bf folio}(G) = \{H~:~|V(H)| \leq \delta \mbox{ and } H \leq_m G\}$. The following lemma states that the $\delta$-folio of a sum can be determined by examining the $\delta$-folios of its terms.

\begin{lemma}\label{lem:boundariedFolioGlue} Let $G_1$ and $G_2$ be $t$-boundaried graphs and $G = G_1 \oplus_\delta G_2$. Then $H \leq_m G$ if and only if there exist $H_1 \leq_m G_1$ and $H_2 \leq_m G_2$ such that $H_1 \oplus_\delta H_2 = H$.
%$|V(H_1)| + |V(H_2)| \leq |V(H)|+t$ and $H \leq_m H_1 \oplus H_2$.
\end{lemma}

\begin{proof} For the reverse direction, observe that $H \leq_m H_1 \oplus H_2 \leq_m G_1 \oplus G_2 = G$ and so $H \leq_m G$. It remains to prove the forward direction, so suppose $H$ is a minor of $G$. Consider a model $(P_1, P_2, \ldots P_{|V(H)|})$ of $H$ in $G$. Notice that each $P_i$ intersects $\delta(G)$ at most once. Furthermore, every $P_i$ that does not intersect $\delta(G)$ lies either entirely in $G_1$ or entirely in $G_2$. 
%Let $S_1 = \{i : P_1 \cap V(G_1) \neq \emptyset\}$ and $S_2 = \{i : P_2 \cap V(G_2) \neq \emptyset\}$. 
Now, consider the collection ${\cal M}_1$ of sets $P_i \cap V(G_1)$ such that $P_i \cap V(G_1) \neq \emptyset$ and the collection  ${\cal M}_2$ of sets $P_i \cap V(G_2)$ such that $P_i \cap V(G_2) \neq \emptyset$. Each set in ${\cal M}_1$ and ${\cal M}_2$ induces a connected subgraph in $G_1$ and $G_2$ respectively. All sets in ${\cal M}_1$ are pairwise disjoint, and the same holds for ${\cal M}_2$. Thus ${\cal M}_1$ and ${\cal M}_2$ are models of graphs $H_1 \leq_m G_1$ and $H_2 \leq_m G_2$ respecitvely. Finally $H_1 \oplus_\delta H_2 = H$ because for every $i$, $j$ there is an edge between $G[P_i]$ and $G[P_j]$ if and only if there is an edge between $G[P_i \cap V(G_1)]$ and $G[P_j \cap V(G_1)]$ or between $G[P_i \cap V(G_2)]$ and $G[P_j \cap V(G_2)]$.
\end{proof}

\begin{lemma}\label{lem:forgetFolio} Let $G'$ be a $t$-boudaried graph, $S \subseteq \delta(G')$ and $G = \forget(G',S)$. Let $H$ be a $t$-boundaried graph. Then $H \leq_m G$ if and only if there is a $t$-boundaried $H' \leq_M G'$ such that $H \leq_m \forget(H',S)$ and $|V(H')| \leq |V(H)| + |S|$. \end{lemma}
\begin{proof} For the reverse direction observe that since $H' \leq_M G'$ and $G = \forget(G',S)$ we have $\forget(H',S) \leq_m \forget(G',S) = G$. In the forward direction, consider a model $(P_1, P_2, \ldots P_{|V(H)|})$ of $H$ in $G$. Start from $G'$ and delete all vertices that are not in a set $P_i$. Then, as long as there is an edge with both endpoints in the same set $P_i$ and at least one endpoint is not boundary vertex of $G'$ contract this edge. Modify the set $P_i$ to contain the vertex resulting from the contraction, rather than the two endpoints. The resulting $t$-boundaried graph $H'$ is a minor of $G'$. Let $P_i'$ denote the vertices corresponding to $P_i$ in $H'$. Note that each $P_i'$ is a single non-boundary vertex and some subset of the boundary vertices. Since $H \leq_m G$, notice that no $P_i'$ contains two boundary vertices from $\delta(G)$. Note that $\delta(G) = \delta(G') \setminus S$, so when we consider the set $P_i'$ in $\forget(H',S)$, it contains at most one boundary vertex. Thus, at this point, $(P_1, P_2, \ldots P_{|V(H)|})$ is a model of $H$ in $H'$ so $H \leq_m H'$.

\begin{figure}[h]
\begin{center}
\includegraphics[scale=0.6]{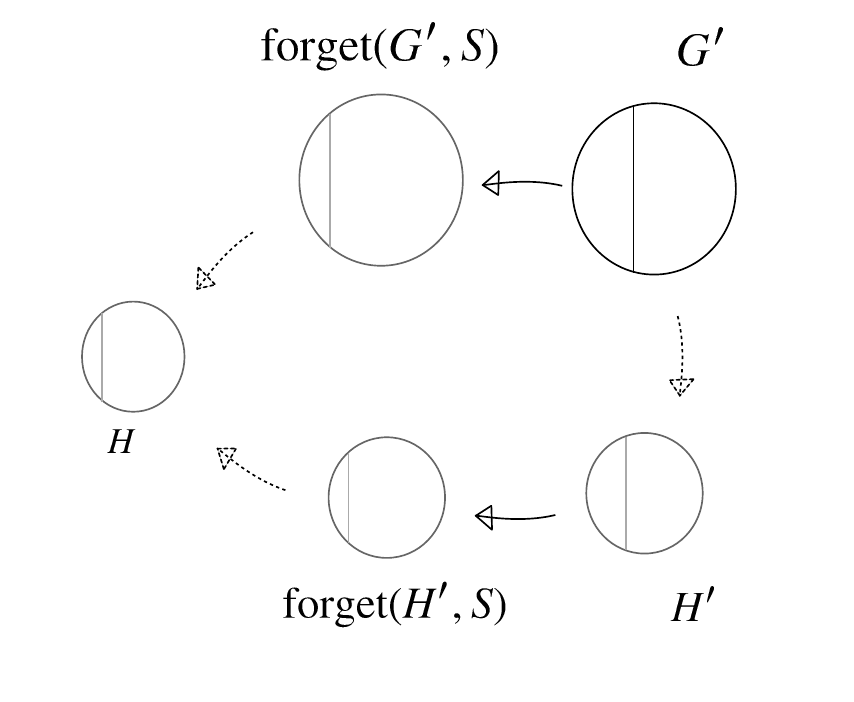}
\end{center}
\caption{The above is a schematic for Lemma~\ref{lem:forgetFolio}. The dotted lines represent minor operations, and the solid arrows represent discardig labels. The lemma shows that a minor $H$ obtained after discarding some set $S$ of labels from $G'$ can be realized by a minor $H'$ of $G'$, whose size is no more than $|H| + |S|$. The graph $H$ can be obtained as a minor of $\forget(H',S)$.}
\end{figure}

We now show that if $|V(H')| > |V(H)| + |S|$ then some $P_i$ contains at least two vertices such that at least one of them is not a border vertex. Suppose not, then every $P_i$ contains at most one vertex that is not a border vertex of $G'$. No $P_i$ contains vertices from $\delta(G') \setminus (\delta(H) \cup S)$ since $(P_1, \ldots, P_{|V(H)|})$ is a minor model of $H$ in $G$ and we can't contract edges between border vertices of $G$. Then every $P_i$ that contains at least one vertex from $\delta(G')$ is a subset of $\delta(H) \cup S$. At most $|V(H)|-|\delta(H)|$ $V_i$'s contain no vertex from $\delta(H) \subseteq \delta(G')$ and each of these $P_i$'s has size $1$. So the total size of the $P_i$'s is at most $|\delta(H)| + |S| + |V(H)| - |\delta(H)| = |V(H)|+|S|$. But every vertex in $H'$ is in some $P_i$, contradicting $|V(H')| > |V(H)| + |S|$. 

Suppose now that $|V(H')| > |V(H)| + |S|$ and consider a $P_i$ that contains at least two vertices such that at least one of them is not a border vertex. Since $H'[P_i]$ is connected there is at least one edge in $H'[P_i]$ that has at least one endpoint which is not on the border of $G'$. This edge should have been contracted, contradicting the construction of $H'$ this proves $|V(H')| \leq |V(H)| + |S|$, concluding the proof.
\end{proof}

The following lemma follows directly from Lemmata~\ref{lem:boundariedFolioGlue} and ~\ref{lem:forgetFolio}. It states that if a graph $G$ has a separator of size $t$, then we can determine whether $G$ contains $H$ as a minor by examining the $(t+|V(H)|)$-folios of the two sides of the separator.

\begin{lemma}\label{lem:folioGlue} Let $G_1$ and $G_2$ be $t$-boundaried graphs and $G = G_1 \oplus G_2$. A graph $H$ is a minor of $G$ if and only if there exist $H_1 \leq_m G_1$ and $H_2 \leq_m G_2$ such that $|V(H_1)| \leq |V(H)|+t$, $|V(H_2)| \leq |V(H)|+t$ and $H \leq_m H_1 \oplus H_2$.
\end{lemma}
\begin{proof} The reverse direction follows since $H \leq_m H_1 \oplus H_2 \leq_m G_1 \oplus G_2 = G$. For the forward direction let $G^* = G_1 \oplus_\delta G_2$. We have that $G = \forget(G^*,\delta(G^*))$ and so by Lemma~\ref{lem:forgetFolio} there exists an $H^* \leq_m G^*$ such that $|V(H^*)| \leq |V(H)|+t$ and $H \leq_m \forget(H^*,\delta(H^*))$. By Lemma~\ref{lem:boundariedFolioGlue} there exist $H_1 \leq_m G_1$ and $H_2 \leq_m G_2$ such that $H_1 \oplus_\delta H_2 = H^*$. But then $|V(H_1)| \leq |V(H^*)| \leq |V(H)|+t$ and identically $|V(H_2)| \leq |V(H^*)| \leq |V(H)|+t$. Finally, since $H_1 \oplus_\delta H_2 = H^*$ we have $H_1 \oplus H_2 = \forget(H^*, \delta(H^*))$ and since $H \leq_m \forget(H^*, \delta(H^*))$ the proof follows.

\begin{figure}[h]
\begin{center}
\includegraphics[scale=0.8]{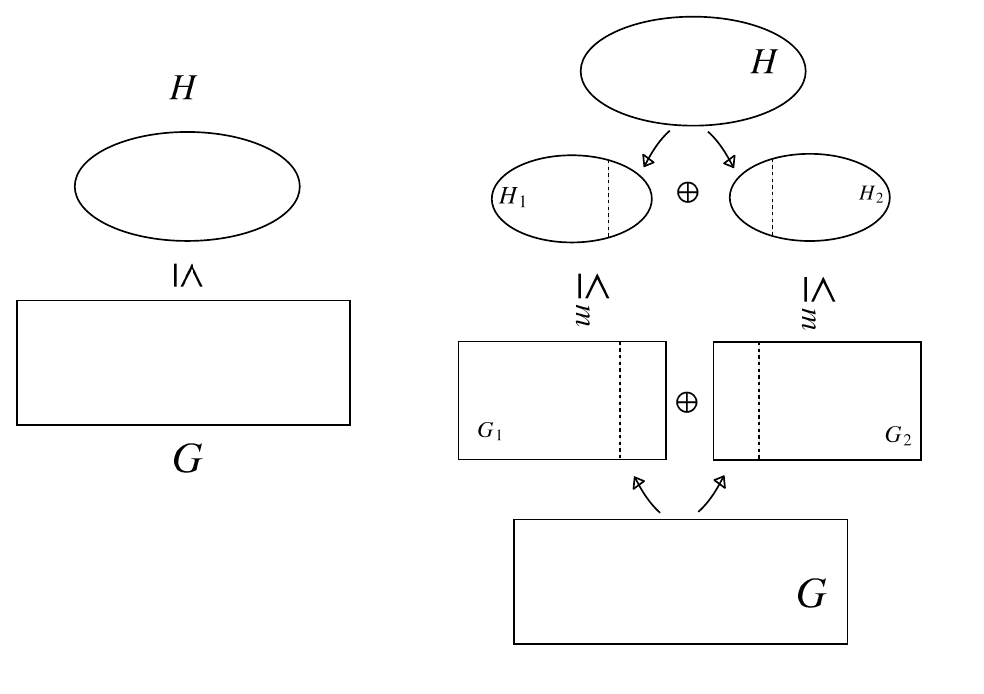}
\caption{A schematic representation of Lemma~\ref{lem:folioGlue}.}
\end{center}
\end{figure}

\end{proof}

%We remark that a stronger variant of Lemma~\ref{folioGlue} with $|V(H_1)| + |V(H_2)| \leq |V(H)|+t$ may be proved directly \todo{check this!} but we do not attempt to optimize constants.

\paragraph{Well-quasi-ordering $t$-boundaried graphs.}
A partial order of a (possibly infinite) set $S$ is a relation $\leq$ on $S$ that is reflexive, transitive and antisymmetric. Two elements $a$ and $b$ of $S$ are {\em comparable} with respect to $\leq$ if $a \leq b$ or $b \leq a$ and {\em incomparable} otherwise. A subset $S'$ of $S$ is an {\em antichain} (of $\leq$) if no two distinct elements of $S'$ are comparable. A partial order $\leq$ is a {\em well-quasi-order} of $S$ if every antichain $S' \subset S$ is finite. The famous graph minor theorem states that graphs are well-quasi-ordered by the minor relation~\cite{RobertsonS-GMXIII}. We will need a weaker version of this theorem translated to $t$-boundaried graphs.

%!TEX root=PlanarFDeletionKernel.tex

\begin{lemma}\label{lem:wqoTBoundaried} For every $t$ and $w$ the set of $t$-boundaried graphs with treewidth at most $w$ is well-quasi-ordered. \end{lemma}

\begin{proof} We prove the lemma by giving an injection $f$ from the set of $t$-boundaried graphs of treewidth at most $w$ to graphs such that for any two uncomparable $t$-boundaried graphs $H$ and $G$ of treewidth at most $w$, $f(H)$ and $f(G)$ are uncomparable as well. Hence an infinite antichain of $t$-boundaried graphs of treewidth at most $w$ would yield an infinite antichain of graphs, contradicting the graph minor theorem. 

For a $t$-boundaried graph $G$, let $L(G)$ denote the set of labels used by the boundary vertices of $G$, that is, $L(G) := \{ l(v) ~|~ v \in \delta(G)\}$. We begin by noticing that it suffices to show the injection $f$ from the set of $t$-boundaried graphs  $G$ for which $L(G) = X$ for some arbitrary but fixed set $X \subseteq [t]$. This is because there are only finitely many subsets of $[t]$, and therefore any infinite antichain of $t$-boundaried graphs of treewidth at most $w$ contains an infinite antichain of $t$-boundaried graphs whose boundary sets use the same set of labels. 

Let $G$ be a $t$-boundaried graph. The graph $f(G)$ is obtained from $G$ by adding some cliques to the vertices of $\delta(G)$. In particular, let $v \in \delta(G)$ and let $l(v) = l_v$. For every $v \in \delta(G)$, in $f(G)$ we introduce a clique $C_v$ on $w + l_v + 1$ vertices and make $v$ adjacent to all vertices in $C_v$. We use $G^*$ to denote $f(G)$.

\begin{figure}[h]
\begin{center}
\includegraphics[scale=0.8]{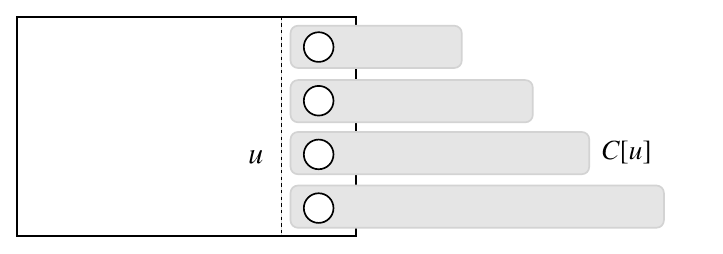}
\caption{A representation of the function $f$. The grey rectangles represent cliques.}
\end{center}
\end{figure}

Recall that $X$ is an arbitrary but fixed subset of $[t]$. Let  $H$ and $G$ be two  $t$-boundaried graphs of treewidth at most $w$ such that $X$ is the set of labels used by the boundary vertices of $H$ and $G$. We claim that if $H^* \leq_m G^*$, then $H \leq_m G$. 

Let $H^* \leq_m G^*$. Let $(P_1, \ldots, P_{V(H^*))})$ be a model of $H^*$ in $G^*$. We abuse notation and use $P_v$ to denote the component of $G^*$ that the vertex $v \in H^*$ is mapped to by the model under consideration. Now, let $u \in \delta(H)$ with label $l_u$. Let $u_g$ be the vertex in $\delta(G)$ that has label $l_u$. Since $H$ is a subgraph of $H^*$, $u \in H^*$, and similarly, $u_g \in G^*$. Let $C[u]$ and $C_g[u]$ denote the vertices of the clique attached to the vertices $u$ and $u_g$ in $H^*$ and $G^*$ (respectively). Further, let $B_G$ denote the set $\delta(G)$ in $G^*$, let $V_G$ denote the vertices of $G \setminus B$ in $G^*$, and let $C_G$ denote the set of all vertices that participate in the cliques introduced in $G^*$ (by construction). 

We first show that for every vertex $z$ in $C[u]$, $P_z$ contains some vertex of $C_g[u]$. We prove this by reverse induction on the set of labels $X$ used in $G$. Let $r$ denote $|X|$, and assume, without loss of generality, that the labels used are from $[r]$. The base case is that $l_u = r$. Consider some $z \in C[u]$, and suppose, for the sake of contradiction, that $P_z \cap C_g[u] = \emptyset$. Note that $P_z$ must contain a vertex whose neighborhood in $G^*$ contains a clique on at least $r+w+1$ vertices. The only such vertices in $G^*$ are $C_g[u] \cup \{u_g\}$. Given our assumption, this implies that $u_g \in P_z$. But now consider $P_u$, which must also contain a vertex whose neighborhood in $G^*$ contains a clique on at least $r+w+1$ vertices, and whose degree is at least $r+w+2$ (since there are no isolated boundary vertices in $H$). Since $u_g \in P_z$, the only remaining candidates for $P_u$ are vertices in $C_g[u]$, but all such vertices have degree exactly $r+w+1$, and therefore, there is no valid choice for $P_u$. This is a contradiction.   

The induction hypothesis states that if $l_u = s+1$, for all $z \in C[u]$, $P_z$ contains some vertex of $C_g[u]$. We now prove that when $l_u = s$, for all $z \in C[u]$, $P_z$ contains some vertex of $C_g[u]$. It is clear that $P_z$ must contain a vertex whose neighborhood in $G^*$ contains a clique on at least $s+w+1$ vertices. The vertices which satisfy this property include $\{ C_g[v] \cup \{v_g\} \}$ for all $v$ for which $l_v \geq s$. Notice that all vertices in $\cup_{\{ v ~|~ l_v > s\} } C_g[v] $ already belong to $P_z$ for some $z \in C[v]$. Since the $P_i$'s are disjoint, the only remaining candidate vertices for $P_z$ are $\{ v_g ~|~ l_v \geq s \} \cup C_g[u]$. For reasons similar to the ones described in the proof of the base case, $P_z$ cannot contain the vertex $v_g$ for any $v \in \delta(G)$. Therefore, $P_z$ must contain some vertex from $C_g[u]$, as desired. This proves the claim. 

We now argue that for any $u \in \delta(H)$, we have that the corresponding vertex in $H^*$ is mapped to a set that contains the boundary vertex with the same label as $u$ in $G^*$. Formally, we claim that $u_g \in P_u$. It is clear that $P_u$ must contain a vertex whose neighborhood in $G^*$ has a clique on at least $l_u + w + 1$ vertices. Since the treewidth of $G$ is at most $w$, we have that $G^*[V_G \cup B_G]$ does not contain any cliques of size $w+2$. Notice also that all the vertices in $C_G$ are shown to belong distinct sets $P_z$. This follows from the claim above and the fact that the set of labels of $H$ and $G$ are identical --- which implies that the sizes of the corresponding cliques introduced in $H^*$ and $G^*$ are the same. Thus, it is clear that only the vertices from $B_G$ have sufficiently large cliques in their neighborhood. It is easy to see that a reverse induction argument similar to the one used in the claim above establishes that $u_g \in P_u$. 

Thus, we have that in any minor model of $H^*$ in $G^*$, the clique vertices introduced in $H^*$ are mapped to the corresponding clique vertices in $G^*$, and the vertices of $\delta(H)$ in $H^*$ are mapped to the corresponding vertices in $\delta(G)$ in $G^*$. Now we simply consider the minor model of $H^*$ in $G^*$ after disregarding the $P_z$'s that contain vertices from $C_G$. This is evidently a minor model of $H$ in $G$, and therefore, we have that $H \leq_m G$, as desired. This concludes the proof.

\end{proof}
%\begin{lemma}\label{lem:wqoTBoundaried} For every $t$ and $w$ the set of $t$-boundaried graphs with treewidth at most $w$ is well-quasi-ordered. \end{lemma}

%\begin{proof} We prove the lemma by giving an injection $f$ from the set of $t$-boundaried graphs to graphs such that for any two uncomparable $t$-boundaried graphs $G_1$ and $G_2$ of treewidth at most $w$, $f(G_1)$ and $f(G_2)$ are uncomparable as well. Hence an infinite antichain of $t$-boundaried graphs of treewidth at most $w$ would yield an infinite antichain of graphs, contradicting the graph minor theorem. \todo{rest of proof!}
%\end{proof}

\subsection{Decomposition into near-Protrusions}
%!TEX root=PFDMerge-KernelAlgo.tex

In this section we prove that any yes-instance $G$ to ${\cal F}$-{\sc Deletion} has a set $D$ of $O(k^3)$ vertices such that every connected component $C$ of $G \setminus D$ is a {\em near-protrusion}. Recall that a $r$-protrusion $C$ in a graph $G$ is a vertex set such that $|\delta(C)| \leq r$ and ${\bf tw}(G[C]) \leq r$. The components of $G \setminus D$ will not necessarily be protrusions, however we will prove that there is a constant $r$ such that if $(G,k)$ is a yes instance, then for any ${\cal F}$-deletion set $S$ of size at most $k$, $C \setminus S$ is a $r$-protrusion of $G \setminus S$. The decomposition algorithm starts by finding a constant factor approximation of the smallest ${\cal F}$-deletion set of $G$ with high probability. This is achieved by running the algorithm given by Theorem~\ref{thm:approx_thm} $n$ times, so as to boost the probability of success to $(1-1/2^n)$.

%{\bf WHAT THE HELL IS THIS...}
%\todo{make sure apx algorithm is actuallty quadratic!}
%\begin{proposition}[\cite{}] For every ${\cal F}$ containing a planar graph, ${\cal F}$-{\sc Deletion} has a constant factor approximation algorithm in time $O(n^2)$. 
%\end{proposition}

By Proposition~\ref{prop:planar_exclude_treewidth} there exists a constant $\eta$ such that every ${\cal F}$-minor-free graph $G$ has treewidth at most $\eta$. Having computed a $c$-approximate ${\cal F}$-deletion set $X$ we check whether $|X| \leq c(k+1)$, and if not\footnote{The kernelization algorithm only requires to check if $|X| \leq ck$, but we choose to use the bound $c(k+1)$ to allow for a more general consequence.}, then the algorithm concludes that $G$ has no ${\cal F}$-deletion set of size $k$. We now compute a set $Y$ disjoint from $X$ as follows. Initially $Y=\emptyset$. For a pair $u$ and $v$ of distinct vertices in $X$, define $G_{u,v}$ to be $(G \setminus (X \cup Y \setminus {u,v})) \setminus uv$. In other words $G_{u,v}$ is obtained from $G$ by removing all vertices in $X$ and $Y$ except for $u$ and $v$, and removing the edge $uv$ if it exists. We check using maximum flow whether there exists a pair of distinct vertices $u$, $v \in X$ such that the number of vertex disjoint paths from $u$ to $v$ in $G_{u,v}$ is less than $\eta+k+3$, but at least one. If such a pair exists then by Menger's theorem there is a set $S \subseteq V(G) \setminus (X \cup Y)$ of size at most $\eta+k+2$ such that there are no paths from $u$ to $v$ in $G_{u,v}$. We add the set $S$ to $Y$. Observe that $|Y| \leq |X|^2 \cdot (\eta+k+2)$ since we add a set of $\eta+k+2$ vertices at most once for each pair of vertices in $X$. 

\begin{figure}[h]
\begin{center}
\includegraphics[scale=1.0]{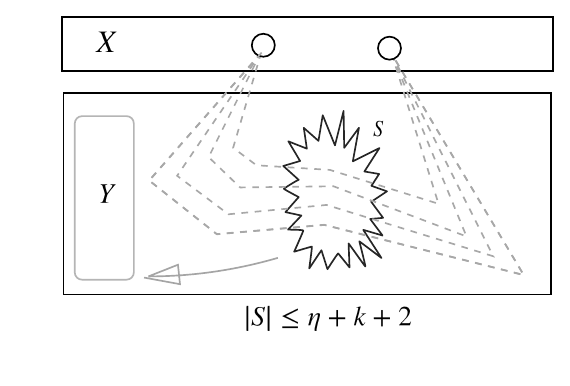}
\end{center}
\caption{This figure demonstrates the construction of $Y$ based on small separating sets for pairs of vertices in $X$.}
\end{figure}

Furthermore, for every connected component $C$ of $G \setminus (X \cup Y)$ it holds that for any two vertices $u$ and $v$ in $N(C) \cap X$ there are at least $\eta+k+3$ vertex disjoint paths from $u$ to $v$. Indeed, if this was not the case then a vertex set separating $u$ and $v$ in $G_{u,v}$ would have been added to $Y$, contradicting the construction of $Y$.

\begin{figure}[h]
\begin{center}
\includegraphics[scale=1.0]{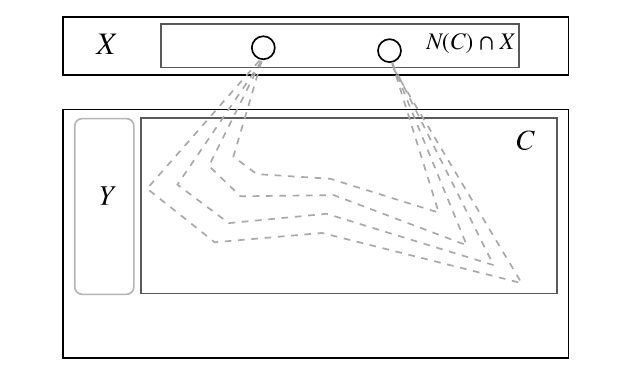}
\end{center}
\caption{After the construction of $Y$, for every component $C$ outside $(X \cup Y)$, every pair of vertices in $X$ in $N(C)$ has a large flow through $G \setminus (X \cup Y)$}
\end{figure}

\begin{figure}[h]
\begin{center}
\includegraphics[scale=0.8]{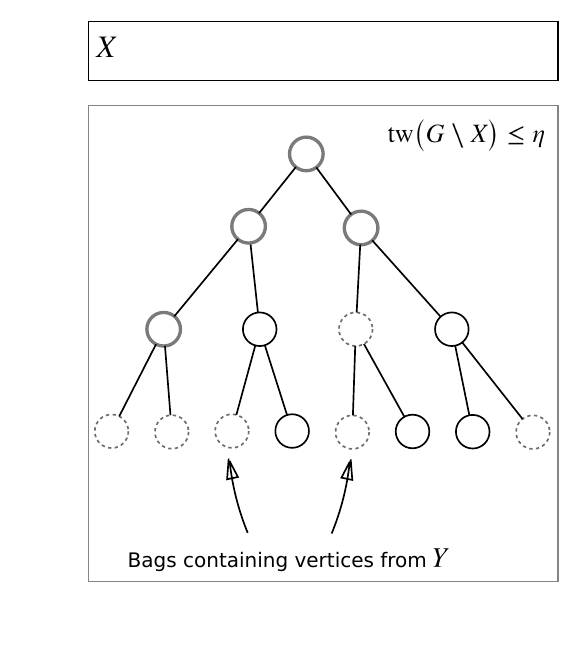}
\end{center}
\caption{This is a schematic for the construction of $Z$ based on $Y$ and a tree decomposition of $G \setminus X$.}
\label{fig:constructionofZ}
\end{figure}

We will now construct a set $Z \supseteq Y$ from $Y$ and $X$ as follows (see Figure~\ref{fig:constructionofZ}). The graph $G \setminus X$ is ${\cal F}$-minor-free and so the treewidth of $G \setminus X$ is at most $\eta$. We compute a nice tree-decomposition $(T,{\cal B}=\{B_1,B_2,\ldots,B_{|V(T)|}\})$ of $G \setminus X$ of width at most $\eta$ in linear time using Bodlaender's algorithm~\cite{Bodlaender96}. For each vertex $v \in Y$ we put one node $t$ of $T$ such that $v \in B_t$ into the set $M'$. Let $M$ be the LCA-closure of $M'$. We refer to $M$ as the set of {\em marked} nodes. By Lemma~\ref{lem:lcaClosure} $|M| \leq 2|M'| \leq 2|Y|$. Set $Z = \bigcup_{t \in M} B_t$. Thus $Y \subseteq Z$, $|Z| \leq 2|Y|(\eta + 1)$ and $Z$ and $X$ are disjoint.

We now prove that for any connected component $C$ of $G \setminus (X \cup Z)$ only sees a constant number of vertices in $Z$. For any connected component $C$ of $G \setminus (X \cup Z)$ there is a component $P$ of $T \setminus M$ such that $C \subseteq (\bigcup_{t \in P} B_t) \setminus Z$. By Lemma~\ref{lem:lcaClosure} we have that in $T$, $|N(P)| \leq 2$. But then $N(C) \cap Z \subseteq \bigcup_{t \in N(P)} B_t$, and hence $|N(C) \cap Z| \leq 2(\eta + 1)$. We can summarize the above discussion in the following result.

\begin{lemma}\label{lem:decomposition} There is a randomized polynomial time algorithm that given an instance $(G,k)$ of ${\cal F}$-{\sc Deletion}, either returns that $(G,k)$ is a no instance, or computes a vertex set $X$ and a set $Z$ disjoint from $X$ such that
\begin{itemize}
\item $|X| = O(k)$ and $|Z| = O(k^3)$.
\item $X$ is an  ${\cal F}$-{\sc Deletion} of $G$.
\item For every connected component $C$ of $G \setminus (X \cup Z)$, $|N(C) \cap Z| \leq 2(\eta+1)$.
\item For every connected component $C$ of $G \setminus (X \cup Z)$, and $u$, $v \in N(C) \cap X$ there are at least $k+\eta+3$ vertex disjoint paths from $u$ to $v$ in $G$.
\end{itemize}
If $(G,k)$ is a yes instance, the algorithm outputs $(X,Z)$ with probability $(1-1/2^n)$, and a no instance is rejected with probability $1$. 
\end{lemma}
In the remainder of the paper the sets $X$ and $Z$ will always refer to the sets $X$ and $Z$ as guaranteed by Lemma~\ref{lem:decomposition}. We now observe that the connected components of $G \setminus (X \cup Z)$ are near-protrusions. 
\begin{lemma}\label{lem:nearProtrusion} For any ${\cal F}$-{\sc Deletion} $S$ of $G$ of size at most $k+1$ and any connected component $C$ of $G \setminus (X \cup Z)$, $|(N(C) \cap X) \setminus S| \leq \eta + 1$. \end{lemma}
\begin{proof}
Let $P = (N(C) \cap X) \setminus S$. Let $(T, {\cal B})$ be a tree-decomposition of $G \setminus S$ of width at most $\eta$. We prove that there is a bag of this tree decomposition that contains $P$, thereby proving $|P| \leq \eta+1$. Since tree-decompositions enjoy the Helly property, it is sufficient to show that for every pair of vertices $u$ and $v$ in $P$ there is a bag of the tree-decomposition that contains both $u$ and $v$~\cite{Kloks1994}. The proof proceeds by contradiction.

\begin{figure}[h]
\begin{center}
\includegraphics[scale=0.8]{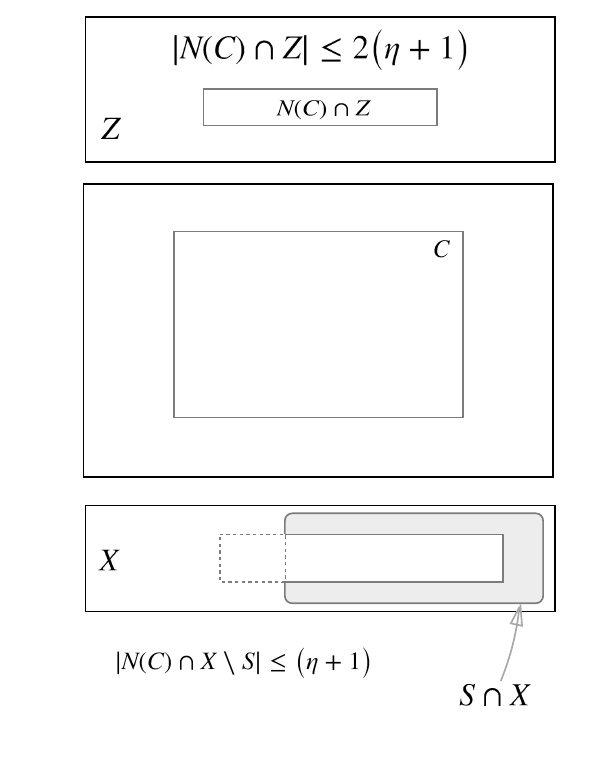}
\end{center}
\caption{This figure shows, for a connected component $C$ in $G \setminus (Z \cup X)$, the bounds on the set $N(C)$ in $G \setminus S$. The neighborhood in $Z$ is bounded based on the construction of $Z$. The fact that the neighborhood in $X \setminus S$ is bounded relies on the design of $C$, which ensured large flows between pairs of vertices in $X$ via $G \setminus (Z \cup X)$ (see Lemma~\ref{lem:decomposition}).}
\end{figure}

Assume that no bag of $T$ contains both $u$ and $v$. Observe that for every pair of vertices $u$ and $v$ in $P$, there are at least $k+\eta+3$ disjoint paths from $u$ to $v$ in $G$ and hence there are at least $\eta + 2$ disjoint paths from $u$ to $v$ in $G \setminus S$. But since no bag of $T$ contains both $u$ and $v$ then $G \setminus S$ has a $u$-$v$ separator of size at most $\eta$, contradicting the existence of  $\eta + 2$ disjoint paths from $u$ to $v$.
\end{proof}

\subsection{Reducing the size of near-Protrusions}
%!TEX root=PlanarFDeletionKernel.tex

We now consider the sets $X$ and $Z$ as guaranteed by Lemma~\ref{lem:decomposition}, and a single component $C$ of $G \setminus (X \cup Y)$. The aim of this section is to prove that if $C$ is ``too large'', then it is ``not doing its job efficiently''. In particular we prove that there exists a constant $\alpha$ depending only on ${\cal F}$ such that if $|C| \geq \alpha \cdot k^\alpha$ then there exists an edge $e$ with at least one endpoint in $C$, an edge $e'$ with both endpoints in $C$, or a vertex $v \in C$ such that deleting $e$, contracting $e'$ or deleting $v$ does not change whether $G$ has an ${\cal F}$-deletion set  of size at most $k$. Let $G'$ be the graph obtained from $G$ by doing this minor operation. If $G$ does have an ${\cal F}$-deletion set  of size at most $k$ then $G'$ does as well, since minor operations can not increase the size of the minimum ${\cal F}$-deletion set . Thus it is sufficient to prove that if $G$ does not have an ${\cal F}$-deletion set  of size at most $k$, then neither does $G'$. We prove this by showing that for any set $S$ on at most $k$ vertices, if $G \setminus S$ contains a copy of a graph $H$ in ${\cal F}$ as a minor, then one of the following two things must happen in $G'$. Either the treewidth of $G'\setminus S$ is more than $\eta$, or the model of $H$ in $G \setminus S$ can be modified such that the minor operation used to obtain $G'$ from $G$ does not destroy it. In both cases this yields a proof that $S$ is not a ${\cal F}$-deletion set  in $G'$. The reduction rule we give in this section is designed so that the plan for the proof of correctness sketched above will go through. 

We will focus our attention on a particular component $C$ of $G \setminus (X \cup Y)$. This component has treewidth at most $\eta$, and so we can compute in linear time a nice tree decomposition $(T, {\cal B})$ of $C$ of width at most $\eta$. For every vertex $v \in C$ let $T_v$ be the subtree of $T$ corresponding to bags that contain $v$. We will say that a set $P$ of vertices in $C$ is {\em interesting} if there exist a subtree $T'$ of $T$ such that $|N(T')| \leq 2$, and $P = \{v \in C~:~T_v \subseteq T'\}$. There are at most $O(n^2)$ interesting sets, and all interesting sets can be listed in polynomial time. Every interesting set $P$ satisfies that $|N(P) \setminus X| \leq 4(\eta + 1)$.

\begin{figure}[h]
\begin{center}
\includegraphics[scale=0.7]{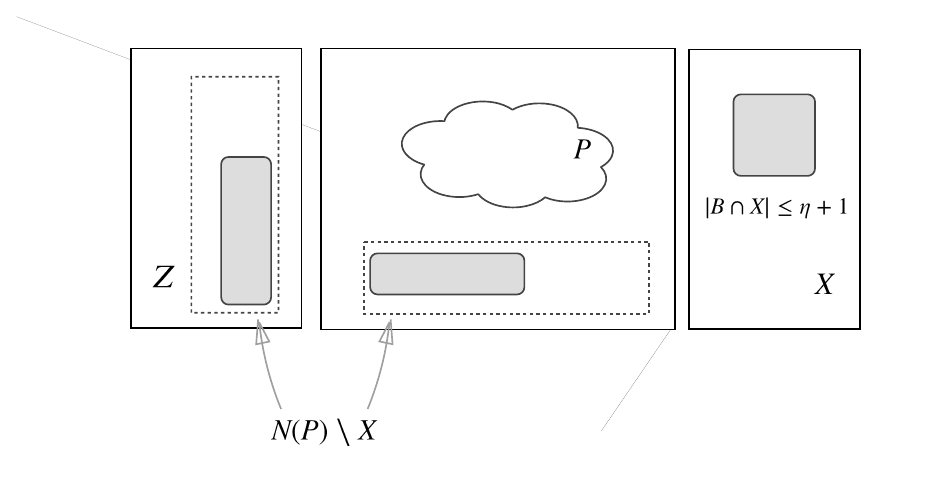}
\end{center}
\caption{A representation of a set in the border collection of $P$. The gray areas show a subset $B$ that would be an element of ${\cal B}_P$. It satisfies that outside $X$, $B$ is contained in $N(P)$, and the intersection of $B$ with $X$ is at most $\eta + 1$. }
\label{fig:border-collection}
\end{figure}

For a vertex set $P$ in $V(G) \setminus (X \cup Y)$, the {\em border collection} of $P$ is the collection ${\cal B}_P$ of all sets $B$ such that $B \setminus X \subseteq N(P) \setminus X$ and $|B \cap X| \leq \eta+1$. (See Figure~\ref{fig:border-collection}.) The {\em signature} of a vertex set $P$ of $G \setminus (X \cup Y)$ is a function $\sigma_P$. The domain of $\sigma_P$ is the set of pairs $(B, {\cal Q})$ where $B \in {\cal B}_P$ while ${\cal Q}$ is a collection of $t$-boundaried graphs on at most $|B| + h$ vertices, and with $B$ as their boundary. Here $h$ is the maximum number of vertices of a graph in ${\cal F}$.
%Every $t$-boundaried graph $Q \in {\cal Q}$ satisfies that $|V(Q)| \leq |B|+h$, where $h$ is the maximum number of vertices of a graph $H \in {\cal F}$. 

\begin{figure}[h]
\begin{center}
\includegraphics[scale=0.7]{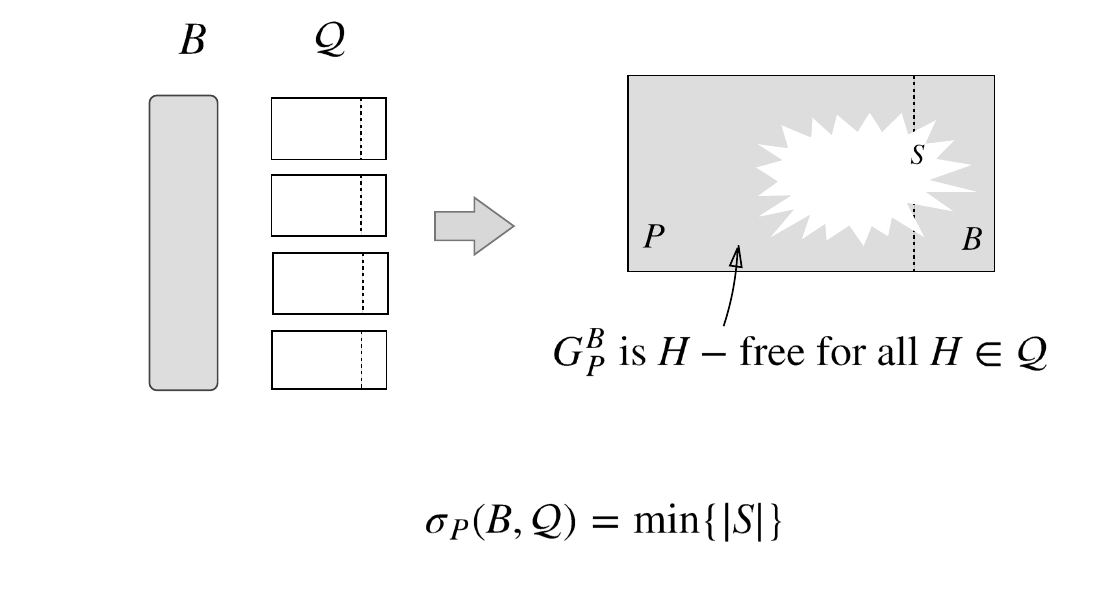}
\end{center}
\caption{The signature function. For a boundary set $B$ and a family $\mathcal{Q}$, the signature function~$\sigma_P$ considers the graph $G_P^B$, that is, the graph with boundary $B$ and $P$ as the internal set. It returns the size of the smallest subset $S$ of $P \cup B$ whose removal makes the graph $H$-minor free for all $H \in \mathcal{Q}$.}
\label{fig:signature}
\end{figure}

For such a pair $(B, {\cal Q})$, $\sigma_P(B, {\cal Q})$ outputs the size of a minimum cardinality set $S \subset P \cup B$ such that {\bf folio}$(G_C^B \setminus S) \cap {\cal Q} = \emptyset$.  (See Figure~\ref{fig:signature}.) Recall that $G_P^B$ is the $t$-boundaried graph $G[P \cup B]$ with boundary $B$. The signature of a vertex set $P$ is taken with respect to the graph $G$ - we will sometimes consider the signature of $P$ with respect to $G'$, a graph obtained from $G$ by a single minor operation. For an integer $r$ the $r$-truncated signature of $P$ is a function $\sigma_P^r$ with the same domain as $\sigma_P$. The function $\sigma_P^r(B, {\cal Q})$ evaluates to $\sigma_P(B, {\cal Q})$ if $\sigma_P(B, {\cal Q}) \leq r$ and $\sigma_P^r(B, {\cal Q}) = \infty$ otherwise.

% In $G'$, $C$ might no longer be a connected component because of a deletion of a vertex or edge, but the function $\sigma$ is still well defined.
%If the size of the minimum such set is at least $k+2$ then $\sigma(B, {\cal Q})$ outputs $\infty$. 

%\todo{check boundary size constants.}

\begin{lemma}\label{lem:computeSig} The signature of an interesting set $P$ can be computed in time $O(n^{O(\eta)})$. \end{lemma}
\begin{proof} Observe that the border collection ${\cal B}_P$ has size $O(k^{O(\eta)}) = O(n^{O(\eta)})$, since $|N(P) \setminus X| \leq 4(\eta + 1)$. For each choice of $B \in {\cal B}_P$, the size of ${\cal Q}$ is given by the number of possible collections of $t$-boundaried graphs with border $B$ and at most $|B|+h$ vertices, which is at most $2^{2^{O(|B|+h)^2}}$, a function of ${\cal F}$. Thus we need to show that we can in polynomial time evaluate $\sigma(B, {\cal Q})$ for any choice of $B$ and ${\cal Q}$. This amounts to finding a minimum cardinality set $S \subseteq P \cup B$ such that {\bf folio}$(G_P^B \setminus S) \cap {\cal Q} = \emptyset$. This problem can be expressed by the following MSO-optimization problem:

\begin{align}
\label{cmso:hitset}
\mbox{Minimize }S\subseteq V(G)[\bigwedge_{H \in \cal Q}\neg \phi(G_P^B \setminus S,H,\delta(G_P^B \setminus S),\delta(H))]
\end{align}

Recall that $\phi$ is as defined in equation~(\ref{cmso:boundariedminor}) (Section~\ref{countmsop}), has length depending only on $h$ and $|B|$. Clearly, the length of this formula depends only on ${\mathcal F}$. Since $G_C^B$ has treewidth at most $5(\eta+1)$ and MSO-optimization problems can be solved in linear time on graphs of bounded treewidth~\cite{BorieParkerTovey1992}, the proof follows.
\end{proof}

\begin{lemma}\label{lem:correctReduce1} Let $P$ be an interesting set and let $G'$ and $P'$ be obtained from $G$ and $P$ respectively by deleting a vertex of $P$, deleting an edge with at least one endponint in $P$, or contracting an edge with both endpoints in $P$. Suppose the $k$-truncated signatures $\sigma_P^{k}$ of $P$ and $\sigma_{P'}^{k}$ of $P'$ are identical, that is $\sigma_P^{k}(B, {\cal Q}) = \sigma_{P'}^{k}(B, {\cal Q})$ for every $(B, {\cal Q})$ with $B \in {\cal B}_P$ and ${\cal Q}$ being a family of t-boundaried graphs of size at most $|B|+h$. Then $G$ has a ${\cal F}$-deletion set  of size $k$ if and only if $G'$ does.
\end{lemma}
\begin{proof}
Since $G'$ is a minor of $G$ it follows that when $G$ has a ${\cal F}$-deletion set  of size $k$ then $G'$ does as well. Suppose now that $G'$ has a ${\cal F}$-deletion set  $S'$ of size at most $k$. $G'$ was obtained from $G$ by deleting a vertex $v \in P$ or deleting or contracting some edge $e$ with an endpoint $v \in P$. Now $S^* = S' \cup \{v\}$ is a ${\cal F}$-deletion set  of $G$ of size at most $k+1$, and so by Lemma~\ref{lem:nearProtrusion} $|N(C) \cap (X \setminus S^*)| \leq \eta+1$. Recall that $C$ is the component of $G \setminus (X \cup Y)$ that contains $P$. Hence  $|N(P) \cap (X \setminus S^*)| \leq \eta+1$ as well. Since $S^* \cap X = S' \cap X$ it follows that $|N(P') \cap (X \setminus S')| \leq \eta + 1$.

\begin{figure}[h]
\begin{center}
\includegraphics[scale=0.9]{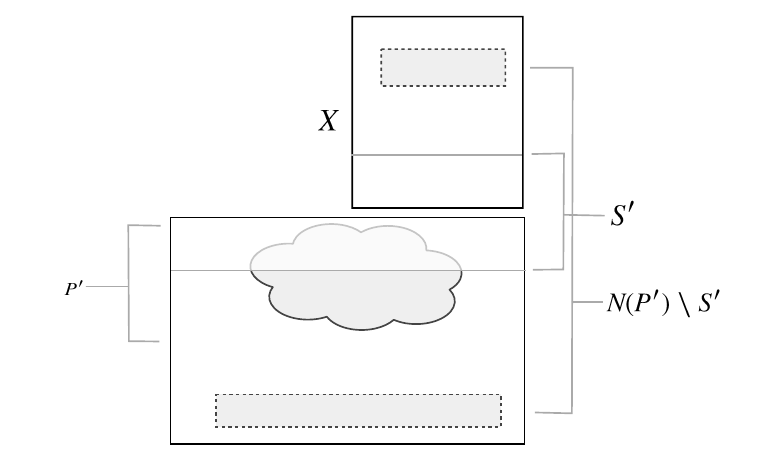}
\end{center}
\caption{A schematic representation of the various subsets involved in the proof of Lemma~\ref{lem:correctReduce1}.}
\end{figure}

%Since $|N(C') \cap Z| \leq 2(\eta + 1)$ we have that $|N_{G'\setminus S'}(C' \setminus S')| \leq 3(\eta + 1)$. 
Set $B = N_{G'\setminus S'}(P' \setminus S')$ and observe that $B \in {\cal B}_P$. Now define $G_{P'\setminus S'}^B$ to be the $t$-boundaried graph $G'[(P' \setminus S') \cup B]$ with boundary $B$. Set ${\cal Q}$ to be the set of all $t$-boundaried graphs of size at most $(5(\eta + 1)+h)$ that are {\em not} minors of $G_{P'\setminus S'}^B$. Observe that $\sigma'_{P'}(B, {\cal Q}) \leq |S' \cap P'|$. Thus there exists a set $S_P$ of size at most $|S' \cap P'|$ such that $G_P^B \setminus S_P$ excludes every graph ${\cal Q}$ as a minor. Set $S = (S' \setminus P') \cup S_P$. Observe that since $|S_P| \leq |S' \cap P'|$ it follows that $|S| \leq |S'| \leq k$. We argue that $S$ is a ${\cal F}$-deletion set  of $G$.

Define $G_R^B$ to be the $t$-boundaried graph $G' \setminus (S' \cup P')$ with boundary $B$. Observe that $G_R^B$ is also equal to $G \setminus (S \cup P)$ with boundary $B$. Thus we have
$$G' \setminus S' = G_R^B \oplus (G_{P'}^B\setminus S')~~~~~~
G \setminus S = G_R^B \oplus (G_P^B \setminus S_P).$$
Since $(5(\eta + 1)+h)$-{\bf folio}$(G_P^B \setminus S) \subseteq (5(\eta + 1)+h)$-{\bf folio}$G_{P'\setminus S'}^B$ and $G_R^B \oplus (G_{P'}^B\setminus S')$ is ${\cal F}$-free, Lemma~\ref{lem:folioGlue} implies that $G_R^B \oplus (G_P^B \setminus S_P) = G \setminus S$ is ${\cal F}$-free as well. Hence $S$ is a ${\cal F}$-deletion set  of $G$ of size at most $k$.
\end{proof}

Lemmata~\ref{lem:computeSig} and~\ref{lem:correctReduce1} naturally lead to the following reduction rule. In particular Lemma~\ref{lem:correctReduce1} guarantees correctness, while Lemma~\ref{lem:computeSig} ensures that we can check in polynomial time whether the rule applies.
\begin{redrule}\label{rule1} 
Let $P$ be an interesting set and let $G'$ and $P'$ be obtained from $G$ and $P$ respectively by deleting a vertex of $P$, deleting an edge with at least one endponint in $P$, or contracting an edge with both endpoints in $P$. If the k-truncated signatures $\sigma_P^{k}$ of $P$ and $\sigma_{P'}^{k}$ of $P'$ are identical, that is $\sigma_P^{k}(B, {\cal Q}) = \sigma_{P'}^{k}(B, {\cal Q})$ for every $(B, {\cal Q})$, then reduce $(G,k)$ to $(G', k)$.
\end{redrule}

We now turn to the analysis of graphs to which Rule~\ref{rule1} can not be applied. Over the rest of this section we will show the following lemma.
\begin{lemma}\label{lem:compsize} There exists a constant $\gamma$ depending only on ${\cal F}$ such that for any graph $G$ on which Rule~\ref{rule1} does not apply, every component of $G \setminus X \cup Y$ has at most $\gamma \cdot k^\gamma$ vertices.
\end{lemma}

We say that an interesting vertex set $P \subseteq V(G) \setminus (X \cup Y)$ has a {\em signature gap} if the following holds: for every set $B \subseteq {\cal B}_P$ and collection ${\cal Q}$ of $t$-boundaried graphs of size at most $|B|+h$ whose boundary is a subset of $B$, either $\sigma_P(B,{\cal Q}) \leq |B|$ or $\sigma_P(B,{\cal Q}) \geq k+2$. We will prove that any large enough interesting set $P$ contains a large interesting subset $P'$ that has a signature gap. To that end we first prove a preliminary lemma. For a graph $H$ and vertex set $P \subseteq C$, define $\zeta_P(H)$ to be the size of the smallest vertex set $S \subseteq P$ such that $G[P \setminus S]$ excludes $H$ as a minor. 

\begin{lemma}\label{lem:fewOrManyMinors} There exist constants $\lambda$ and $\gamma$ depending only on ${\cal F}$ such that for any interesting set $P$ of size at least $\gamma k^\gamma$ there exists an interesting set $P' \subseteq P$ of size at least $\frac{|P|}{\lambda \cdot k^\lambda}$ and for every graph $H$ of size at most $5(\eta+1)+h$, either $\zeta_P(H) = 0$ or $\zeta_P(H) \geq k+2$.
\end{lemma}

\begin{proof}
Consider the nice tree-decomposition $(T, {\cal B} = \{B_t : t \in V(T)\})$ of width at most $\eta$ of $G[C]$, where $C$ is the component of $G \setminus (X \cup Z)$ that contains $P$. Let $Q$ be the subtree of $T$ such that $P = \{v \in V(G) : T_v \subseteq Q\}$. Since $P$ is an interesting set such a subtree $Q$ exists, and $|N(T_P)| \leq 2$. Now, set $M = N(Q)$. The set $M$ refers to {\em marked} nodes in $T$. We run and analyze an algorithm that will modify $Q$ and $M$. For a $Q$ define $P(Q)$ to be $\{v \in V(G) : T_v \subseteq Q\}$. We will maintain the following invariants: $N_T(Q) = M$, $|M| \leq 2$ and $Q$ induces a connected subtree of $T$. It is easy to verify that initially $Q$ and $M$ satisfy these invariants.

The algorithm proceeds as follows. If there does not exist a graph $H$ on at most $5(\eta+1)+h$ vertices such that $0 < \zeta_{P(Q)}(H) < k+2$ output $P(Q)$. Otherwise there exists a set $S \subseteq P(Q)$ of size at most $k+1$ such that $G[P(Q)] \setminus S$ excludes $H$ as a minor. We build the set $M'$ by starting from $M$ and then for each vertex $v \in S$ select a node $t \in Q$ such that $v \in B_t$ and add it to $M'$. Then, set $M^* = \lcac{M'}$. Let $Q'$ be the component of $Q \setminus M^*$ that maximizes $|P(Q')|$. Clearly $Q'$ is a connected subtree of $T$ and by Lemma~\ref{lem:lcaClosure}, $|N(Q')| \leq 2$. The algorithm changes $Q$ to $Q'$ and $M = N(Q')$, maintaining the invariants. Since the size of $Q$ drops in every iteration the algorithm eventually terminates, and when it terminates it satisfies that $P' = P(Q)$ is an interesting set and that for every graph $H$ of size at most $5(\eta+1)+h$, either $\zeta_{P'}(H) = 0$ or $\zeta_{P'}(H) \geq k+2$. Thus it remains to show that $|P'| \geq \frac{|P|}{\lambda \cdot k^\lambda}$.

For every iteration of the algorithm the set $P(Q)$ shrinks to a subset of itself. Hence if $\zeta_{P(Q)}(H) = 0$ for a graph $H$ at some step it will remain $0$ in all future steps. Furthermore the graph $H$ which had $0 < \zeta_{P(Q)}(H) < k+2$ before the iteration satisfies $\zeta_{P(Q)}(H) = 0$ after. Thus in every iteration the number of graphs $H$ of size at most $5(\eta+1)+h$ with $\zeta_P(H) = 0$ increases by at least one. Since the number of graphs on at most $5(\eta+1)+h$ vertices is at most $2^{(5(\eta+1)+h)^2}$, the algorithm must terminate after this many iterations.

Finally, in every iteration of the algorithm the size of the set $M'$ is at most $k+4$ and hence by Lemma~\ref{lem:lcaClosure} $M^*$ is at most $2k+8$. Since every vertex in $T$ has degree at most $3$ the number of components of $T[Q \setminus M^*]$ is at most $6k+24$. Hence 
$$P(Q') \geq \frac{|P(Q)| - (\eta+1)(2k+8)}{6k+24}.$$
As long as $|P(Q)| \geq 2(\eta+1)(2k+8)$ and $k \geq 1$ this is lower bounded by $\frac{|P(Q)|}{60k}$. Hence $|P(Q)|$ drops at most by a factor $60k$ in every iteration. Thus there exist constants $\lambda$ and $\gamma$ depending only on ${\cal F}$ such that for any interesting set $P$ of size at least $\gamma \cdot k^\gamma$, the size of $P'$ is at least $\frac{|P|}{\lambda \cdot k^\lambda}$.
\end{proof}

\begin{lemma}\label{lem:signatureGap}  There exist constants $\lambda$ and $\gamma$ depending only on ${\cal F}$ such that for any interesting set $P$ of size at least $\gamma k^\gamma$ there exists an interesting set $P' \subseteq P$ of size at least $\frac{|P|}{\lambda \cdot k^\lambda}$ such that $P'$ has a signature gap.
\end{lemma}

\begin{proof}
Set the constants $\gamma$ and $\lambda$ as guaranteed by Lemma~\ref{lem:fewOrManyMinors}, and let $P' \subseteq P$ be a set of size at least $\frac{|P|}{\lambda \cdot k^\lambda}$ as guaranteed by Lemma~\ref{lem:fewOrManyMinors}. That is, for every graph $H$ of size at most $5(\eta+1)+h$, either $\zeta_{P'}(H) = 0$ or $\zeta_{P'}(H) \geq k+2$. We prove that $P'$ has a signature gap.

If there is at least one graph $H \in {\cal Q}$ that has no boundary and $\zeta_{P'}(H) \geq k+2$ then $\sigma({\cal Q}, B) \geq k+2$ since any set $S$ of size less than $k+2$ can't hit all copies of $H$ as a minor in $G[{P'}]$, and hence can't hit all copies of $H$ as a $t$-boundaried minor of $G_{P'}^B$. On the other hand if no such graph $H$ exists then every $H \in {\cal Q}$ that has no boundary does not appear in $G_{P'}^B$ as a minor because $H$ is not a minor of $G[{P'}]$ and a model of $H$ in $G_{P'}^B$ can not contain boundary vertices. Furthermore, the set $B$ hits all $t$-boundaried graphs $H$ that contain at least one boundary vertex. Hence $G_{P'}^B \setminus B$ is ${\cal Q}$-free and so ${P'}$ has a signature gap.
\end{proof}

We now show that if $P$ is an interesting set with a signature gap and some vertex $x \in X$ has too many neighbours in $X$, then there exists an edge $e$ from $x$ to $P$ such that deleting $e$ does not affect the $k$-truncated signature of $P$. We first observe that we only need care about preserving the signature value for the pairs $(B, {\cal Q})$ such that $\sigma_P(B, {\cal Q}) \leq |B|$.

\begin{lemma}\label{lem:preserveSmall} Let $G$ be a graph and $P$ be an interesting set in $G$ with a signature gap. Let $G'$ be obtained from $G$ by deleting an edge $e$ with at least one endpoint in $P$, contracting an edge $e'$ with both endpoints in $P$ or deleting a vertex in $P$. Let $\chi_P$ and $\chi_P^k$ be the signature and the $k$-truncated signature of $P$ in $G'$ respectively. If $\sigma_P(B,{\cal Q}) = \chi_P(B,{\cal Q})$ whenever $\sigma_P(B,{\cal Q}) \leq |B|$, then $\sigma_P^k(B,{\cal Q}) = \chi_P^k(B,{\cal Q})$ for all $(B,{\cal Q})$.
\end{lemma}

\begin{proof}
Cleary $\sigma_P^k(B,{\cal Q}) \geq \chi_P^k(B,{\cal Q})$ for all $(B,{\cal Q})$ so it suffices to show that $\sigma_P(B,{\cal Q}) > |B|$ implies $\sigma_P^k(B,{\cal Q}) \leq \chi_P^k(B,{\cal Q})$. If $\sigma_P(B,{\cal Q}) > |B|$ we have that $\sigma_P(B,{\cal Q}) \geq k+2$ since $P$ has a signature gap. Then $\sigma_P^k(B,{\cal Q}) = \infty$ and $\chi_P(B,{\cal Q}) \geq k+1$ so  $\chi_P^k(B,{\cal Q}) = \infty$ as well, concluding the proof.
\end{proof}

Now we need a few definitions. For a pair of integers $t$, $\beta$ and finite set ${\cal Q}$ of $t$-boundaried graphs, define the class $\Pi_{\cal Q}^{(t,\beta)}$ to be the set of all $t$-boundaried graphs of treewidth at most $t$, such that every graph $G \in \Pi_{\cal Q}^{(t,\beta)}$ has a vertex set $S$ of size at most $\beta$ such that $G \setminus S$ is ${\cal Q}$-free. For every choice of $t$, $\beta$ and ${\cal Q}$ the corresponding class $G \in \Pi_{\cal Q}^{(t,\beta)}$ is closed under minors, since taking minors can not increase treewidth nor increase the size of the smallest ${\cal Q}$-deletion set in $G$. Thus, by Lemma~\ref{lem:wqoTBoundaried} for each such class $\Pi_{\cal Q}^{(t,\beta)}$ there is a finite set ${\cal F}_{\cal Q}^{(t,\beta)}$ such that a $t$-boundaried graph $G$ is in $\Pi_{\cal Q}^{(t,\beta)}$ if and only if it is ${\cal F}_{\cal Q}^{(t,\beta)}$-free. 

We also introduce the notion of a vital set in a minor model. Let ${\mathcal P}_H := \{P_1, \ldots, P_{|V(H)|} \}$ be a minor model of $H$ in $G$, and let $h := |V(H)|$. The vital trees of ${\mathcal P}_H$ are given by $\{T_1, \ldots, T_{|V(H)|} \}$, where $T_i \subseteq E(P_i)$ induces a spanning tree of $P_i$. Let ${\mathcal E}_H := \{e_i ~|~ 1 \leq i \leq |E(H)| \}$ denote a set of edges in $G$ which witness the adjacencies of $H$ in ${\mathcal P}_H$, that is, if $(i,j) \in E(H)$, then $e_i$ represents some $uv \in E(G)$ such that $u \in P_i$ and $v \in P_j$.  be The vital edges of $P_H$ are  defined as $\bigcup_{1 \leq i \leq |V(H)|} T_i \cup {\mathcal E}_H$. The following observation is easy to make:

\begin{observation}
\label{obs:vitalEdges}
Let ${\mathcal P}_H := \{P_1, \ldots, P_{|V(H)|} \}$ be a minimal minor model of $H$ in $G$, where $h := |V(H)|$. Let $\{T_1, \ldots, T_{|V(H)|} \}$ be the vital trees of ${\mathcal P}_H$, and let ${\mathcal E}_H$ be the set of vital edges. Then every $T_i$ has at most $h$ leaves, and no vertex in any $P_i$ is incident to more than $h$ vital edges.
\end{observation}

\begin{proof}
We first show that the vital trees of a minimal minor model have at most $h$ leaves. Suppose not, and let $T_i$ be the vital tree with more than $h$ leaves. This implies that there is at least one leaf vertex $v$ which is not adjacent to any vertex outside $P_i$. Note that $\{P_1, \ldots, P_i \setminus \{v\}, \ldots, P_{|V(H)|}\}$ continues to be a minor model of $H$, since deleting a leaf vertex from a spanning tree of $P_i$ does not affect the connectivity of $P_i$ and the adjacencies of $H$ are maintained since $v$ was not adjacent to any vertex outside $P_i$. This contradicts the assumption that  ${\mathcal P}_H$ was a minimal minor model of $H$.

Since $T_i$ has at most $h$ leaves, the degree of the internal vertices of $T_i$ (and hence, $P_i$) is bounded by $h$. A leaf vertex in $P_i$ has one neighbor in $P_i$ and is adjacent at most $(h-1)$ vertices outside $P_i$ (note that in a minimal minor model, a leaf of $T_i$ is never adjacent to more than one vertex in $P_j$, $j \neq i$). Therefore, every vertex in $P_i$ is adjacent to at most $h$ vital edges. 
\end{proof}

We are now ready to show that in a irreducible graph a vertex in $X$ can't have too many neighbours in an interesting set with a signature gap.

\begin{lemma}\label{lem:degreeReduce}
There exists a constant $\mu$ depending only on ${\cal F}$ such that if $x \in X$ has at least $\mu \cdot k^\mu$ neighbors in an interesting set $P$ that has a signature gap, then there is an edge $e$ from $x$ to $P$ such that the $k$-truncated signature $\sigma_P$ remains unchanged when $e$ is deleted from $G$.
\end{lemma}

\begin{proof}
Recall that $\sigma_P$ is the signature and $\sigma_P^k$ is the $k$-truncated signature of $P$ respectively. For an edge $e$ let $G' = G \setminus e$ and let $\chi_P$ be the signature of $P$ in $G'$. Similarly, let $\chi_P^k$ be the $k$-truncated signature of $P$ in $G'$. By Lemma~\ref{lem:preserveSmall} it is sufficient to find an edge $e$ incident to $x$ such that $\sigma_P(B,{\cal Q}) = \chi_P(B,{\cal Q})$ whenever $\sigma_P(B,{\cal Q}) \leq |B|$.

For every pair $(B, {\cal Q})$ such that $\sigma_P^{k}(B,{\cal Q}) \leq |B|$, set $\beta = \sigma_P^{k}(B,{\cal Q})$. We have that $G_P^B \in \Pi_{\cal Q}^{(\eta,\beta)}$, but also that $G_P^B \notin \Pi_{\cal Q}^{(\eta,\beta-1)}$. Hence $G_P^B$ contains a graph $H \in {\cal F}_{\cal Q}^{(\eta,\beta-1)}$ as a minor. Mark a vital set of edges of a minimal model of $H$ in $G_P^B$. Do this marking for every pair $(B, {\cal Q})$ such that $\sigma_P^{k}(B,{\cal Q}) \leq |B|$. The size of any graph $H$ in ${\cal F}_{\cal Q}^{(\eta,\beta-1)}$ is upper bounded by a function depending only on ${\cal F}$. Thus, for each pair $(B, {\cal Q})$, by Observation~\ref{obs:vitalEdges} the number of edges incident to $x$ that are marked is bounded by a constant depending only on ${\cal F}$. The number of pairs $(B, {\cal Q})$ is bounded by $O(k^{5(\eta+1)})$ and hence, the exists a constant $\mu$ depending only on ${\cal F}$ such that if $x \in X$ has at least $\mu \cdot k^\mu$ neighbors $P$, then at least one edge $e$ is left unmarked by the process above. We claim that $\sigma_P(B,{\cal Q}) = \chi_P(B,{\cal Q})$ whenever $\sigma_P(B,{\cal Q}) \leq |B|$.

Suppose for contradiction that $\sigma_P(B,{\cal Q}) \leq |B|$ but $\sigma_P(B,{\cal Q}) > \chi_P(B,{\cal Q})$ for some pair $(B,{\cal Q})$. Set $\beta =  \sigma_P^{k}(B,{\cal Q})$. We have that $G_P^B \notin \Pi_{\cal Q}^{(\eta,\beta-1)}$. Hence $G_P^B$ contains a graph $H \in {\cal F}_{\cal Q}^{(\eta,\beta-1)}$ as a minor. Furthermore, all vital edges of a model of $H$ in $G_P^B$ are marked. Since the edge $e$ is unmarked $H$ is also a minor of $G_P^B \setminus e$, contradicting that $\chi_P^{k}(B,{\cal Q}) \leq \beta-1$.
\end{proof}

Lemma~\ref{lem:signatureGap} allows us to find a large interesting set $P$ with a signature gap. Lemma~\ref{lem:degreeReduce} shows that in a graph where Reduction Rule~\ref{rule1} does not apply, every vertex in $X$ has few neighbours inside $P$. We now that if $P$ is a large interesting set and $X$ has few neoghbours in $P$ then $P$ contains a large interesting subset $P'$ that has a signature gap and contains no neighbours of $X$.

\begin{lemma}\label{lem:noXNeighbours} There exist constants $\rho$ and $\tau$ such that for any interesting set $P$ of size at least $\rho k^\rho \cdot (|P \cap N(X)|)$ there is an interesting set $P' \subseteq P$ of size at least $\frac{|P|}{\tau \cdot k^\tau \cdot |P \cap N(X)|}$, such that $P'$ has a signature gap and $P' \cap N(X) = \emptyset$.
\end{lemma}
\begin{proof}
Since $P$ is interesting there is a subtree $Q$ of $T$ such that $|N(Q)| \leq 2$ and $P = \{v \in C~:~T_v \subset Q\}$. Initially, set $M = N(Q)$ and then, for each $v \in N(X) \cap P$ we insert a node in $T_v$ into $M$. We let $M^* = \lcac{M}$ and let $Q^*$ be the component of $Q \setminus M$ that maximizes $|\{v \in C: T_v \subseteq Q^*\}|$. Set $P^* = \{v \in C: T_v \subseteq Q^*\}$. We have that $|M| \leq |P \cap N(X)|$ and by Lemma~\ref{lem:lcaClosure} $M^* \leq 2|M|$. Thus the number of connected components of $Q \setminus M$ is at most $6|P \cap N(X)|$. From this is follows that $|P^*| \geq \frac{|P| - (\eta+1)2|P \cap N(X)|}{6|P \cap N(X)|} \geq \frac{|P|}{12|P \cap N(X)|}$. Now, $P^*$ is an interesting set, as evidenced by $Q^*$ and the fact that $|N(Q^*)| \leq 2$. Furthermore, $P^* \cap N(X) = \emptyset$. If $\rho$ is chosen appropriately then we can invoke Lemma~\ref{lem:signatureGap} and obtain an interesting set $P' \subseteq P$ with a signature gap. The size of $P'$ is lower bounded by $\frac{|P|}{\tau \cdot k^\tau \cdot |P \cap N(X)|}$ for a constant $\tau$.
\end{proof}

Now we show that Reduction Rule~\ref{rule1} will reduce large enough interesting sets with a signature gap and no neighbors in $X$.

\begin{lemma}\label{lem:reduceNoNeighbors} There exists a constant $\omega$ depending only on ${\cal F}$ such that for any interesting set $P$ such that $|P| > \omega$, $P$ has a signature gap and $P \cap N(X) = \emptyset$, there is an edge $e$ with both endpoints in $P$ such that deleting or contacting $e$ does not change the k-truncated signature of $P$, or a vertex $v$ such that deleting $v$ does not change the $k$-truncated signature of $P$.
\end{lemma}

\begin{proof}
We show the existence of a minor operation inside $P$ that produces a graph $G'$ with the following properties. Let $\chi_P$ and $\chi_P^k$ BE the signature and the $k$-truncated signature of $P$ in $G'$ respectively. Then, if $\sigma_P(B,{\cal Q}) \leq |B|$ then $\sigma_P(B,{\cal Q}) = \chi_P(B,{\cal Q})$. By Lemma~\ref{lem:preserveSmall} this implies that the minor operation in fact preserves the $k$-truncated signature of $P$.

For each pair $(B, {\cal Q})$ such that $\sigma_P(B,{\cal Q}) \leq |B| \leq 5(\eta + 1)$, let $\beta = \sigma_P(B,{\cal Q})$. We have that $G_P^B$ contains a boundaried graph $H_{B,{\cal Q}} \in {\cal F}_{\cal Q}^{(t,\beta-1)}$ as a minor. Thus $G_P^{N(P)}$ contains $H_{B,{\cal Q}}$ as a minor as well. Let ${\cal R}$ be the collection of all graphs $H_{B,{\cal Q}}$ for pairs $(B, {\cal Q})$ such that $\sigma_P(B,{\cal Q}) \leq |B|$. If $G_P^{N(P)}$ contains all graphs in ${\cal R}$ even after the minor operation, then $\sigma_P(B,{\cal Q}) = \chi_P(B,{\cal Q})$ for all $(B, {\cal Q})$ such that $\sigma_P(B,{\cal Q}) = \chi_P(B,{\cal Q})$. 

Consider the collection ${\cal S}$ of $5(\eta + 1)$ boundaried graphs of treewidth at most $\eta$, such that each graph $J \in {\cal S}$ contains all graphs in ${\cal R}$ as a minor, but no minor of $J$ contains all graphs in ${\cal R}$ as a minor. No two graphs in ${\cal S}$ are minors of each other and hence by Lemma~\ref{lem:wqoTBoundaried} ${\cal S}$ is finite. Let $\omega$ be the size of the largest graph in ${\cal S}$. Since $\omega$ only depends on the values of $\sigma_P(B,{\cal Q})$, each value of $\sigma_P(B,{\cal Q})$ is bounded by $5(\eta+1)$ and the number of choices for $(B,{\cal Q})$ is a function of $\eta$ only, $\omega$ is upper bounded by a function of ${\cal F}$. Furthermore, $G_P^{N(P)}$ contains all graphs in ${\cal R}$ as a minor, and if $|P| > \omega$, $G_P^{N(P)}$ has a minor that also contains all graphs in ${\cal R}$ as minors. Thus there is a minor opreration that ensures that if $\sigma_P(B,{\cal Q}) \leq |B|$ then $\sigma_P(B,{\cal Q}) = \chi_P(B,{\cal Q})$, proving the lemma.
\end{proof}

We are now in position to tie the results of the section together and prove Lemma~\ref{lem:compsize}.
\begin{proof}[Proof of Lemma~\ref{lem:compsize}] If $|C| \geq \gamma \cdot k^\gamma$ then by Lemma~\ref{lem:signatureGap} there is an interesting set $P$ of size $\frac{\gamma}{\lambda} \cdot k^{\gamma - \lambda}$ with a signature gap. If some vertex $x \in X$ has at least $\mu \cdot k^\mu$ neigbours in $P$ then by Lemma~\ref{lem:degreeReduce}, Rule~\ref{rule1} applies to $P$. If no vertex $x \in X$ has at least $\mu \cdot k^\mu$ neigbours in $P$ then by Lemma~\ref{lem:noXNeighbours} $P$ has an interesting subset $P'$ with a signature gap such that $|P'| \geq \frac{\gamma}{\lambda \cdot \mu \cdot \tau} \cdot k^{\gamma - \lambda - \mu - \tau}$ and $P \cap N(X) = \emptyset$. For a sufficiently large choice of $\gamma$ (still depending only on ${\cal F}$), we have that $|P'| > \omega$ and hence by Lemma~\ref{lem:reduceNoNeighbors}, Rule~\ref{rule1} applies on $P'$.
\end{proof}

\subsection{Reducing the number of near-Protrusions}
%!TEX root=PFDMerge-KernelAlgo.tex

If Rule~\ref{rule1} can not be applied but the input graph $G$ is still large, the number of components of $G \setminus X$ must be large. We now show that if the number of components is too large then we can remove a vertex from one of them. For a component $C$, set $B \in {\cal B}_C$ and a boundaried graph $H$ with $B$ as boundary, we say that $C$ {\em realizes} $(B,H)$ if $H \leq_m G_C^B$. We say that a pair $(B, H)$ is {\em rich} if there are at least $|X|+|Z|+k+(h+3(\eta+1))^2+2$ components. We prove the following lemma, which naturally leads to a reduction rule.

\begin{lemma}\label{lem:reduceNumComponents}
Let $C$ be a component of $G \setminus (X \cup Z)$ such that every pair $(B, H)$ that $C$ realizes is rich. Then, for any vertex $v \in C$, $G \setminus v$ has an {\cal F}-deletion set of size at most $k$ if and only if $G$ does.
\end{lemma}

\begin{proof}
If $G$ has a ${\cal F}$-deletion set  $S$ of size at most $k$, then $S \setminus \{v\}$ is a ${\cal F}$-deletion set  of $G$, hence it suffices to show the reverse direction. Assume now that $G \setminus v$ has an ${\cal F}$-deletion set  $S$ of size at most $k$. We argue that $S$ is in fact a ${\cal F}$-deletion set  of $G$ as well. Suppose not, then $S \cap v$ is an ${\cal F}$-deletion set  of $G$ of size at most $k+1$ and hence by Lemma~\ref{lem:nearProtrusion}, $|(N(C) \cap X) \setminus S| \leq \eta + 1$. Set $B = N(C) \setminus S$ and $R = V(G) \setminus (C \cup S \cup B)$. We have that $|B| \leq 3(\eta + 1)$ (and hence $B \in {\cal B}_C$) and that $G \setminus S = G_C^B \oplus G_R^B$. Since $G \setminus S$ is not ${\cal F}$-free, by Lemma~\ref{lem:folioGlue} there are boundaried graphs $H_1$ and $H_2$ on at most $h+3(\eta + 1)$ vertices each such that $H \leq_m H_1 \oplus H_2$, $H_1 \leq_m G_C^B$ and $H_2 \leq_m G_R^B$.

Suppose that we show the existence of a component $C' \neq C$ of $G \setminus (X \cup Z)$ such that $C' \subseteq R$ such that $C' \cap S = \emptyset$, $H_1 \leq_m G_{C'}^B$ and $H_2 \leq_m (G_R^B \setminus C')$. Then Lemma~\ref{lem:folioGlue} would yield that $H$ is a minor of $G_{C'}^B \oplus (G_R^B \setminus C') = G \setminus (S \cup C)$, contradicting that $S \cup v$ is an ${\cal F}$-deletion set  of $G$. It remains to find such a component $C'$. We prove the lemma assuming the following claim.

\begin{claim}\label{clm:modelComponents} Any minimal minor model of $H_2$ in $G_R^B$ has a non-empty intersection with at most $|X|+|Z|+(h+3(\eta+1))^2$ components of $G \setminus (X \cup Z)$. \end{claim}

Since $C$ realizes $(B,H_1)$, we have that $(B,H_1)$ is rich and so there are at least $|X|+|Z|+k+(h+3(\eta+1))^2+2$ components of $G \setminus (X \cup Z)$ that realize $(B,H_1)$. Consider any minimal minor model of $H_2$ in $G_R^B$. By Claim~\ref{clm:modelComponents} there are at least $k+2$ components of $G \setminus (X \cup Z)$ that realize $(B,H_1)$ and are disjoint from this model of $H_2$. At least two of these are disjoint from $S$, and at least one of these two is not $C$. Let this component be $C'$. We have that $C' \neq C$,  $C' \cap S = \emptyset$, $H_1 \leq_m G_{C'}^B$ and $H_2 \leq_m (G_R^B \setminus C')$. This completes the proof of the lemma, up to proving Claim~\ref{clm:modelComponents}.

\begin{proof}{Proof of Claim~\ref{clm:modelComponents}.} 
Consider a minimal model $(P_1, P_2, \ldots, P_\ell)$ of $H_2$ in $G_R^B$. Here $\ell = |V(H_2)| \leq h+3(\eta+1)$. For every $i \leq \ell$ select a spanning tree $T_i$ of $G[P_i]$. Since $(P_1, P_2, \ldots, P_\ell)$ is a minimal model, each tree $T_i$ has at most $\ell$ leaves. Consider the connected components of $T_i \setminus (X \cup Z)$. Each component has at least one edge to $X \cup Z$ and the components that do not contain leaves have at least two edges to $X \cup Z$. Counting the total number of vertices and edges incident to each component and using that $E(T_i) = V(T_i) + 1$ yields that the number of components of  $T_i \setminus (X \cup Z)$ is at most $|V(T_i) \cap (X \cup Z)| - 1 + \ell - 1$. Since the intersection of $T_i$ with different components of $G \setminus (X \cup Z)$ must be in different components of $T_i \setminus (X \cup Z)$ it follows that the number of components that intersect with $T_i$ is at most $|V(T_i) \cap (X \cup Z)| - 1 + \ell - 1$ as well. Hence the number of components with non-empty intersection with the model of $H_2$ is at most $\sum_{i \leq \ell} |V(T_i) \cap (X \cup Z)| + \ell - 2 \leq |X \cup Z| + \ell^2$. Since $\ell \leq h+3(\eta+1)$ this proves the claim.
\end{proof}
\end{proof}

For each component $C$ of $G \setminus (X \cup V)$ and set $B$ of size at most $3(\eta + 1)$ the boundaried graph $G_C^B$ has constant treewidth. For every boundaried graph $H$ on at most  $h+3(\eta + 1)$ vertices we can encode whether $G_C^B$ contains $H$ as a minor in monadic second order logic. Thus for each choice of $C$, $B \in {\cal B}_C$ and $H$ we can test whether $C$ realises a pair $(B, H)$ in linear time.  Since the number of pairs $B \in {\cal B}_C$ is bounded by a polynomial and the number of boundaried graphs on at most  $h+3(\eta + 1)$ vertices is constant this means we can test in polynimial time whether there exists a component $C$ of $G \setminus (X \cup Z)$ such that every pair $(B, H)$ that $C$ realizes is rich, and find such a component if it does. This, together with Lemma~\ref{lem:reduceNumComponents} yields the following reduction rule.

\begin{redrule}\label{rule2} If there is a component $C$ of $G \setminus (X \cup Z)$ such that every pair $(B, H)$ that $C$ realizes is rich, remove an arbitrary vertex $v$ from $C$. \end{redrule}

Lemma~\ref{lem:reduceNumComponents} shows that Rule~\ref{rule2} is safe, it remains to show that if the number of components of $G \setminus (X \cup Y)$ is too large, then Rule~\ref{rule2} applies.

\begin{lemma}\label{lem:boundComponentNumber} There exists a constant $\theta$ depending only on ${\cal F}$ such that if the number of components of $G \setminus (X \cup Z)$ is at least $\theta \cdot k^\theta$ then Rule~\ref{rule2} applies. \end{lemma}
\begin{proof}
If Rule~\ref{rule2} does not apply then every component $C$ can realize at least one pair $(B, H)$ that is not rich. The number of choices for $B$ is at most $|X \cup Z|^{3(\eta+1)}$ while the number of choices for $H$ is at most $2^{3(\eta+1)+h \choose 2}$. For each choice of $(B, H)$ that is poor, at most $|X|+|Z|+k+(h+3(\eta+1))^2+1$ components realize that pair. Thus by the pigeon hole principle the number of components is at most $\theta \cdot k^\theta$ for some constant $\theta$ depending only on ${\cal F}$.
\end{proof}

Lemmata~\ref{lem:decomposition}, \ref{lem:compsize} and~\ref{lem:boundComponentNumber} together imply that every \fd{} problem has a polynomial kernel. Specifically, Lemma~\ref{lem:decomposition} proves that in a reduced graph $|X \cup Z| \leq O(k^3)$, Lemma~\ref{lem:compsize} shows that the size of each component of $G \setminus (X \cup Z)$ is $k^{O(1)}$ while Lemma~\ref{lem:boundComponentNumber} shows that the number of components of  $G \setminus (X \cup Z)$ is $k^{O(1)}$. This proves the following theorem.

\begin{theorem}\label{thm:kernelmain} Every \fd{} problem has a randomized polynomial kernel. \end{theorem}

We now turn to two interesting consequences of Theorem~\ref{thm:kernelmain}. The first concerns a bound on the obstructions for the family  $ {\cal{G}}_{{\cal{F}},k}$, which contains all yes instances of the  \fd{} problem with parameter $k$. We first make the following observation:

\begin{observation}
\label{obs:approx}
For a graph $G \notin  {\cal{G}}_{{\cal{F}},k}$, let $H$ be some graph in the obstruction set of $ {\cal{G}}_{{\cal{F}},k}$, and let $H^*$ be a minimal minor model of $H$ in $G$. Then, $H^* \in {\cal{G}}_{{\cal{F}},k+1}$. 
\end{observation}

\begin{proof}
Clearly, $H^* \notin {\cal{G}}_{{\cal{F}},k}$. If $H^* \notin {\cal{G}}_{{\cal{F}},k+1}$, then $H^* \in {\cal{G}}_{{\cal{F}},j}$ for some $j > k+1$, and this contradicts the minimality of $H^*$. 
\end{proof}

The first step in the kernelization algorithm used the approximation algorithm to reject all instances where the size of the optimal solution was at least $(k+2)$. Therefore, by Observation~\ref{obs:approx}, all minimal minor models of graphs in the obstruction set of  ${\cal{G}}_{{\cal{F}},k}$ are present in any non-trivial reduced instance. The other two reduction rules performed by the kernelization algorithm involve only minor operations, which clearly do not destroy minimal minor models of the obstruction graphs. Therefore, we have the following corollary.

\begin{corollary} When  $\cal{F}$ contains a planar graph,   every minimal obstruction for   $ {\cal{G}}_{{\cal{F}},k}$ is of size  polynomial in $k$. 
 \end{corollary}

By standard arguments that translate a kernelization algorithm into a FPT algorithm, we also have the following consequence.

\begin{corollary}\label{cor:fptalgo} Every \fd{} problem admits an $O(k^k)$ FPT algorithm. \end{corollary}

Finally a note about computability; the implementation of the decomposition algorithm and the two rules is entirely constructive. That is, there is an algorithm that given $G$, $k$ and ${\cal F}$, runs in time $O(n^c)$ and outputs a reduced instance. Here the constant $c$ depends on ${\cal F}$ and is upper bounded by $2^{2^{h^{10}}}$, where $h$ is the size of the largest graph in ${\cal F}$. A slightly tighter analysis reveals that the running time is actually $n^2 \cdot k^c$ which of course is still horrendous, but at least it is quadratic for fixed $k$. The {\em size} of the kernel, however, is not explicit. Several of the constants that go into the proof of Lemma~\ref{lem:compsize} depend on the size of the largest graph in certain antichains in a well-quasi-order and thus we dont know what the (constant) exponent bounding the size of the kernel is. We leave it to future work to make also the size of the kernel explicit.

 \section{Conclusions and open problems}
The techniques of fast protrusion reductions developed for \fd{}  have a broader spectrum of applications which we mention briefly. 
  By combining results from  \cite{FominLST10} with fast protrusion reducers, we have that 
 kernelization algorithms on apex-free  and $H$-minor free graphs  for
 \emph{all} bidimensional problems from  \cite{FominLST10}  can be implemented in linear time if we use randomized protrusion reducer  and in time $O(n\log^{2}{n})$ when we use deterministic reducer.  This gives randomized linear time  linear kernels for a multitude of problems.

 In the framework for obtaining EPTAS  on $H$-minor-free graphs in \cite{FominLRS11}, the running time
of approximation algorithms for many problems is   $f(1/\varepsilon) \cdot n^{O(g(H))}$, where $g$ is some function of $H$ only.  The only bottleneck for improving polynomial time dependence in  \cite{FominLRS11} is Lemma~4.1, which gives a constant factor approximation algorithm for  {\sc Treewidth $\eta$-Deletion} or  {\sc
$\eta$-Transversal} of running time  $n^{O(g(H))}$. Now instead of that algorithm, we can use the algorithm from 
Theorem~\ref{thm:approx_thm}, which runs in time $O(n^2)$.  Therefore each EPTAS from   \cite{FominLRS11} runs in time   $O(f(1/\varepsilon) \cdot n^{2})$. For the same reason, PTAS for many problems on unit disc and map graphs from 
\cite{FominLS12} become EPTAS. 

 %\medskip
 
Finally, an  interesting   direction for further research is to investigate  \fd{} when none of the graphs in $\cal{F}$ is planar. The most interesting case here is when  ${\cal{F}} =\{K_5, K_{3,3}\}$ aka the  \textsc{Vertex Planarization} problem. Surprisingly, we are not aware even of  a single case of  \fd{}  with ${\cal{F}} $ containing no planar graph admitting   either  constant factor approximation, or   polynomial kernelization, or parameterized single-exponential  algorithms.  
%
%Even for very special case 
% ${\cal{F}}=K_5$ we do not know if   constant factor approximation or polynomial kernelization algorithms are  possible. 
  It is tempting to conjecture that the line of tractability is determined by whether  ${\cal{F}} $ contains a planar graph or not. 
\bibliographystyle{abbrv}
{\bibliography{kernels,references}}

\end{document}